\documentstyle[preprint,aps,prd,eqsecnum,tighten,amsfonts,amssymb,epsfig]{revtex} 
\newcommand{\Mpl}{\text{$M_{\text{{\tiny P}}}$}}

\newcommand{\phih}{\text{$\hat{\phi}$}}
\newcommand{\phihp}{\text{$\hat{\phi}_{+}$}}
\newcommand{\phihm}{\text{$\hat{\phi}_{-}$}}
\newcommand{\phihpm}{\text{$\hat{\phi}_{\pm}$}}
\newcommand{\vphi}{\text{$\varphi$}}
\newcommand{\vecphi}{\text{$\vec{\phi}$}}

\newcommand{\sqmg}{\text{$\sqrt{-g}$}}
\newcommand{\sqmgp}{\text{$\sqrt{-g'}$}}

\begin{document}
\draft
\preprint{UMDPP\#97-084}
\title{Nonequilibrium inflaton dynamics and reheating:\\
Back reaction of parametric particle creation and
curved spacetime effects}
\author{S. A. Ramsey\thanks{Electronic address: 
{\tt sramsey@physics.umd.edu}} and 
B. L. Hu\thanks{Electronic address: 
{\tt hu@umdhep.umd.edu}}}
\address{Department of Physics, University of Maryland,
College Park, Maryland 20742-4111}
\date{\today}
\maketitle
\begin{abstract}
We present a detailed and systematic analysis of the nonperturbative,
nonequilibrium dynamics of a quantum field in the reheating phase of
inflationary cosmology, including full back reactions of the quantum field
on the curved spacetime, as well as the fluctuations on the mean field.
We use the O$(N)$ field theory with unbroken symmetry
in a spatially flat Friedmann-Robertson-Walker (FRW) universe to study
the dynamics of the inflaton in the post-inflation, preheating stage.
Oscillations of the inflaton's zero mode induce parametric amplification
of quantum fluctuations, resulting in a rapid transfer of energy to the
inhomogeneous modes of the inflaton 
field.  The large-amplitude oscillations 
of the mean field, as well as stimulated emission effects
require a nonperturbative formulation of the quantum dynamics, while the
nonequilibrium evolution 
requires a statistical field theory treatment.
We adopt the coupled nonperturbative equations for the mean field and
variance derived in a preceding paper \cite{ramsey:1997a} by means of
a two-particle-irreducible (2PI), closed-time-path (CTP) effective action
for curved spacetimes while specialized to a dynamical FRW background,
up to leading order in the $1/N$ expansion. Adiabatic regularization is
employed to yield a covariantly conserved, renormalized energy-momentum tensor.
The renormalized dynamical equations are evolved numerically from initial data
which are generic to the end state of slow roll in many inflationary
cosmological scenarios. The initial conditions consist of
a large-amplitude, quasiclassical, oscillating mean field $\langle\Phi\rangle$
with variance $\langle\Phi^2\rangle - \langle \Phi \rangle^2$ around the
de~Sitter-invariant vacuum. We find that for sufficiently large initial
mean-field amplitudes $\gtrsim \Mpl/300$ (where $\Mpl$ is the Planck mass)
in this model, the parametric resonance effect alone (in a collisionless
approximation) is not an efficient means to ``preheat'' the quantum field.
For small initial mean-field amplitude, damping of the mean field 
via parametric amplification of quantum fluctuations 
is seen to occur, and in this case 
can be adequately described by prior analytic studies with approximations
based on field theory in Minkowski spacetime.
Our results indicate that the self-consistent dynamics of spacetime plays
an important role in determining the physics of the post-inflationary Universe.
This study calls into question the validity of general claims made without full
consideration of the self-consistent dynamics of spacetime and quantum fields.
\end{abstract}
\newpage

\section{Introduction and summary}
\label{sec-intro}
The inflationary Universe 
\cite{guth:1981a,sato:1981a,albrecht:1982b,linde:1982a,linde:1985a,starobinsky:1986a,bardeen:1987a,brandenberger:1985a,abbott:1986a,kolb:1990a,linde:1990a,linde:1996a} 
has for over a decade been the new
paradigm for addressing many basic issues in cosmology such as the 
flatness-entropy problem, the horizon-homogeneity problem, and the
fluctuation-structure formation problem. 
The linkage between observations, especially
those from the recent Cosmic Background Explorer (COBE) 
data, and theory, based on grand unified theories
(GUT's) and Friedmann-Robertson-Walker-- (FRW--)de~Sitter 
models, has been pursued in earnest, but most
theoretical discussions to date are largely phenomenological and somewhat
utilitarian in nature 
\cite{steinhardt:1984a,lidsey:1995a}. This lack of rigor and precision
is understandable for at least two reasons: the precise physical
conditions between the Planck and GUT scales
(when the most cosmologically significant inflationary evolutions
are believed to have taken place) have not been clearly understood, and the
theoretical framework for the treatment of processes affecting the inception
and completion of inflation were not well developed. As stressed  by
one of us earlier \cite{hu:1986d,hu:1986a}, the important physical
processes which can determine whether inflation can occur,
sustain, and finish with the necessary features are affected by at least
three aspects: the geometry, topology, and dynamics of the spacetime
\cite{birrell:1982a},
the quantum field theory aspects pertaining to the analysis of infrared
behavior, and the statistical mechanical aspects pertaining to
nonequilibrium processes.
These quantum and statistical processes include phase transition, particle
creation, entropy generation, fluctuation or stochastic dynamics, and structure
formation  \cite{hu:1993b,hu:1995b}.
 Most of these invoke the quantum field and statistical mechanical
aspects, and for processes occurring at the Planck scale (which are 
instrumental in starting certain models of inflation, such as proposed in
\cite{linde:1985a,starobinsky:1980a,hu:1986b}), 
also the geometry and topology of spacetime.
Two important problems involving field theory in curved spacetime
\cite{birrell:1982a}, namely, the back reaction of cosmological particle
creation 
\cite{parker:1969a,sexl:1969a,zeldovich:1970a,zeldovich:1971a,hu:1974a} 
on the structure and dynamics of spacetimes
\cite{starobinsky:1980a,zeldovich:1971a,zeldovich:1977a,hu:1977a,hu:1978a,fischetti:1979a,hartle:1979a,hartle:1980a,hu:1981a,anderson:1984a},
and the effects of geometry and topology of spacetime on cosmological
phase transitions
\cite{hu:1986a,hu:1986b,oconnor:1983a,shen:1985a,hu:1987b,berkin:1992a},
were investigated systematically and comprehensively in the 1970s and 1980s.
The statistical mechanical aspect has not been considered with equal mastery.

The statistical mechanical aspect enters into all three stages of inflationary
cosmology:
(i) At the inception: What conditions would be most conducive to starting
inflation? Do there exist metastable states for the Higgs boson field which can
generate inflation \cite{hu:1986c}? 
 Can thermal or quantum
fluctuations assist the inflaton in hopping or tunneling out of the
potential barrier in the spinodal or nucleation pictures? Most depictions so
far have been based on the finite temperature effective potential, 
which assumes an unrealistic equilibrium condition and a constant background 
field.  However, when asking such questions in critical dynamics one
should be using a Langevin or Fokker-Planck equation (a generalized
time-dependent Landau-Ginzberg equation \cite{calzetta:1997a})
incorporating dynamic dissipation and intrinsic noise consistently.
(ii) During inflation, the dynamics of the inflaton field can be more
easily understood
in terms of a Kadanoff-Migdal exponential scaling transform \cite{hu:1993d}.
The reason why the inflaton evolves as a classical stochastic field 
\cite{guth:1985a,cornwall:1988a,buryak:1996a} at late times involves the
process of decoherence,
caused by noise and fluctuations from environmental fields \cite{hu:1996a};
this necessitates statistical mechanical considerations. 
The evolution of the classical density contrast (containing the seedings
of structures) from quantum fluctuations of the inflaton also requires
both quantum and stochastic field theory considerations
\cite{guth:1982a,starobinsky:1982a,hawking:1982a,bardeen:1983a,brandenberger:1983a,mukhanov:1992a,deruelle:1992a,hu:1993a,calzetta:1995a,matacz:1996a}.
(iii) In the reheating epoch, particle creation induces
dissipation of the inflaton field, and the interaction of quantum fields is the source
for reheating the Universe. This last epoch is the focus of much recent work,
as we shall detail below.

The construction of a viable theoretical framework for treating quantum
statistical processes in the  early Universe has been the aim of research
of one of us for the past decade (for a review, see \cite{hu:1994a}).
This framework has now been successfully established,
and its application to the problems mentioned above has just begun.
The cornerstones are the Schwinger-Keldysh closed-time-path (CTP)
\cite{schwinger:1961a,bakshi:1963a,keldysh:1964a,niemi:1981a,niemi:1984a,landsman:1987a,zhou:1985a,su:1988a,dewitt:1986a,jordan:1986a,calzetta:1987a,calzetta:1988b,calzetta:1989a}
effective action and the Feynman-Vernon influence functional 
\cite{feynman:1963a,feynman:1965a,caldeira:1983a,grabert:1988a,hu:1992a,hu:1993c,calzetta:1994a}
formalisms.
They are useful for treating particle creation back reaction 
\cite{calzetta:1987a,calzetta:1989a},
fluctuation or noise, and dissipation or entropy problems 
\cite{hu:1993a,calzetta:1995a,calzetta:1994a}.
Other essential ingredients include the Wigner function 
\cite{wigner:1932a,hillery:1984a}, the $n$-particle-irreducible 
($n$PI) effective action
\cite{cornwall:1974a,hu:1987b,calzetta:1988b,ramsey:1997a},
and the correlation hierarchy
\cite{calzetta:1993a,calzetta:1995b}
for treating kinetic theory processes
\cite{calzetta:1988b,calzetta:1988a} and phase transition problems
\cite{calzetta:1989b,calzetta:1995a}.
In this and three following papers we apply these techniques to the
problems of inflaton dissipation due to parametric particle creation
\cite{ramsey:1997c,ramsey:1997d}
and reheating due to particle interaction \cite{ramsey:1997b}
(in the third epoch depicted above). In parallel,
these newly developed methods in statistical field theory are now being applied
to derive the classical  stochastic dynamics of the inflaton (in the second
epoch) \cite{hu:1996a},
and the statistical field theory of spinodal
decomposition (in the first epoch) \cite{calzetta:1997a}.

\subsection{Background and issues}
\label{sec-backissu}
All inflationary cosmologies share the feature of a period of cosmic
expansion driven by a nearly constant vacuum energy density $\rho$
(a ``vacuum-dominated'' era with effective equation of state, $p = -\rho$):
In a Friedmann-Robertson-Walker (FRW) spacetime,
the scale factor expands exponentially in cosmic time, resulting in 
extreme redshifting of the energy density of all other
forms of matter and fields. As long as the interaction time scale of any 
physical process involving given fields is longer than the cosmic
expansion time $H^{-1}$, the fields will remain in disequilibrium.
This condition can prevail in all three stages of inflation,
and one should use a fully nonequilibrium, nonperturbative treatment of
the dynamics of the inflation field.
The physics of the reheating epoch is important because it directly
determines several important cosmological parameters which are relevant
to later evolution of the Universe, and in principle verifiable by
observational data. For example, the reheating temperature is a vital link
between the inflationary Universe scenario and GUT scale baryogenesis
\cite{kolb:1996a}.

It is generally believed that at the end of inflation, the state of the
inflaton field can be approximately described by a condensate
of zero-momentum particles
undergoing coherent quasioscillations about the true minimum
of the effective potential \cite{kolb:1990a,linde:1990a,traschen:1990a}.  
The reheating problem involves describing the processes  by which the many 
light fields coupled to the inflaton become populated with quanta, and
eventually thermalize.  It is commonly 
believed that if the fields interact sufficiently rapidly and strongly,  
the Universe thermalizes and turns into the radiation-dominated condition 
described by the standard Friedmann solution, but this has not been proven
satisfactorily.

There has been a great deal of work over the past 15 years on the reheating
problem, and in attempting to understand reheating, a wealth of interesting
physics has been revealed (see, e.g., \cite{kofman:1996a}).
To date, the work on
particle production during reheating largely follows two distinct approaches,
each pursued in two stages.

In the first stage of work on the reheating problem
(group 1A, \cite{abbott:1982a,albrecht:1982a,dolgov:1982a,dolgov:1990a}),
time-dependent perturbation theory was used to compute the rate of particle
production into light fields (usually fermions) coupled to the
inflaton.  Particle production rates were computed in flat space assuming
an eternally sinusoidally oscillating inflaton field.  The inflaton evolution
in FRW spacetime was modeled with a phenomenological c-number 
equation involving the Hubble parameter $H$ and the classical inflaton
amplitude $\phi$,
\begin{equation} 
\frac{d^2\phi}{dt^2} + m^2 \phi + (\Gamma + 3H) \frac{d\phi}{dt} = 0,
\label{eq-phenom}
\end{equation}
where $\Gamma$, given by the imaginary part of the self-energy of $\phi$, is
the total perturbative decay rate. Bose enhancement of particle production
into the spatial Fourier modes of the inflaton fluctuation field $\vphi$
(and light Bose fields coupled to the inflaton) was not taken into account.  

In the second stage of this first approach to the reheating problem
(group 1B, \cite{kofman:1994a,shtanov:1995a,dolgov:1994a,allahverdi:1997a})
Eq.\ (\ref{eq-phenom}) was still utilized to model the
mean-field dynamics, but with $\Gamma$ computed  beyond first-order in
perturbation theory.  In the work of Shtanov, Traschen, and 
Brandenberger \cite{shtanov:1995a} and Kofman, Linde, and Starobinsky (KLS)
\cite{kofman:1994a}, $\Gamma$ was computed for a real self-interacting scalar 
inflaton field $\phi$ which was both Yukawa-coupled to
a spinor field $\psi$, and bi-quadratically coupled to a scalar field $\chi$
(KLS studied both the $\phi \rightarrow -\phi$ symmetry-breaking and
unbroken symmetry cases).  From the one-loop equations for the quantum
modes of the $\chi$, $\vphi$, and $\psi$ fields (in which the mean field
$\phih$ appears quadratically as an effective mass), approximate expressions
for the growth
rate of occupation numbers were derived, assuming a quasi-oscillatory
mean field $\phih$. For bosonic decay-product fields, it was found that
first-order time-dependent perturbation theory drastically underestimates
the particle production rate for modes which are in an instability band (due to
parametric resonance).
Parametric amplification of quantum fluctuations in Bose decay-product fields
can result in rapid out-of-equilibrium transfer of energy from the inflaton
mean field to the (spatially) inhomogeneous inflaton modes and light Bose
fields coupled to the inflaton.  This phenomenon was called 
{\em preheating\/} by
KLS.  It has been suggested that exponential growth of 
quantum fluctuations can in some cases lead to out-of-equilibrium 
(nonthermal) symmetry restoration in the ``new'' inflation models with a 
spontaneously broken symmetry \cite{kofman:1996c,tkachev:1996a}.
(See, however, the work of Boyanovsky
{\em et al.,\/} which reached a different conclusion on the possibility of 
nonequilibrium symmetry restoration \cite{boyanovsky:1996b}.)
This may have interesting implications for baryogenesis,
defect formation, and generation of primordial density perturbations 
\cite{kofman:1996a,kofman:1994a,tkachev:1996a}.

In both stages of this first approach, the back reaction of the variance
of the inflaton on the mean-field dynamics, and of the variance on the
quantum mode functions, were not treated self-consistently.
The effect of spacetime dynamics was either excluded
entirely, or not included self-consistently using the semiclassical Einstein
equation.  
Due to the potentially large initial inflaton amplitude at the onset of 
reheating, particularly in the case of chaotic inflation \cite{linde:1990a},
the effect of cosmic expansion on quantum particle production
needs to be included.  Since the mean field and variance (mean-squared
fluctuations) are coupled,
the back reaction of particle production on the mean-field dynamics
must be accounted for in a self-consistent manner. 

In the decade before the advent of inflationary cosmology, there was active
research on quantum processes in curved spacetimes. An important class of
problems is vacuum particle creation
\cite{parker:1969a,sexl:1969a,zeldovich:1970a,zeldovich:1971a,hu:1974a}
and its effect on the dynamics and structure of the early Universe 
\cite{zeldovich:1971a,zeldovich:1977a,hu:1977a,hu:1978a,fischetti:1979a,hartle:1979a,hartle:1980a,hu:1981a,anderson:1984a} at
the Planck time. The effect of spacetime dynamics and the importance of
parametric amplification on cosmological particle creation
were realized very early \cite{parker:1969a,zeldovich:1970a,hu:1974a}.
Most of the effort in the latter part of the 1970s
was focused on obtaining both a regularized energy-momentum tensor and a 
viable formalism for the treatment of back reaction effects. The wisdom 
gained from work in that period before the inflationary cosmology program was
initiated is particularly relevant to the
reheating problem. Simply put, for obtaining a finite energy-momentum 
tensor for a quantum field in a cosmological spacetime, the adiabatic
\cite{hu:1974a,parker:1974a,fulling:1974a,fulling:1974b} and dimensional
\cite{brown:1977a} regularization methods are the most useful. 
For studying the back reaction of particle creation,
the Schwinger-Keldysh (CTP, ``in-in'') effective action formalism
\cite{schwinger:1961a,keldysh:1964a,zhou:1985a,su:1988a,dewitt:1986a,jordan:1986a,calzetta:1987a,calzetta:1989a} is more appropriate than the usual 
Schwinger-DeWitt (``in-out'') method \cite{schwinger:1951a,dewitt:1965a}.

The second approach to the post-inflationary reheating problem is built upon
the body of earlier work on cosmological particle creation.
Following the application of closed-time-path techniques to nonequilibrium
relativistic field theory problems 
\cite{calzetta:1987a,calzetta:1988b},
several authors (which we call group 2A) derived perturbative
mean-field equations for a scalar inflaton with cubic \cite{calzetta:1989a}
and quartic \cite{paz:1990a} self-couplings, as well as for
a scalar inflaton Yukawa-coupled to fermions \cite{stylianopoulos:1991a}.
The closed-time-path method yields a real and causal mean-field 
equation with back reaction from quantum particle
creation taken into account.  For the case
of Bose particle production, perturbation theory in the coupling constant is
known to break down for sufficiently large occupation numbers, which occurs
on the time scale $\tau_1$ for parametric resonance effects to become
important \cite{boyanovsky:1995a,son:1996a}.  It is, therefore, necessary
to employ nonperturbative techniques in order to study reheating in most 
inflationary models.

The second stage of work in this second approach to the reheating problem
used the closed-time-path
method to derive self-consistent mean-field equations for
an inflaton coupled to lighter quantum fields (group 2B, 
\cite{boyanovsky:1995b,boyanovsky:1995c,boyanovsky:1995d,baacke:1997a,kaiser:1996a,mazzitelli:1989a,khlebnikov:1997a}).
In the first of these studies
\cite{boyanovsky:1995b,boyanovsky:1995c,boyanovsky:1995d,baacke:1997a}, 
the coupled one-loop mean-field and mode-function 
equations were solved numerically in Minkowski space, implicitly carrying out
an {\em ad hoc\/} nonperturbative resummation in $\hbar$.  In the 
one-loop equations, the variances
for the inflaton $\langle\varphi^2\rangle$ and light Bose fields
$\langle\chi^2\rangle$
do not back-react on the mode functions directly.  However, mean-field
equations were derived for an O$(N)$-invariant linear $\sigma$ model (with a
$\lambda \Phi^4$ self-interaction) at leading order in the large-$N$ 
approximation by Boyanovsky {\em et al.\/} \cite{boyanovsky:1995a}.
In this approximation, the variance 
does back-react on the quantum mode functions.  At leading order in 
the $1/N$ expansion, the unbroken symmetry dynamical equations for the
quartic O$(N)$
model are formally similar to the dynamical equations for a single
$\lambda \Phi^4$ field theory in the time-dependent Hartree-Fock 
approximation 
\cite{cornwall:1974a}.  The nonequilibrium dynamics of the quartically 
self-interacting O$(N)$ field theory in Minkowski space has been numerically
studied at leading order in the $1/N$ expansion in both the unbroken symmetry 
\cite{boyanovsky:1996b,boyanovsky:1995a,cooper:1994a}
and symmetry-broken 
\cite{boyanovsky:1996b,boyanovsky:1995a,cooper:1997b} cases.
Some analytic work has been done on the self-consistent Hartree-Fock 
mean-field equations for a quartic scalar field in Minkowski space
\cite{son:1996a}.
In addition, the Hartree-Fock equations for a $\lambda \Phi^4$ field in the 
slow-roll regime have been studied numerically in Minkowski space 
\cite{boyanovsky:1993a} and in FRW spacetime \cite{boyanovsky:1994a}.
However, the effect of spacetime dynamics on reheating in the O$(N)$ field
theory has not (to our knowledge) been studied using the coupled,
self-consistent semiclassical Einstein equation and matter-field dynamical
equations, though some simple analytic work has been done on curvature
effects in reheating \cite{dolgov:1994a,kaiser:1996a}.  The semiclassical
equations for one-loop reheating in FRW spacetime were derived in 
\cite{mazzitelli:1989a}.  The $\phi^2 \chi^2$ theory has been studied
in FRW spacetime by \cite{hu:1993a,khlebnikov:1997a,zhang:1991a}.
In addition, numerical work has been done on
symmetry-breaking phase transitions in both
a $\lambda \Phi^4$ scalar field in de~Sitter spacetime \cite{boyanovsky:1996a},
and an  O$(N)$ theory in FRW spacetime \cite{mazenko:1986a,boyanovsky:1996c}.
The group 2B studies represent the state of the art on the preheating problem.

\subsection{Our work -- how it differs from others}
In the present work we study the nonperturbative, out-of-equilibrium 
dynamics of a minimally coupled scalar O$(N)$ field theory 
(with quartic self-interaction) in a spatially flat FRW universe whose
dynamics is given self-consistently by the semiclassical Einstein equation.
The purpose of this study is to
understand the preheating period in inflationary cosmology,
with particular emphasis on the effect of spacetime dynamics on the 
phenomenon of particle production via parametric amplification of 
quantum fluctuations.  Of primary interest is obtaining the dynamics
of the inflaton (including back reaction from created particles) using rigorous
methods of nonequilibrium field theory in curved spacetime 
\cite{calzetta:1988b}.  We have chosen to focus in this paper on
parametric amplification of quantum fluctuations 
because this phenomenon can be the dominant effect in the 
preheating stage of unbroken symmetry inflationary scenarios, among which
the chaotic inflation scenarios most directly necessitate (through
initial conditions) considerations of Planck-scale physics.
``New'' inflationary scenarios which involve a spontaneously broken symmetry
often contain additional subtleties (e.g., infrared divergences, spinodal 
instabilities), and are the subject of ongoing investigation 
\cite{calzetta:1997a,ramsey:1997d}.
The results of our work are, therefore, particularly relevant to chaotic
inflation scenarios \cite{linde:1983a}.  
The additional interactions which should be included to treat the 
broken-symmetry case are discussed in \cite{ramsey:1997a}.

In the preceding paper \cite{ramsey:1997a} we have derived the evolution
equations for the mean field 
$\langle \Phi_{\text{{\tiny H}}} \rangle$ (subscript H denotes
the Heisenberg field operator) and mean-squared fluctuations
(variance) $\langle \Phi_{\text{{\tiny H}}}^2 
\rangle - \langle \Phi_{\text{{\tiny H}}} \rangle^2$  using the
closed-time-path (CTP), two-particle-irreducible (2PI) effective
action \cite{cornwall:1974a} in a fully covariant form.
Here we use these results for the case of spatially flat FRW spacetime.
The quantum state for the field theory (in the case of FRW spacetime) 
consists of a coherent state for the spatially homogeneous field mode, and the
adiabatic vacuum state for the spatially inhomogeneous modes.
At conformal past infinity, the spacetime is assumed to be asymptotically
de~Sitter, and the mean field is chosen to be asymptotically constant.

In this paper we study the O$(N)$ field theory using the
$1/N$ expansion, which yields nonperturbative dynamics in the regime of
strong mean field.  This is particularly important for chaotic inflation
scenarios \cite{linde:1985a},
in which the inflaton mean-field amplitude can be as large as $\Mpl/3$ at the
end of the slow-roll period \cite{linde:1990a,linde:1994a}.  Treatments of
the reheating problem which rely on time-dependent perturbation theory do not
apply to such cases where the inflaton undergoes large-amplitude oscillations,
in contrast with nonperturbative methods such as large $N$.

We now summarize the principal distinctions between our work and previous
treatments of  preheating in inflationary cosmology.  Our work
improves on the group 1A methodology by including parametric resonance
effects.  As it is based on first-order, time-dependent perturbation theory,
the group 1A approach cannot correctly describe the inflaton dynamics
with large initial mean-field amplitude.
In addition, our work improves on both the group 1A and group 1B
studies by including the effect of back reaction from quantum particle
creation on both the mean field and the inhomogeneous modes.
We are treating the inflaton dynamics from first principles, without
assuming a phenomenological equation (with a damping term
$\Gamma \dot{\phi}$ put in by hand) for the
mean field.  In our approach the damping of the mean field is due
to back reaction from quantum particle production in the self-consistent
equations for the mean field and its variance.  In contrast, the analytic 
results of
the group 1A and 1B work are based on the assumption of either large-amplitude
mean-field oscillations ($\lambda \phih^2/2 \gg m^2$) or harmonic oscillations
($m^2 \gg \phih^2/2$), and, therefore, cannot describe the interesting case
of inflaton dynamics in which neither term dominates the tree-level
effective mass, i.e., $m^2 \sim \lambda \phih^2$. Furthermore, our work
improves on group 1A, group 1B, and group 2A in that the closed-time-path
effective action is computed in {\em curved spacetime\/} without assuming that 
$H^{-1} \gg \tau_0$ (where $\tau_0$ is the period of mean-field oscillations).
In our work, the dynamics of the two-point function
(which reflects quantum particle production) is formulated in curved 
spacetime assuming only that semiclassical gravity is valid,  
i.e., $\Mpl \gg H$.

Most significantly, our work improves on all the previous treatments 
in that it includes curved spacetime effects
systematically using the coupled, {\em self-consistent\/}
semiclassical Einstein equation and matter field equations.
Among the group 2B studies of preheating dynamics, inflaton
dynamics has been studied primarily in {\em fixed\/} background spacetimes:
Minkowski space \cite{boyanovsky:1996b,boyanovsky:1995a}, de~Sitter space
\cite{boyanovsky:1996a,boyanovsky:1996c,boyanovsky:1997c}, and in 
radiation-dominated, spatially flat FRW spacetime \cite{boyanovsky:1996c}.
In the present work, the spacetime is {\em dynamical,\/} with the renormalized
trace of the semiclassical Einstein equation governing the dynamics of the 
scale factor $a$.  This permits quantitative study of the transition of the 
spacetime from the (slow-roll) period of vacuum-dominated expansion to the 
radiation-dominated (``standard'') FRW cosmology.  In particular,
our method yields the spacetime dynamics naturally, 
without making reference to
an ``effective Hubble constant'' (which has been used
in calculations on a fixed background spacetime \cite{boyanovsky:1997c}).

With additional couplings (see \cite{ramsey:1997a}), our
method may also be used to study preheating in ``new'' inflationary scenarios
\cite{linde:1982a}.  In new inflation, the vacuum-dominated expansion of
the Universe is typically driven by the classical potential energy of the
mean field as it rolls towards the symmetry-broken ground
state.  In one of the group 2B studies (Boyanovsky {\em et al.\/} 
\cite{boyanovsky:1997c}),
a quench-induced phase transition is studied with small initial mean-field
amplitude, in which the classical terms in the mean-field equation are 
{\em dominated\/} by spinodal fluctuations.  As a result, the mean field in
their model does not oscillate about the symmetry-broken ground 
state as is generally expected in a new inflation preheating scenario (this
point was emphasized in \cite{kofman:1996b}).
The initial conditions studied in \cite{boyanovsky:1997c} are more
appropriate to a study of defect formation in a 
quench-induced phase transition
than preheating dynamics in new inflation.

In addition, the renormalization
scheme employed in \cite{boyanovsky:1997c}
is not generally covariant (as can be seen by comparing it with 
\cite{bunch:1978a}), and covariant conservation of the renormalized 
energy-momentum tensor is put in by hand.  The regularization scheme
employed here is the well-tested adiabatic regularization 
\cite{hu:1974a,parker:1974a,fulling:1974a,fulling:1974b},
which is simple to use and physically intuitive. It also ensures both covariant
conservation of the regularized energy-momentum
tensor and agreement with manifestly covariant regularization procedures such
as point splitting \cite{bunch:1978a}.  

A related difference between our approach and that of some of the 
group 2B studies is the choice of vacuum state.  
The choice of initial conditions for the quantum mode functions in most
studies of reheating in FRW spacetime
\cite{khlebnikov:1997a,boyanovsky:1994a,khlebnikov:1996b}
has been to instantaneously diagonalize the matter-field Hamiltonian at the
initial-data hypersurface.  However, as has been pointed out long ago
\cite{fulling:1989a}, this method does not correspond
to the vacuum state which registers the least particle flux on 
a comoving detector.  In our work we use the de~Sitter-invariant
(or Bunch-Davies) vacuum, obtained via the adiabatic construction; the
adiabatic vacuum most closely aligns with an intuitive notion of vacuum state 
in a cosmological spacetime \cite{birrell:1982a}.

In many of the group 2B
studies \cite{boyanovsky:1995a,boyanovsky:1996c,boyanovsky:1997c} the
large-$N$ equations for the mean field and variance are derived using a
factorization method which does not readily 
generalize to next-to-leading order in the $1/N$ expansion.  
In all of the group 2B studies of nonperturbative inflaton 
dynamics of which we are aware, the equations for the mean field and variance
are not derived using methods which encompass higher-order correlations
in the Schwinger-Dyson hierarchy. As found in earlier studies of
phase transitions \cite{hu:1987b,calzetta:1989b}, this is necessary in
order to derive the correct infrared behavior of a quantum field
in a study of critical phenomena. In the present work, we use the result of
the preceding paper \cite{ramsey:1997a} in which the CTP
two-particle-irreducible (2PI) formalism is derived.  It has a direct
generalization in terms of the $n$-particle-irreducible ($n$PI) ``master''
effective action \cite{calzetta:1995b}.  The master effective action
can be used to derive a self-consistent truncation of the Schwinger-Dyson 
equations to arbitrary order in the correlation hierarchy 
\cite{calzetta:1995b}.  The techniques
employed here are, therefore, most readily generalized to the study of phase 
transitions in curved spacetime, where higher-order correlation functions 
can become important \cite{calzetta:1997a}.

In summary, our approach to the inflaton dynamics problem
has the following advantages:  
it is nonperturbative and fully covariant; it is
based on rigorous methods of nonequilibrium field theory in curved spacetime;
we use the correct adiabatic vacuum construction; and 
we employ an approximation scheme which can be systematically 
generalized beyond leading order, within a fully covariant and
self-consistent theoretical framework.

\subsection{Summary of major findings}
Our results are obtained by solving the mean-field
and spacetime dynamics self-consistently using the coupled matter-field
and semiclassical Einstein equations in a FRW spacetime,
including the effect of back reaction
of the variance on the mean field.  Within the leading-order, large $N$
approximation used here, 
we find that (using the conventional value for the self-coupling,
$\lambda = 10^{-14}$) for sufficiently large initial mean-field amplitude,
parametric amplification of quantum fluctuations is not an efficient mechanism
of energy transfer from the mean field to the inhomogeneous field modes.
In this case the energy
density of the inhomogeneous modes remains negligible in comparison to the 
mean-field energy density for all times.  This can be understood from the time
scales for the competing processes of parametric resonance and cosmic
expansion.  When the time scale for parametric amplification of
quantum fluctuations $\tau_1$ is of the same order as (or greater than)
the time scale for cosmic expansion $H^{-1}$, cosmic expansion redshifts 
the energy density of the inhomogeneous modes faster than it increases
due to parametric resonance.  We find that this occurs when 
$\phih \gtrsim \Mpl/300$, for the model and coupling studied here.  

In many chaotic inflation scenarios, the mean-field amplitude at the
end of the slow-roll period can be as large as
$\Mpl/3$ \cite{linde:1990a,linde:1994a}.
In light of our result, in such models, it is
clearly essential to include the effect of spacetime dynamics in order to 
study mean-field dynamics and resonant 
particle production during reheating.  In addition, our result indicates
that for the case of a minimally coupled $\lambda\Phi^4$ inflaton with
unbroken symmetry, parametric amplification of its own quantum fluctuations
is not a viable mechanism for reheating, unless the coupling is significantly
strengthened (see \cite{calzetta:1995a,matacz:1996a}).
Parametric amplification of quantum fluctuations may still
play a dominant role in the reheating of chaotic inflaton models with an
inflaton coupled to other fields, e.g., a $\phi^2 \chi^2$ model.  This
is the subject of a forthcoming paper \cite{ramsey:1997d}.

For more moderate cosmic expansion, where $H^{-1} \gtrsim 100 \; \tau_1$,
parametric
amplification of quantum fluctuations is an efficient mechanism of energy
transfer to the inhomogeneous modes, and the asymptotic effective equation
of state is found to agree with the prediction of a two-fluid model consisting
of the elliptically oscillating mean field and relativistic
energy density contained in the inhomogeneous mode occupations. In a
collisionless approximation, the mean field eventually decouples from the
mean-squared fluctuations (variance) and at late times undergoes asymptotic
oscillations which are damped solely by cosmic expansion \cite{linde:1994a}.
For the case when cosmic expansion is subdominant, $H^{-1} \gg \tau_1$, the 
mean-field dynamics and the growth of quantum fluctuations are in agreement
with results of studies of preheating in Minkowski space 
\cite{boyanovsky:1996b}.  In particular, the total adiabatically regularized
energy density is found to be constant (to within the limits of 
numerical precision) for the case of $H^{-1} \rightarrow \infty$, in agreement
with the predictions of field theory in Minkowski space.

While there has been a large volume of work on the preheating period and
particle production, the thermalization of inflationary models has not
yet been understood from first principles.
From the nonperturbative inflaton dynamics, Boyanovsky {\em et al.\/} 
claimed that \cite{boyanovsky:1996b}
the Boltzmann equation is inadequate for studying collisional
thermalization at the end of the preheating stage.
In particular, in the leading-order, large-$N$ approximation employed here
(and in the one-loop approximation which it contains), 
this model also does not thermalize.  However, it still may approach
a radiation-dominated effective equation of state
in Minkowski space as found in \cite{boyanovsky:1996b}.
Clearly, a first-principles analysis of thermalization is necessary.
Continuing the early work of kinetic field theory \cite{calzetta:1988b},
and the recent work on correlation hierarchy \cite{calzetta:1995b},
we know that such a first-principles analysis should involve at a minimum
the full two-loop, two-particle-irreducible effective action
(or alternatively, next-to-leading order in the large-$N$ approximation).
Since it  represents a rigorous truncation of
the full Schwinger-Dyson hierarchy, in this sense it is the most natural
generalization of the collisionless approximations used
previously to study reheating. However, the equations derived from it for
the mean-field and gap equation are nonlocal and hence difficult to solve
even numerically \cite{cooper:1994a}.
This paper is, therefore, concerned only 
with preheating via parametric resonance particle creation;
work on thermalization is in progress\cite{ramsey:1997b}.

\subsection{Organization and notation}
This paper is organized as follows.
In Sec.~\ref{sec-lpffrw} we present the general
theory of nonequilibrium dynamics of a scalar field in curved spacetime,
including a summary discussion of reheating in inflationary cosmology.
In Sec.~\ref{sec-onfrw} we specialize to the case of spatially
flat FRW spacetime, and derive the dynamical equations.
The results of numerically solving the evolution equations are contained
in Sec.~\ref{sec-results}.
Discussion and conclusions follow in Sec.~\ref{sec-discussion}.

Throughout this paper we use units in which 
$c = 1$. Planck's constant $\hbar$ is shown explicitly 
(i.e., not set equal to 1) except in those sections where noted. 
In these units, Newton's constant is $G = \hbar \Mpl^{-2}$, where
$\Mpl$ is the Planck mass.
We work with a four-dimensional spacetime manifold, and follow
the sign conventions\footnote{In the classification
scheme of Misner, Thorne, and Wheeler \cite{misner:1973a}, the 
sign convention of Birrell and Davies \cite{birrell:1982a} is 
classified as $(+,+,+)$.}  
of Birrell and Davies \cite{birrell:1982a}
for the metric tensor $g_{\mu\nu}$, 
the Riemann curvature tensor $R_{\mu\nu\sigma\rho}$, and
the Einstein tensor $G_{\mu\nu}$.  We use greek letters to 
denote spacetime indices. The beginning latin letters 
$a,b,c,d,e,f$ indicate the time branch (see \cite{ramsey:1997a},
Sec.~II),
and the middle latin letters $i,j,k,l,m,n$ are reserved as indices in the
O$(N)$ space (see Sec.~\ref{sec-onlna}). 
Einstein summation convention over repeated indices is employed.

\section{$\lambda \Phi^4$ Inflaton Dynamics in FRW spacetime}
\label{sec-lpffrw}
\subsection{$\lambda \Phi^4$ quantum fields in curved spacetime}
\label{sec-lpfcst}
As a simple model of inflation, let us consider a scalar $\lambda \Phi^4$
field in semiclassical gravity, where the matter field is quantized on a
classical, dynamical background spacetime.  The classical action has the form
\begin{equation}
S[\phi,g^{\mu\nu}] = S^{\text{{\tiny G}}}[g^{\mu\nu}] + S^{\text{{\tiny F}}}[
\phi,g^{\mu\nu}],
\label{eq-lpfcsta}
\end{equation}
where $S^{\text{{\tiny F}}}$ is the matter field action, which for 
the scalar $\lambda \Phi^4$ theory takes the form
\begin{equation}
S^{\text{{\tiny F}}}[\phi,g^{\mu\nu}] = -\frac{1}{2} \int \! d^{\, 4} x 
\sqmg \left[
\phi (\square + m^2 + \xi R) \phi + \frac{\lambda}{12}\phi^4\right],
\label{eq-lpfcstca}
\end{equation}
and for a renormalizable theory, the gravity action 
$S^{\text{{\tiny G}}}$ must have the form \cite{birrell:1982a,dewitt:1975a}
\begin{equation}
S^{\text{{\tiny G}}}[g^{\mu\nu}] = \frac{1}{16 \pi G} \int \! d^{\, 4} x 
\sqmg \left[
R - 2 \Lambda + c R^2 + b R^{\alpha\beta} R_{\alpha\beta} +
a R^{\alpha\beta\gamma\delta} R_{\alpha\beta\gamma\delta} \right].
\label{eq-lpfcstga}
\end{equation}
In Eqs.\ (\ref{eq-lpfcstca}) and (\ref{eq-lpfcstga}), 
$\lambda$ is the coupling constant (with dimensions of 
inverse mass times inverse length), $m$ is the ``mass'' (with dimensions of
inverse length), $\xi$ is the dimensionless coupling to gravity, 
$G$ is Newton's constant (with dimensions of length divided by mass), and
$a$, $b$, and $c$ are constants with dimensions of length squared.  The
symbol $\square$ denotes the Laplace-Beltrami operator in terms of the
covariant derivative $\nabla_{\mu}$, $R$ is the scalar curvature,
$R_{\mu\nu}$ is the Ricci tensor,  $R_{\alpha\beta\gamma\delta}$ is the
Riemann tensor, and $\sqrt{-g}$ is the square root of minus the determinant of
the metric tensor $g_{\mu\nu}$.  

The inflaton field $\phi$ is then quantized on the classical
background spacetime; we denote the Heisenberg field operator by 
$\Phi_{\text{{\tiny H}}}$, and the quantum state by $|\phi\rangle$.  
Of particular interest in a study of inflaton dynamics are the mean field
\begin{equation}
\phih(x) \equiv \langle \phi | \Phi_{\text{{\tiny H}}}(x) | \phi \rangle,
\end{equation}
the fluctuation field
\begin{equation}
\varphi_{\text{{\tiny H}}}(x) \equiv \Phi_{\text{{\tiny H}}}(x) - \phih(x),
\label{eq-dff2}
\end{equation}
and the mean-squared fluctuations, or variance
\begin{equation}
\langle \phi | \varphi_{\text{{\tiny H}}}^2 (x) | \phi \rangle = 
\langle \phi | \Phi_{\text{{\tiny H}}}^2(x) | \phi \rangle - 
\langle \phi | \Phi_{\text{{\tiny H}}}(x) | \phi \rangle^2.
\end{equation}
In a previous paper \cite{ramsey:1997a}, a systematic procedure was presented
for deriving real and causal evolution equations for the mean field, 
two-point function, 
and the metric tensor in semiclassical gravity.  Assuming a globally hyperbolic
spacetime, one can evolve the coupled evolution equations forward from initial 
data specified at an initial Cauchy hypersurface.  The evolution equations
follow from functional differentiation (and subsequent field identifications)
of the closed-time-path (CTP) two-particle-irreducible (2PI) effective action,
$\Gamma[\phih_{\pm},G_{\pm\pm},g^{\mu\nu}_{\pm}]$.  The CTP-2PI effective
action is a functional of the mean field $\phih$, two-point function $G$, and 
metric tensor $g^{\mu\nu}$, which now carry not only spacetime labels but
also {\em time branch\/} labels, which have an index set $\{+, -\}$.  The 
evolution equations for $\phih$, $\langle \varphi^2 \rangle$, and $g_{\mu\nu}$
then follow from 
\begin{mathletters}
\begin{eqnarray}
\label{eq-lpfsee}
\left. \frac{\delta ({\mathcal S}^{\text{{\tiny G}}}[g^{\mu\nu}_{\pm}] + 
\Gamma[\phihpm,G_{\pm\pm},g^{\mu\nu}_{\pm}])}{\delta g^{\mu\nu}_a}\right|_{ 
\phihp = \phihm = \phih ;\;\;\;
g^{\mu\nu}_{+} = g^{\mu\nu}_{-} = g^{\mu\nu}} &=& 0, \\
\label{eq-lpfseeb}
\left.
\frac{\delta \Gamma[\phihpm,G_{\pm\pm},g^{\mu\nu}_{\pm}]}{\delta \phih_a}
\right|_{
\phihp = \phihm = \phih ;\;\;\;
g^{\mu\nu}_{+} = g^{\mu\nu}_{-} = g^{\mu\nu}} &=& 0, \\
\label{eq-lpfseec}
\left.
\frac{\delta \Gamma[\phihpm,G_{\pm\pm},g^{\mu\nu}_{\pm}]}{\delta G_{ab}}
\right|_{ 
\phihp = \phihm = \phih ;\;\;\; 
g^{\mu\nu}_{+} = g^{\mu\nu}_{-} = g^{\mu\nu}} &=& 0.
\end{eqnarray}
\end{mathletters}
The variance is related to the coincidence limit of 
CTP two-point function as follows:
\begin{equation}
\langle \varphi_{\text{{\tiny H}}}^2(x) \rangle = \hbar G_{ab}(x,x)
\end{equation}
for all $a,b \in \{ +,-\}$.  The energy-momentum tensor $\langle T_{\mu\nu}
\rangle$ is defined by
\begin{equation}
\langle T_{\mu\nu} \rangle = \left. \frac{2}{\sqmg} \left( \frac{\delta
\Gamma[\phihpm,G_{\pm\pm},g^{\mu\nu}_{\pm}]}{\delta g^{\mu\nu}_{+}} \right)
\right|_{
\phihp = \phihm = \phih; \;\;\; g^{\mu\nu}_{+} = g^{\mu\nu}_{-} =
g^{\mu\nu}},
\label{eq-lpfcstemt}
\end{equation}
which (after renormalization) enters as the source of the semiclassical 
Einstein field equation:
\begin{equation}
G_{\mu\nu} = -8 \pi G \langle T_{\mu\nu} \rangle.
\label{eq-seesimp}
\end{equation}

Eqs.\ (\ref{eq-lpfsee})--(\ref{eq-lpfseec}) 
constitute a set of coupled, nonlocal, nonlinear
equations for the mean field, two-point function, and metric tensor.
The renormalized versions are what enter into the description of inflaton
dynamics.  The CTP-2PI effective action can be computed using 
diagrammatic methods described in the previous paper, where a covariant 
expression for $\Gamma$ was computed in 
a general curved spacetime (truncated at two loops).

\subsection{Inflaton dynamics in FRW spacetime}
\label{sec-idfrw}
We now consider a spatially
flat Friedmann-Robertson-Walker (FRW) spacetime, which
is spatially homogeneous, isotropic, and conformally flat.
Its line element can be written in the form
\begin{equation}
ds^2 = a(\eta)^2 \left[d\eta^2 - \sum_{i=1}^3 (dx^i)^2\right],
\label{eq-frwcrd}
\end{equation}
where $a$ is the scale factor,  $x^i$ ($i \in \{1,2,3\}$) are the physical
position coordinates on the spatial hypersurfaces
of constant conformal time $\eta$ (related to the cosmic time $t$ by
$\eta = \int dt/a$).
The Hubble parameter, which measures the rate of cosmic expansion, is
\begin{equation}
H(\eta) = \frac{\dot{a}}{a},
\end{equation}
where the over-dot denotes differentiation with respect to cosmic time $t$.
Given our choice of sign convention and metric signature, the Ricci tensor 
in the FRW coordinates is given by
\begin{mathletters}
\begin{eqnarray}
R_{00} &=& 3\left[\frac{a''}{a} - \frac{(a')^2}{a^2}\right],  \\
R_{ij} &=& -\left[\frac{a''}{a} + \frac{(a')^2}{a^2}\right]\delta_{ij},
\end{eqnarray}
\end{mathletters}
where the prime denotes differentiation with respect
to $\eta$, and $R_{00}$ is the component of the Ricci tensor
proportional to $d\eta \otimes d\eta$. The scalar curvature is
\begin{equation}
R  = \frac{6a''}{a^3}, 
\end{equation}
and the Einstein tensor is
\begin{mathletters}
\begin{eqnarray}
G_{00} &=& -\frac{3(a')^2}{a^2}, \\
G_{ij} &=& \left[\frac{2 a''}{a} - \frac{(a')^2}{a^2}\right]\delta_{ij}.
\end{eqnarray}
\end{mathletters}
Finally, the volume form on $M$ is
\begin{equation}
\bbox{\epsilon}_{\text{{\tiny M}}} = a^4 (d\eta \wedge dx^1 \wedge 
dx^2 \wedge dx^3).
\end{equation}
The higher-order (e.g., $R^2$)
geometric terms in the geometrodynamical field equation are not
shown because the renormalized constants $b$ and $c$ are set to zero
in Sec.~\ref{sec-rsee}.

In restricting the spacetime to be a spatially flat FRW,
we are reducing the number of degrees of freedom in the metric:
\begin{equation}
g_{\mu\nu} \rightarrow a(\eta)^2 \eta_{\mu\nu}.
\end{equation}
This reduction should not be carried out in the 2PI generating functional
$\Gamma[\phih,G,g^{\mu\nu}]$, but
only in the equations of motion
(\ref{eq-lpfsee})--(\ref{eq-lpfseec}).  
This is because functional differentiation 
of $\Gamma[\phih,G,a^{-2}\eta^{\mu\nu}]$ with respect to the scale factor 
$a$ gives only the {\em trace\/} 
of the energy-momentum tensor,
$a^{-2} \eta^{\mu\nu} \langle T_{\mu\nu} \rangle$,
and not the additional constraint equation which the initial data must 
satisfy.

The spatial homogeneity and isotropy of FRW spacetime permits only
two algebraically independent components of the energy-momentum tensor,
which in the FRW coordinates of Eq.\ (\ref{eq-frwcrd}) are given by
$\langle T_{00} \rangle$ and $\langle T_{ii}\rangle$; all other components
are zero.  These must be functions of $\eta$ only (due to spatial
homogeneity).
For the purpose of numerically solving the semiclassical Einstein equation,
it is convenient to work with the trace 
\begin{equation}
{\mathcal T} = g^{\mu\nu}\langle T_{\mu\nu}\rangle = a^{-2} \eta^{\mu\nu}
\langle T_{\mu\nu} \rangle,
\end{equation}
instead of $\langle T_{ii} \rangle$.  The trace
${\mathcal T}$ enters into the dynamical equation for $a(\eta)$, and
$\langle T_{00} \rangle$ enters into the constraint equation.

Another consequence of the spatial symmetries of FRW spacetime is
the restriction on the generality with which we may specify initial data
for dynamical evolution.  
Let us choose to specify initial data on a Cauchy
hypersurface $\Sigma_{\eta_0}$ of constant conformal time $\eta_0$.
In the Heisenberg picture,\footnote{As discussed in
Sec.~\ref{sec-initcond}, for our purposes it is sufficient to consider only
the case of a pure state.  The analysis can, however, be easily extended
to encompass a mixed state with density matrix $\bbox{\rho}$.}
for consistency with spatial homogeneity, the quantum state 
$|\phi\rangle$ must satisfy 
\begin{mathletters}
\begin{eqnarray}
\langle \phi | \Phi_{\text{{\tiny H}}}(\eta_0,\vec{x}) | \phi \rangle &=& 
\phih(\eta_0), 
\label{eq-symcdmf} \\
\langle \phi | \Phi_{\text{{\tiny H}}}'(\eta_0,\vec{x}) | \phi \rangle &=& 
\phih'(\eta_0), 
\label{eq-symcdmfb}
\end{eqnarray}
\end{mathletters}
for all $\vec{x} \in {\mathbb R}^3$, where $\Phi_{\text{{\tiny H}}}$ is
the Heisenberg field operator for the scalar field.  The values of 
$\phih(\eta_0)$ and $\phih'(\eta_0)$ constitute initial data for the 
mean field.  In addition, the quantum state must satisfy  
\begin{mathletters}
\begin{eqnarray}
\langle \phi | \vphi_{\text{{\tiny H}}}(\eta_0,\vec{x})
\vphi_{\text{{\tiny H}}}(\eta_0,\vec{x}') | \phi \rangle
&=& F(\eta_0,|\vec{x}-\vec{x}'|), 
\label{eq-symcdgf}  \\
\frac{\partial}{\partial \eta}_{|\eta_0}\!\langle\phi | 
\vphi_{\text{{\tiny H}}}(\eta,\vec{x})
\vphi_{\text{{\tiny H}}}(\eta,\vec{x}')|
\phi\rangle &=& F'(\eta_0,|\vec{x}-\vec{x}'|), 
\label{eq-symcdgfb}
\end{eqnarray}
\end{mathletters}
in terms of an equal-time correlation function $F(\eta_0,|\vec{x}-\vec{x}'|)$ 
which is invariant under simultaneous translations and rotations of
$\vec{x}$ and $\vec{x}'$.  As defined in Eq.\ (\ref{eq-dff2}), 
$\vphi_{\text{{\tiny H}}}$ denotes the Heisenberg field operator for the 
fluctuation field.  The spatial Fourier transform of $F$ is 
related to the power spectrum of quantum fluctuations at $\eta_0$ for
the quantum state $|\phi\rangle$.  Alternatively, we may say that 
$F(\eta_0,r)$ and $F'(\eta_0,r)$ give initial data for the evolution of
the two-point function $G_{++}$ via the gap equation (\ref{eq-lpfseec}).
The symmetry conditions (\ref{eq-symcdmf}), (\ref{eq-symcdmfb}), 
(\ref{eq-symcdgf}), (\ref{eq-symcdgfb}),
along with the spatial symmetries of the classical action
in FRW spacetime, guarantee that the mean field and two-point function
satisfy spatial homogeneity and isotropy for {\em all time,\/} i.e.,
\begin{mathletters}
\begin{eqnarray}
\langle \Phi_{\text{{\tiny H}}}(x)\rangle &=& \phih(\eta),
\label{eq-gfssym} \\
G_{++}(x,x') &=& G_{++}(\eta,\eta',|\vec{x}-\vec{x'}|),
\label{eq-gfssymb}
\end{eqnarray}
\end{mathletters}
for all $x \in M$.
The conditions (\ref{eq-gfssym}), (\ref{eq-gfssymb})
permit a formal solution of the gap equation (\ref{eq-lpfseec})
for $G_{++}$ in terms of homogeneous mode functions, via a Fourier
transform in comoving momentum $\vec{k}$, as shown in
Sec.~\ref{sec-resfrw}.  By rotational invariance, the Fourier transform
depends only on the magnitude $k \equiv \sqrt{\vec{k}\cdot\vec{k}}$.
Of course, the quantum state $|\phi\rangle$ is not uniquely defined by
the spatial symmetries; a unique choice of the initial conditions for
$\phih$ and $G_{ab}$ at $\Sigma_{\eta_0}$ is (in the Gaussian wave-functional
approximation) equivalent to choosing $|\phi\rangle$.
The choice of quantum state depends on the physics of the problem we wish
to study.

As a consequence of covariant conservation of the energy-momentum tensor
\begin{equation}
\nabla^{\mu} \langle T_{\mu\nu} \rangle = 0,
\label{eq-emtcc}
\end{equation}
the functions $\langle T_{00}(\eta) \rangle$ and 
$\langle T_{ii}(\eta) \rangle$ satisfy 
\begin{equation} 
\frac{d}{d\eta} \Bigl( a \langle T_{00} \rangle \Bigr) = 
- \frac{\langle T_{ii} \rangle}{a^2} \frac{d}{d\eta} \left( a^3 \right),
\end{equation}
which comes from taking the $\nu = 0$ component of Eq.\ (\ref{eq-emtcc}).
In analogy with the continuity relation for a classical perfect fluid in FRW
spacetime,
\begin{equation}
\frac{d}{d\eta} \left( a^3 \rho \right) = -p\frac{d}{d\eta}\left( a^3 \right),
\label{eq-fltd}
\end{equation}
we may define the energy density $\rho$ and pressure $p$, by
\begin{mathletters}
\begin{eqnarray}
\rho(\eta) &=& \frac{1}{a^2}\langle T_{00}(\eta) \rangle, \\
p(\eta) &=& \frac{1}{a^2}\langle T_{ii}(\eta) \rangle.
\label{eq-frwrho} 
\end{eqnarray}
\end{mathletters}
However, the quantity $p$ should not be interpreted
as the true hydrodynamic pressure until a perfect-fluid
condition is shown to exist;
otherwise, bulk viscosity corrections can
enter into Eq.\ (\ref{eq-frwrho}) \cite{weinberg:1972a}.
The effective 
equation of state is defined as a time average (over the time scale
$\tau_1$ for the matter field dynamics, to be discussed in
Sec.~\ref{sec-reheating}) of the ratio $p/\rho$,
\begin{equation}
\bar{\gamma} \equiv \frac{p}{\rho}.
\label{eq-defeos}
\end{equation}
The effective 
equation of state $\bar{\gamma}$ (where the bar denotes a time average)
is an important quantity
in differentiating between the various stages of inflationary cosmology.

Several solutions to the semiclassical Einstein equation (\ref{eq-seesimp})
for idealized equations of state are of particular interest in cosmology.
The effective equation of state $\bar \gamma = -1$ (eternally 
``vacuum dominated'')
leads to a solution  $a(\eta) = -1/(H\eta)$, for $-\infty < \eta < 0$, 
where $H = \sqrt{8 \pi G \rho/3}$ and $\rho$ is a
constant.  This solution
corresponds to the ``steady-state'' coordinatization covering 
one-half of the 
de~Sitter manifold \cite{birrell:1982a}.
The effective 
equation of state $\bar{\gamma} = 0$ 
corresponds to nonrelativistic matter, in which case the scale factor
conformal-time dependence is $a \propto \eta^2$.  The 
effective equation of state
$\bar{\gamma} = 1/3$ corresponds to relativistic matter, and its 
scale factor conformal-time dependence is $a \propto \eta$.

\subsection{Initial conditions for post-inflaton dynamics}
\label{sec-initcond}
In most realizations of inflationary cosmology, the Universe evolves through
a period in which a dominant portion of the energy density $\rho$ comes from
a quantum field $\Phi_{\text{{\tiny H}}}$, the {\em inflaton field,\/} whose 
effective equation of state [defined as in Eq.\ (\ref{eq-defeos})] is  
$\bar{\gamma} \simeq -1$.  
In chaotic inflation, this condition
is due to the fact that the inflaton field is in a quantum state $|\phi\rangle$
in which the Heisenberg field operator $\Phi_{\text{{\tiny H}}}$ acquires a 
large (approximately spatially homogeneous) expectation value, defined by
\begin{equation}
\phih(\eta) \equiv \langle \phi | \Phi_{\text{{\tiny H}}}(x) | \phi \rangle.
\end{equation}
A requirement for chaotic inflation is that the potential energy $V(\phih)$ 
of the expectation value $\phih$ dominates over both the spatial
gradient energy [coming from $\langle (\nabla \vphi_{\text{{\tiny H}}})^2 
\rangle$] and kinetic energy for the inflaton field, and the energy
density of all other quantum fields coupled to the inflaton.
The potential energy $V(\phih)$ gives a contribution
to the energy-momentum tensor satisfying precisely $\gamma = -1$.
During inflation, the scale factor grows by a factor of approximately
$\exp(H \Delta t)$, where $\Delta t$ is the 
interval of inflation in cosmic time, typically larger than
$60 H^{-1}$. While the Universe is inflating, the
expectation value $\langle \Phi_{\text{{\tiny H}}} \rangle$
is slowly rolling toward the true minimum of the effective 
potential. (In reality, the situation is much more complicated than
this.  The effective potential is an inadequate tool for
studying out-of-equilibrium mean-field dynamics \cite{hu:1986d,mazenko:1985a}.)
Assuming the Universe was in local thermal
equilibrium prior to inflation, the temperature during inflation decreases
in proportion to $1/a$.  The energy density of any relativistic
(nonrelativistic) fields coupled to the inflaton is proportional to
$1/a^4$ ($1/a^3$).
The contribution to the quantum energy density from spatial gradients
of fluctuations about the inflaton field
is proportional to $1/a^4$ (see Sec.~\ref{sec-onfrw} below).
Most importantly, any inhomogeneous modes $\delta \phih_k$ of the 
{\em mean field\/} which might exist at the onset of inflation are
redshifted. The physical momentum of a quantum mode,
$k_{\text{{\tiny phys}}} = k/a,$
decreases as $1/a$ relative to the comoving momentum $k$.  
The quantum state of any field coupled to the inflaton 
at the end of inflation is, therefore, approximately given by
the vacuum state. The inflaton field is well approximated by
a spatially homogeneous mean field, with vacuum fluctuations around the 
mean-field configuration.   The mean field
can be thought of as representing the coherent oscillations of a
condensate of zero-momentum inflaton particles.

Let us consider the case of inflation driven by a 
single self-interacting scalar field $\phi$ (with
unbroken symmetry) in spatially flat FRW spacetime. 
The above arguments imply that one can model post-inflationary physics with a
quantum state $|\phi\rangle$ which at $\eta_0$ corresponds
to a coherent state for the field operator $\Phi_{\text{{\tiny H}}}$
[in which $\langle \phi| \Phi_{\text{{\tiny H}}}(x) | \phi \rangle = 
\phih(\eta)$], and the fluctuation field $\vphi_{\text{{\tiny H}}} \equiv 
\Phi_{\text{{\tiny H}}} - \phih$ is very nearly
in the vacuum state.\footnote{
Though this is a pure quantum state, the methods
employed in this study can be used to treat a quantum field theory in
a mixed state  (for example, a system initially in thermal equilibrium
with a heat bath).} 
Then for $\eta < \eta_0$, $\langle T_{00} \rangle$ is dominated by
the classical energy density of the mean field $\phih$.
The 00 component of the Einstein equation then yields
\begin{equation}
\frac{a'}{a^2} = \sqrt{\frac{8 \pi G \rho_{\text{{\tiny C}}}}{3}},
\end{equation}
where $\rho_{\text{{\tiny C}}}$ is the classical energy density of the mean
field, defined by
\begin{equation}
\rho_{\text{{\tiny C}}} = \frac{1}{2 a^2} (\phih')^2 + V(\phih).
\label{eq-dfrc}
\end{equation}
The mean field $\phih$ satisfies the classical equation
\begin{equation}
\phih'' + \frac{2 a'}{a}\phih' + a^2 V'(\phih) = 0,
\end{equation}
where $V(\phih)$ denotes the classical potential.
For the $\lambda \Phi^4$ theory, the potential is [from the
Minkowski-space limit of Eq.\ (\ref{eq-lpfcsta})]
\begin{equation}
V(\phih) = \frac{1}{2} m^2 \phih^2 + \frac{\lambda}{24}\phih^4.
\end{equation}
The assumption that the Universe is inflating (i.e., $\bar{\gamma} \simeq -1$)
for $\eta < \eta_0$ requires that the energy density $\rho_{\text{{\tiny C}}}$
be potential dominated,
\begin{equation}
V(\phih) \gg \frac{1}{2 a^2}(\phih')^2,
\label{eq-pdas}
\end{equation}
and that the mean field satisfies the slow-roll condition,
\begin{equation}
\phih'' \ll \frac{2 a'}{a} \phih'.
\label{eq-sras}
\end{equation}
Given Eqs.\ (\ref{eq-pdas}) and (\ref{eq-sras}),  
an approximate ``0th adiabatic order''
solution to the Einstein equation can be obtained [normalized
to $a(\eta_0) = 1$],
\begin{equation}
a(\eta) \simeq \frac{1}{1 + H(\eta)(\eta - \eta_0)},
\label{eq-asol}
\end{equation}
where $H$ is a slowly varying function of $\eta$, given by
\begin{equation}
H(\eta) = \sqrt{\frac{8\pi G \rho_{\text{{\tiny C}}}(\eta)}{3}}.
\label{eq-defhinv}
\end{equation}
From Eq.\ (\ref{eq-defhinv}), we can evaluate the expansion rate 
nonadiabaticity
parameter $\bar{\Omega}_H$ \cite{hu:1993d} for 
$\eta < \eta_0$ using Eq.\ (\ref{eq-defhinv}). 
During slow-roll it follows from conditions (\ref{eq-pdas}) and
(\ref{eq-sras}) that
\begin{equation}
\bar{\Omega}_H \equiv \frac{H'}{H^2} 
= \frac{V'(\phih) \phih'}{\sqrt{\frac{32 \pi G}{3} V(\phih)^3}} \ll 1.
\end{equation}
The solution (\ref{eq-asol}) for $a(\eta)$ is exact in the limit of constant
$H$ (corresponding to a constant $\phih$ at the tree level).  For simplicity,
let us assume that $\phih$ goes to a constant value
$\gtrsim \Mpl$ in the asymptotic past, $\eta \rightarrow -\infty$.  
The spacetime is then asymptotically
de Sitter, with the scale factor having an asymptotic cosmic-time 
dependence $a(t) \simeq \exp(H t)$.
Because the enormous cosmic expansion during the slow-roll period redshifts
away all nonvacuum energy in the Universe, it is reasonable to assume
that the quantum state $|\phi\rangle$ would register no particles for
a comoving detector coupled to the fluctuation field $\vphi$ 
at conformal-past infinity; i.e., that the fluctuation field $\vphi$ is in
the vacuum state at $\eta \rightarrow -\infty$.  This would mean that
$a \simeq 1/(H\eta)$ at $\eta \rightarrow -\infty$.  This spacetime is
{\em not\/} asymptotically static in the past, but it is conformally static
with a conformal factor whose nonadiabaticity parameter vanishes at
conformal-past infinity.  Therefore, the best approximation to a
``no-particle'' state for a comoving detector in the asymptotic past is
given by the adiabatic vacuum \cite{parker:1974a} constructed via matching
at $\eta \rightarrow -\infty$.  All higher-order adiabatic vacua will in this
case agree at conformal past infinity.

To construct the $n$th-order adiabatic vacuum matched at an
equal-time hypersurface $\Sigma_{\eta_m}$, one first exactly solves the 
conformal-mode function equation for the quantum field 
[see Eq.\ (\ref{eq-onhmfe}) below].  Since the mode-function equation is 
second order, each $k$ mode has two independent
solutions, which can be represented as $u_k$ and $u_k^{\star}$.  A particular
solution consists of a linear combination of $u_k$ and $u_k^{\star}$.
The adiabatic vacuum is constructed by choosing (for each $k$) a linear
combination which smoothly matches the $n$th-order positive frequency WKB
mode function at $\Sigma_{\eta_m}$.
The resulting orthonormal basis of mode functions is then used to expand
the Heisenberg field operator $\Phi_{\text{{\tiny H}}}(x)$ in terms of $a_k$
and $a_k^{\dagger}$.
The vacuum state is defined by $a_k |\text{vac}\rangle = 0$ for all $k$,
which can be shown to correspond (in the $\eta_m \rightarrow -\infty$ limit)
to the de~Sitter-invariant, adiabatic (Bunch-Davies) vacuum.

\subsection{Post-inflation preheating}
\label{sec-reheating}
Inflation ends when the mean field has rolled down to the point where 
condition (\ref{eq-sras}) ceases to be valid, which we assume occurs at 
conformal time $\eta_0$.  The inflaton mean field then begins to 
oscillate about the true minimum of the effective potential, leading to a 
change in the effective equation of state.  A harmonically
oscillating scalar mean field ($m^2 \gg \lambda \phih^2/6$) 
has an effective equation of state $\bar{\gamma} = 0$, and
a scalar inflaton undergoing extreme large-amplitude oscillations
($\lambda \phih^2/6 \gg m^2$) has an effective equation of state 
$\bar{\gamma} = 1/3$ \cite{kolb:1990a}.  
In realistic models, the inflaton field is coupled to 
various lighter fields consisting of fermions and/or bosons.  
These couplings, as well as the inflaton's self-coupling,
provide mechanisms for damping of the mean-field oscillations
via back reaction from quantum particle production, and energy transfer
to the lighter fields and the inflaton's inhomogeneous modes.

Let us consider the scalar $\lambda \Phi^4$ 
field theory with unbroken symmetry in Minkowski space 
[with classical action given by the Minkowski-space limit of 
Eq.\ (\ref{eq-lpfcsta})],
and suppose that the mean field $\phih$ oscillates about the stable
equilibrium configuration $\phih = 0$  with initial amplitude $\phih_0$.  
For the moment we are neglecting the effect of spacetime dynamics, 
i.e., assuming $a(\eta)=1$.  The
time scale for oscillations of the mean field is given by 
\cite{boyanovsky:1996b}
\begin{equation}
\tau_0 = \frac{4 K(k)}{m \sqrt{1 + f^2}},
\label{eq-deftau0}
\end{equation}
where $f$ and $k$ are defined by
\begin{mathletters}
\begin{eqnarray}
f &=& \sqrt{\frac{\lambda}{6}} \frac{\phih_0}{m}, 
\label{eq-deff}  \\
k &=& \frac{f}{\sqrt{2(1 + f^2)}},
\end{eqnarray}
\end{mathletters}
and $K(k)$ is the complete elliptic integral of the first kind 
\cite{gradshteyn:1964a}. For harmonic oscillations where
\begin{equation}
\frac{\lambda}{6} \phih_0^2 \ll m^2,
\label{eq-tdptas1}
\end{equation}
time-dependent perturbation theory was used in the group 1A studies
(see Sec.~\ref{sec-intro}) to compute the damping
rate $\Gamma$ for the mean field in the $\lambda \Phi^4$ model.   At lowest
order in $\lambda$, the damping rate for the mean field $\phih$
corresponds to the rate for four zero-momentum, free-field excitations of the
inflaton to decay into a $\vphi$ (fluctuation field) particle pair, due to the
$\lambda \phi^4$ self-coupling \cite{shtanov:1995a},
\begin{equation}
\Gamma_{\phi} \simeq O(1) \frac{(\lambda \phih_0)^2}{4\pi m},
\label{eq-tdptdr}
\end{equation}
with vacuum initial state for $\vphi$.
The symbol $O(1)$ denotes a constant of order unity.
In addition to the assumption (\ref{eq-tdptas1}), 
there is another crucial assumption in the derivation
of Eq.\ (\ref{eq-tdptdr}), namely, that the dominant contribution to the decay
rate is given by the lowest order, $|\text{vac}\rangle \rightarrow
|1_{\vec{k}}, \; 1_{-\vec{k}} \rangle$ process, where the occupation numbers  
are for the fluctuation field $\vphi$.
It can be shown  \cite{parker:1969a,hu:1987a} that for this (bosonic) case, the
perturbative decay rate into the $\vec{k}$ momentum shell for the
fluctuation field $\vphi$ is enhanced by $1+2n$ when the occupation of
the $\vec{k}$ shell is $n$.  This is a stimulated 
emission effect due to Bose statistics.\footnote{
In contrast with the case with Bose fields, 
the use of time-dependent perturbation theory to study 
inflaton decay into fermions via a Yukawa coupling does not require the
condition $n_{\vec{k}} \ll 1$, because of the Pauli exclusion principle.
It is still necessary, however, to assume weak coupling (or small 
mean-field amplitude)
in order to use perturbation theory
\cite{dolgov:1990a,shtanov:1995a}.}
The use of Eq.\ (\ref{eq-tdptdr}) to estimate the damping rate thus implicitly
assumes that for all $\vec{k}$, the fluctuation field occupation numbers
are small, i.e., $n_{\vec{k}} \ll 1$.   This is because
time-dependent perturbation theory in terms of the $\lambda \Phi^4$
interaction corresponds to an expansion of the field theory around
the vacuum configuration.  Equivalently, it corresponds to an amplitude
expansion (in powers of the ``classical field'' $\phihpm$) of the 1PI
closed-time-path effective action $\Gamma[\phih_{+},\phih_{-}]$,
which is defined in Eq.\ (2.11) in Ref.\ \cite{ramsey:1997a}. 
When $\lambda \phih_0^2$ is sufficiently large
at $\eta_0$, or on a time scale for $n_{\vec{k}}$ to grow to order unity,
the perturbative expansion in $\lambda$ breaks down.

In many inflationary scenarios, condition (\ref{eq-tdptas1}) does not hold
at $\eta_0$.
A correct analysis of the dynamics of the inflaton field must, therefore,
be nonperturbative, if the inflaton is self-interacting and/or coupled to
Bose fields.  Again, of interest in ``preheating'' 
is the time scale for damping of the
mean field $\phih$ due to back reaction from particle production
into the inhomogeneous modes of the fluctuation field.  
This quantum particle production is known
to occur by parametric amplification of quantum vacuum fluctuations, for
the zero-temperature, unbroken symmetry system under study here.
Boyanovsky {\em et al.\/} \cite{boyanovsky:1996b} have
obtained an approximate analytic expression (in Minkowski space)
for the time scale $\tau_1$ for the variance 
$\langle \vphi_{\text{{\tiny H}}}^2 \rangle$ to
grow to the point where $\lambda \langle \vphi_{\text{{\tiny H}}}^2 \rangle/2$
is of the same order of magnitude as the tree-level effective mass
$m^2 + \lambda \phih^2/6$,
\begin{equation}
\tau_1 = \frac{m^{-1}}{B(f)} \text{ln} \left( \frac{ (1 + 
f^2/2)}{\lambda \sqrt{B(f)}/(8 \pi^2)} \right).
\label{eq-deftau1}
\end{equation}
The function $B(f)$ is of order unity, and in terms of the asymptotic value
of $f$ at $\eta \rightarrow \infty$, $B[f(\eta\rightarrow \infty)] \simeq
0.285 \, 953$.
Their result is valid in flat space and based on
a solution of the one-loop dynamics which neglects the back reaction
of particle production on the mode functions.
The essential feature of the time scale $\tau_1$ is that it depends on
the $\text{ln}(\lambda^{-1})$.  As a consequence of the analytic solution
to the classical mean-field equation and the estimate for $\tau_1$,
it is possible to estimate (for the case of Minkowski space) the effective
equation of state $\bar{\gamma}_{\text{{\tiny C}}}$
for the mean field \cite{boyanovsky:1996b},
\begin{equation}
\bar{\gamma}_{\text{{\tiny C}}} 
\equiv \left(\overline{\frac{p_{\text{{\tiny C}}}}{
\rho_{\text{{\tiny C}}}}}\right) =
\frac{ -\frac{1}{6} f_0^2 \left[ 1 - \frac{1}{2} f_0^2 \right] + \frac{2}{3}
(1 + f_0^2) \left[ 1 - \frac{E(k)}{K(k)} \right]}{\frac{1}{2} f_0^2 \left[
1+ \frac{1}{2} f_0^2 \right]},
\label{eq-aseos}
\end{equation}
where $E(k)$ is the complete elliptic integral of the second kind 
\cite{gradshteyn:1964a}, $p_{\text{{\tiny C}}}$ is the pressure of
the mean field, and $\rho_{\text{{\tiny C}}}$ is the energy density of
the mean field, defined in Eq.\ (\ref{eq-dfrc}).  The late-time effective equation
of state can
be studied using an idealized two-fluid model consisting of the classical
mean-field oscillations $\bar{\gamma}_{\text{{\tiny C}}}$ and a relativistic
component corresponding to the energy density in the quantum modes
$\rho_{\text{{\tiny Q}}}$ [defined in Eq.\ (\ref{eq-rhoqdef}) below].

The physical processes discussed above neglect collisional scattering
of excitations of the inhomogeneous modes due to the
$\lambda \Phi^4$ self-interaction, for example, binary
scattering.  These scattering processes ultimately lead to thermalization
of the system.  
A quantitative understanding of the time scales for such processes
in the nonperturbative regime studied here within a
rigorous field-theoretic framework is at present lacking.
A perturbative treatment of collisional thermalization of the system using the
Boltzmann equation assumes a separation of time scales for collisionless 
processes ($\tau_1$) and thermalization.  However, due to the 
nonperturbatively large occupation numbers which arise in the resonance
band of the inhomogeneous field modes on the time scale $\tau_1$, such a
naive approach would predict that the time scale for thermalization is on
the order of (or earlier than) the preheating time scale $\tau_1$.  A 
nonperturbative approach to studying the collisional thermalization of the
system is, therefore, required.  However, within the $1/N$ expansion 
(to be discussed in Sec.~\ref{sec-onlna}), the collisional scattering 
processes are subleading order in $1/N$, and thus the separation of time 
scales is assured within this controlled expansion \cite{boyanovsky:1996b}.
Let us denote the time scale for scattering by $\tau_2$.  

In typical inflationary scenarios, the self-coupling $\lambda$ of the 
inflaton is very weak \cite{kolb:1990a}, in the range $10^{-12}$--$10^{-14}$
(see, however, \cite{calzetta:1995a,matacz:1996a}).  In our numerical
work, $f$ is initially unity, in which case the three time scales
$\tau_0$, $\tau_1$ and $\tau_2$ separate dramatically,
\begin{mathletters}
\begin{eqnarray}
&& \tau_1/\tau_0  \simeq O\left[\text{ln}\left(\frac{1}{\lambda}
\right)\right],
\label{eq-adtau1}\\
&& \tau_2/\tau_1  \simeq O(N).
\end{eqnarray}
\end{mathletters}
The period leading up to $\tau_1$ is called {\em preheating\/}, because (i)
the energy transfer from the mean field is entirely nonequilibrium in origin,
and (ii) the occupation numbers of the fluctuation field are extremely
nonthermal.  In this regime, since $\tau_2 \gg \tau_1$, collisional effects
can be neglected.  In a collisionless approximation, the damping of the mean
field is due to energy transfer into the inhomogeneous quantum modes, 
a process similar to Landau damping in plasma physics \cite{cooper:1994a}.

So far in this section we have not included 
the effect of spacetime dynamics on the
particle production and back reaction processes.  Cosmic expansion introduces
an additional time scale $H^{-1}$, where $H$ is the Hubble parameter
defined in Eq.\ (\ref{eq-defhinv}).
In typical chaotic inflation scenarios, the initial inflaton amplitude
can be as large as $\Mpl/3$, leading to
\begin{equation}
H^{-1} \simeq \frac{3 \tau_0}{\sqrt{2\pi}}
\end{equation} 
at the onset of reheating.  In this case, $H^{-1} \ll \tau_1$
when $\lambda$ is very small.  Clearly, for sufficiently large initial inflaton
amplitude, it is necessary to include the effect of spacetime dynamics
in a systematic study of preheating dynamics of the inflaton field.

\section{O$(N)$ inflaton dynamics in FRW spacetime}
\label{sec-onfrw}
In this section, we study the nonequilibrium dynamics of a quartically
self-interacting, minimally coupled, O$(N)$ field theory (with unbroken
symmetry) in spatially flat FRW spacetime.  We use the covariant evolution 
equations derived in \cite{ramsey:1997a}, in order to study the dynamics of 
the mean field, variance, and the spacetime, at leading order in the $1/N$
expansion.

\subsection{The O$(N)$ model in the $1/N$ expansion}
\label{sec-onlna}
The classical action for the unbroken symmetry O$(N)$ model in a general curved
spacetime is
\begin{equation}
S^{\text{{\tiny F}}}[\phi^i, g_{\mu\nu}] = 
-\frac{1}{2} \int_M \! d^{\, 4} x \sqmg
\left[ \vecphi \cdot (\square + m^2 + \xi R ) \vecphi + \frac{\lambda}{4 N} 
( \vecphi \cdot \vecphi )^2 \right], \label{eq-onsm} 
\end{equation}
where the O$(N)$ inner product is defined by\footnote{
In our index notation, the latin letters
$i,j,k,l,m,n$  are used to designate O$(N)$ indices 
(with index set $\{1,\ldots, N\}$), while the latin letters
$a,b,c,d,e,f$ are used below to designate CTP indices 
(with index set $\{+,-\}$).}
\begin{equation}
\vecphi \cdot \vecphi = \phi^i \phi^j \delta_{ij}.
\end{equation}
As in Eq.\ (\ref{eq-lpfcsta}), $\lambda$ is a coupling constant with dimensions
of $1/\hbar$, and $\xi$ is the dimensionless coupling to gravity (and is 
necessary in order for the quantized theory to be renormalizable).

In \cite{ramsey:1997a}, the covariant mean-field equation, gap equation, 
and geometrodynamical field equation were computed for this model at leading
order in the $1/N$ expansion.  The evolution equations follow from 
Eqs.\ (\ref{eq-lpfsee})--(\ref{eq-lpfseec}), 
with the 2PI, CTP effective action truncated at
leading order in the $1/N$ expansion.  At leading order in the $1/N$ expansion,
we need only keep track of one component of the CTP two-point function
$G_{ab}(x,x')$; we choose $G_{++}(x,x')$, which is the Green function with
Feynman boundary conditions.  The covariant gap equation for $G_{++}$ at 
leading order in the $1/N$ expansion is
\begin{equation}
\left( \square_x + m^2 + \xi R(x) + \frac{\lambda}{2}\phih^2(x) + 
\frac{\hbar \lambda}{2} G_{++}(x,x) 
\right) G_{++}(x,x')  = 
\delta^4(x-x') \frac{-i}{\sqmgp},
\label{eq-onige}
\end{equation}
plus terms of $O(1/N)$.  The covariant $\delta$ function is defined
in Ref.\ \cite{birrell:1982a}.
The mean-field equation is, at leading order in $1/N$,
\begin{equation}
\left(\square + m^2 + \xi R + \frac{\lambda}{2} \phih^2 + 
\frac{\hbar \lambda}{2}
G_{++}(x,x) \right) \phih(x) = 0,
\label{eq-onmflm}
\end{equation}
where we note that $G_{++}(x,x) = G_{ab}(x,x)$ for all $a,b \in \{+,-\}$.
The coincidence limit $G_{++}(x,x)$ is divergent in four spacetime dimensions,
and the regularization method is described in Sec.~\ref{sec-onrenorm} below.
The geometrodynamical field equation is
\begin{equation}
G_{\mu\nu} + \Lambda g_{\mu\nu} + 
c\; ^{(1)\!}H_{\mu\nu} + b\; ^{(2)\!}H_{\mu\nu} = -8 \pi G
\langle T_{\mu\nu} \rangle
\label{eq-lpfcstsee2}
\end{equation}
in terms of the (unrenormalized) energy-momentum tensor computed at leading 
order in the $1/N$ expansion, which is shown in 
Eqs.\ (5.37) and (5.38) in Ref.\ \cite{ramsey:1997a}.

\subsection{Restriction to FRW spacetime}
\label{sec-resfrw}
Let us now specialize to the spatially flat FRW universe, with initial
conditions appropriate to post-inflation dynamics of the inflaton field.  
As discussed in Sec.~\ref{sec-initcond}, initial Cauchy data for
$\phih$, $G_{++}$, and $a$ are specified on a spacelike hypersurface
$\Sigma_{\eta_0}$ (at conformal time $\eta_0$).  
The spatial symmetries of $\phih$ and $G_{++}$ for a quantum state 
$|\phi\rangle$ consistent with a spatially homogeneous and isotropic 
cosmology are given in (\ref{eq-gfssym}--b).  As a consequence of these
symmetries, both the mean field $\phih$ and variance
$\langle \vphi_{\text{{\tiny H}}}^2 \rangle$ are spatially homogeneous,
i.e., functions of conformal time only. 

Eq.\ (\ref{eq-onige}) for $G_{++}$ in spatially flat FRW spacetime
has the formal solution
\begin{eqnarray}
G_{++}(x,x') = a(\eta)^{-1} a(\eta')^{-1} \int \frac{d^{\,3}k}{(2\pi)^3}
e^{i \vec{k} \cdot (\vec{x} - \vec{x}')} \bigl[
\Theta(\eta' - \eta) & & \tilde{u}_k(\eta)^{\star} \tilde{u}_k(\eta')  
\nonumber \\ & & 
+ \Theta(\eta - \eta') \tilde{u}_k(\eta')\tilde{u}_k(\eta)^{\star} \bigr],
\end{eqnarray}
in terms of conformal-mode functions $\tilde{u}_k$ which satisfy a
harmonic oscillator equation with conformal-time-dependent effective
frequency,
\begin{equation}
\left( \frac{d^2}{d\eta^2} + \Omega_k^2(\eta) \right) \tilde{u}_k = 0.
\label{eq-onhmfe}
\end{equation}
The fact that $\tilde{u}_k(\eta)$ depends only on $\eta$ and $k$ (where $k$
is comoving momentum) implies that $G_{++}$ is invariant under simultaneous
spatial translations and rotations of $\vec{x}$ and $\vec{x}'$.
The effective frequency $\Omega_k(\eta)$ appearing in Eq.\ (\ref{eq-onhmfe}) is
defined by
\begin{mathletters}
\begin{eqnarray}
\Omega_k^2(\eta) &=& k^2 + a^2 {\frak M}^2(\eta), 
\label{eq-oneffrq}  \\
{\frak M}^2(\eta) &=& M^2(\eta) + \left( \xi - \frac{1}{6} \right) R(\eta),
\label{eq-oneffrqb}  \\
M^2(\eta) &=& m^2 + \frac{\lambda}{2}\phih^2(\eta) + 
\frac{\lambda}{2}\langle \vphi_{\text{{\tiny H}}}^2(\eta) \rangle. 
\label{eq-oneffrqc} 
\end{eqnarray}
\end{mathletters}
Initial conditions for the positive frequency conformal mode functions
$\tilde{u}_k(\eta)$ must be specified (for all $k$) at $\eta_0$.  A choice
of initial conditions corresponds to a choice of quantum state 
$|\phi\rangle$ for the fluctuation field $\vphi_{\text{{\tiny H}}}$; initial
conditions are discussed in Sec.~\ref{sec-onbc} below.
The (bare) variance $\langle \vphi_{\text{{\tiny H}}}^2 \rangle$ 
has a simple representation in terms of the conformal-mode functions:
\begin{equation}
\langle \phi | \vphi_{\text{{\tiny H}}}(x)^2 
| \phi \rangle = \hbar G_{++}(x,x) 
 = \frac{\hbar}{a^2} \int \frac{d^3k}{(2\pi)^3} | \tilde{u}_k(\eta) |^2.
\label{eq-frwfluc}
\end{equation}
It should be noted that this expression is divergent, in
consequence of our having computed the variance in terms
of the bare (unrenormalized) constants of the theory.  In terms of a
physical upper momentum cutoff $K$, $G_{++}(x,x)$ diverges like
$K^2$; there is additionally a logarithmic dependence on $K$.
In addition, the mode functions $\tilde{u}_k$ depend on $\langle 
\vphi_{\text{{\tiny H}}}^2 \rangle$ through Eq.\ (\ref{eq-oneffrqc}).
The leading-order, large-$N$, mean-field equation in spatially flat FRW 
spacetime becomes 
\begin{equation}
\phih'' + \frac{2 a'}{a} \phih' + a^2 M^2(\eta)
\phih = 0,
\label{eq-bmfeqfrw}
\end{equation}
where the time-dependent bare effective mass $M(\eta)$ is given by 
Eq.\ (\ref{eq-oneffrqc}).  For simplicity of notation, we will henceforth 
write $M$ instead of $M(\eta)$, and similarly for ${\frak M}(\eta)$.

Finally, we can express the bare energy-momentum tensor in terms of
the conformal-mode functions $\tilde{u}_k(\eta)$.  As discussed in
Sec.~\ref{sec-idfrw}, it is convenient to work with the 00 
component and the trace of the energy-momentum tensor.  
The components of the classical part of the energy-momentum 
tensor are spatially homogeneous, and given by
\begin{mathletters}
\begin{eqnarray}
T_{00}^{\text{{\tiny C}}}(\eta) &=&  \frac{1}{2} (\phih')^2
- \frac{3 \xi}{2}  \left( \phih'' + \frac{2 a'}{a} \phih' \right) \phih 
+ \frac{1}{2} a^2 \left( m^2 + \frac{\lambda}{4} \phih^2 
+ \frac{3 \xi (a')^2}{2 a^4} \right) \phih^2,
\label{eq-onemtcla}  \\
{\mathcal T}^{\text{{\tiny C}}}(\eta) &=& 
\frac{1}{a^2} \left\{(6\xi-1)(\phih')^2 + 6\xi\left( \phih'' 
+ \frac{2a'}{a}\phih' \right) \phih \right\} 
+ 2 \left( m^2 + \frac{\lambda}{4}\phih^2 + \frac{\xi}{2}
R \right) \phih^2.
\label{eq-onemtclb} 
\end{eqnarray}
\end{mathletters}
The quantum energy-momentum tensor components are also spatially homogeneous.
We find for the 00 component,
\begin{eqnarray}
T^{\text{{\tiny Q}}}_{00}(\eta) = \frac{\hbar}{2 a^2} 
\int \frac{d^{\,3}k}{(2\pi)^3} \Biggl[  | \tilde{u}_k'|^2  +
\Bigl( k^2 + a^2 M^2 + (1 && -6\xi) \frac{(a')^2}{a^2} \Bigr) |
\tilde{u}_k|^2 \nonumber \\ 
&& + (6 \xi - 1) \frac{a'}{a}\Bigl[ 
(\tilde{u}_k')^{\star}\tilde{u}_k
+ \tilde{u}_k'\tilde{u}_k^{\star} \Bigr] \Biggr],
\label{eq-onfrwetc}
\end{eqnarray}
and for the trace,
\begin{eqnarray}
{\mathcal T}^{\text{{\tiny Q}}}(\eta) = \frac{\hbar}{a^4} 
\int \frac{d^{\,3}k}{(2\pi)^3} \Biggl[  (6\xi - 1)\biggl\{
| \tilde{u}_k' |^2 - (k^2 && + a^2 M^2) |\tilde{u}_k|^2 
- \frac{a'}{a}\Bigl[ (\tilde{u}_k')^{\star}
\tilde{u}_k + \tilde{u}_k'\tilde{u}_k^{\star}\Bigr] \nonumber \\ & &
+ \biggl( \frac{(a')^2}{a^2} - \xi a^2 R \biggr) | \tilde{u}_k|^2 \biggr\}
+ a^2 M^2 | \tilde{u}_k|^2 \Biggr].
\label{eq-onfrwetcp}
\end{eqnarray}
It can be shown by asymptotic analysis that, in terms of
a physical upper momentum cutoff $K$, the bare $T_{00}^{\text{{\tiny Q}}}$
is quartically divergent, i.e., $O(K^4)$, and that (for minimal coupling)
${\mathcal T^{\text{{\tiny Q}}}}$ is quadratically divergent.
In addition, the components of the bare energy-momentum
tensor contain the effective mass $M^2$, which contains the divergent
variance $\langle \vphi_{\text{{\tiny H}}}^2 \rangle$.  The energy density 
$\rho_{\text{{\tiny Q}}}$ of quantum modes of the $\vphi$ field
is defined in terms of $T^{\text{{\tiny Q}}}_{00}$ by
\begin{equation}
\rho_{\text{{\tiny Q}}} = \frac{1}{a^2} T^{\text{{\tiny Q}}}_{00}
- \frac{\lambda}{8}\langle \vphi_{\text{{\tiny H}}}^2 \rangle^2.
\label{eq-rhoqdef}
\end{equation}
We shall also refer to $\rho_{\text{{\tiny Q}}}$ as the energy density of the
``fluctuation field.'' 

\subsection{Renormalization of the dynamical equations}
\label{sec-onrenorm}
The variance $\langle \vphi_{\text{{\tiny H}}}^2 \rangle$ and
quantum energy-momentum tensor components $T_{00}^{\text{{\tiny Q}}}$
and ${\mathcal T}^{\text{{\tiny Q}}}$ are divergent in four spacetime 
dimensions, and must be regularized within the context of a systematic, 
covariant renormalization procedure.  In the ``in-out'' formulation of 
quantum field theory, renormalization may be carried out via addition
of counterterms to the effective action, which amounts to renormalization 
of the constants in the classical action \cite{ramond:1990a}.
The closed-time-path formulation of the effective dynamics is renormalizable
provided the theory is renormalizable in the ``in-out'' formulation
\cite{hartle:1979a,calzetta:1987a}, as is the case with the O$(N)$ field
theory in curved spacetime 
\cite{ramsey:1997a,toms:1982a,mazzitelli:1989b}.
For our purposes it is convenient (in this model) to carry out
renormalization in the leading-order, large-$N$, evolution equations,
rather than in the CTP effective action \cite{jordan:1986a}.

In this study we employ the adiabatic regularization method of
Parker, Fulling, and Hu \cite{parker:1974a,fulling:1974a}.  The idea
is to define an adiabatic approximation to the conformal mode function, 
and then to construct a regulator for the integrands of the bare 
energy-momentum tensor and
variance from the adiabatic mode functions \cite{fulling:1989a}.   
Renormalization occurs when we define the renormalized variance
and energy-momentum tensor to be the difference between the bare expressions
and the regulators and simultaneously replace the bare quantities
$m,$ $\lambda,$ $G,$ $b,$ $c,$ $\Lambda,$ and $\xi$ by their renormalized
counterparts.  The equivalence of this procedure to other manifestly
covariant methods (such as dimensional continuation) is well established 
\cite{bunch:1980a}.
We implement renormalization as a two-step process: First, we  adiabatically
regularize the variance and renormalize $\xi$, $m$, and
$\lambda$; Second, we adiabatically regularize the energy-momentum tensor
and renormalize the semiclassical geometrodynamical field equation.

We define the adiabatic order of a conformal mode function as follows: 
let $\Omega_k(\eta) \rightarrow \Omega_k(\eta/T)$, where $T$ is introduced
as a time scale which is formally taken to be unity at the end of the
calculation.  Then the adiabatic order of an expression involving
derivatives of $\Omega_k$ is simply the inverse power of $T$, of the
leading-order term in an asymptotic expansion about $T \rightarrow \infty$.
However, in order for the
adiabatically regulated energy-momentum tensor for an interacting scalar
field theory to agree with the renormalized energy-momentum tensor
obtained by manifestly covariant methods (e.g., covariant point splitting 
\cite{bunch:1978a}), it is 
necessary to define the adiabatic order of expressions involving $\lambda$ and
derivatives with respect to $\eta$,
such as $\lambda (\phih^2)''$, as the sum of the exponent of $\phih$ and
the number of conformal time differentiations \cite{paz:1988a}.  Therefore,
$\lambda \langle \vphi_{\text{{\tiny H}}}^2 \rangle''$ is considered fourth
adiabatic order, as is $\lambda (\phih^2)''$.  

Having defined adiabatic order, we now construct the 
adiabatic mode functions. It is well known that the WKB ansatz
\begin{equation}
\tilde{u}_k(\eta) = \frac{1}{\sqrt{2 W(\eta)}} \exp \left(
i \int^{\eta} d\eta' W(\eta') \right)
\end{equation}
turns the harmonic oscillator equation (with time-dependent frequency 
$\Omega_k$) into a nonlinear differential equation for $W$,
\begin{equation}
W(\eta)^2 = \Omega_k^2(\eta) + \frac{3 [W'(\eta)]^2}{4 W^2(\eta)} -
\frac{W''(\eta)}{2 W(\eta)}.
\end{equation}
Starting with the lowest-order ansatz $W^{(0)}(\eta) = \Omega_k(\eta)$,
one can iterate this equation; the $n$th-order iteration
yields the $n$th-order WKB approximation for $\tilde{u}_k$.
For the free field theory, the $n$th-order WKB approximation gives
an expression for $\tilde{u}_k$ which is of adiabatic order $2n$.
In the interacting case, the above definition of adiabatic order calls for 
removing terms such as $\lambda (\phih^2)''''$ at 4th adiabatic order.
Thus we have a method of deriving expressions for 
$T^{\text{{\tiny Q}}}_{00}$, ${\mathcal T}^{\text{{\tiny Q}}}$,
and $\langle \vphi_{\text{{\tiny H}}}^2 \rangle$ at fourth, fourth,
and second adiabatic orders, respectively.  One then sets $T=1$ in the
truncated expression.  
We can thus obtain a fourth-order adiabatic approximation to the quantum
energy-momentum tensor $(T_{\mu\nu}^{\text{{\tiny Q}}})_{\text{{\tiny ad4}}}$,
and a second-order adiabatic approximation to the variance 
$\langle \vphi_{\text{{\tiny H}}} \rangle_{\text{{\tiny ad2}}}$.
By subtracting $(T_{\mu\nu}^{\text{{\tiny Q}}})_{
\text{{\tiny ad4}}}$ from the divergent 
$T_{\mu\nu}^{\text{{\tiny Q}}}$ and
$\langle \vphi_{\text{{\tiny H}}}^2 \rangle_{\text{{\tiny ad2}}}$ from
the divergent
$\langle \vphi_{\text{{\tiny H}}}^2 \rangle$, finite expressions for the
renormalized energy-momentum tensor and variance are obtained.

First we regularize the variance $\langle \vphi_{
\text{{\tiny H}}}^2 \rangle$, and carry out a renormalization 
of $\lambda$, $m$, and $\xi$.
In the leading-order, large-$N$ approximation, no terms appear 
in the mode-function equation (\ref{eq-onhmfe}) which would permit
addition of counterterms; therefore, $\Omega_k$ must be finite 
\cite{root:1974a}.  The effective frequency $\Omega_k$ which appears in 
Eq.\ (\ref{eq-oneffrq})
is the ``bare'' effective frequency, which we denote by
$(\Omega_k)_{\text{{\tiny B}}}$.
In conjunction with the adiabatic regularization procedure,
we fix the renormalization scheme by demanding equivalence of the bare and 
renormalized effective mass \cite{cooper:1994a},
\begin{equation}                                             
(\Omega_k^2)_{\text{{\tiny R}}} = (\Omega_k^2)_{\text{{\tiny B}}},
\label{eq-rencst}
\end{equation}
where the ``R'' subscripted quantities are renormalized.  Using
Eqs.\ (\ref{eq-oneffrq}) and (\ref{eq-oneffrqb}), we have
\begin{equation}
\xi_{\text{{\tiny R}}} R + M^2_{\text{{\tiny R}}} =
\xi_{\text{{\tiny B}}} R + M^2_{\text{{\tiny B}}},
\label{eq-rencst2}
\end{equation}
where $M^2_{\text{{\tiny B}}}$ is defined in Eq.\ (\ref{eq-oneffrqc}),
\begin{equation}
M^2_{\text{{\tiny B}}} = m_{\text{{\tiny B}}}^2 +
\frac{\lambda_{\text{{\tiny B}}}}{2} \phih^2 + 
\frac{\lambda_{\text{{\tiny B}}}}{2} \langle \vphi_{\text{{\tiny H}}}^2
\rangle_{\text{{\tiny B}}},
\end{equation}
and $M^2_{\text{{\tiny R}}}$ is defined similarly,
\begin{equation}
M^2_{\text{{\tiny R}}} = m_{\text{{\tiny R}}}^2 +
\frac{\lambda_{\text{{\tiny R}}}}{2} \phih^2 + 
\frac{\lambda_{\text{{\tiny R}}}}{2} \langle \vphi_{\text{{\tiny H}}}^2
\rangle_{\text{{\tiny R}}}.
\label{eq-reneff}
\end{equation}
Now, $\lambda_{\text{{\tiny B}}}$, $m_{\text{{\tiny B}}}$,
and $\xi_{\text{{\tiny B}}}$ are the bare constants of the theory
which appeared (without B's) in the classical action (\ref{eq-onsm}).
The renormalized quantities in Eq.\ (\ref{eq-oneffrq}) are defined below.
The bare 
$\langle \vphi_{\text{{\tiny H}}}^2\rangle_{\text{{\tiny B}}}$ 
is a conformal-time-dependent function defined by Eq.\ (\ref{eq-frwfluc}),
\begin{equation}
\langle \vphi_{\text{{\tiny H}}}^2(\eta) \rangle_{\text{{\tiny B}}} =
\frac{\hbar}{a^2} \int \frac{d^{\,3}k}{(2\pi)^3} | \tilde{u}_k(\eta) |^2,
\label{eq-renbfle}
\end{equation}
where the conformal-mode functions $\tilde{u}_k(\eta)$ obey 
Eq.\ (\ref{eq-onhmfe}).  Now we demand that the renormalized
theory be minimally coupled, i.e., we set $\xi_{\text{{\tiny R}}} = 0$.
Because of Eq.\ (\ref{eq-rencst}), we can formally use
$(\Omega_k^2)_{\text{{\tiny R}}}$ in computing the adiabatic regulator for
the variance $\langle \vphi_{\text{{\tiny H}}}^2 \rangle_{\text{{\tiny B}}}$.
Computing the asymptotic series (in $1/T$) of the
quantity $|\tilde{u}_k(\eta)|^2$ to $O(1/T^2)$, where
$\Omega_k^2(\eta/T)$ is the effective frequency, we obtain
(after setting $T=1$)
\begin{equation}
\langle \vphi_{\text{{\tiny H}}} \rangle_{\text{{\tiny ad2}}}
= \frac{\hbar}{2 C} \int \frac{d^{\,3}k}{(2\pi)^3}
\left[ \frac{1}{\tilde{\omega}_k}
- \frac{(C')^2 - 2C C''}{8 C^2 \tilde{\omega}_k^3} + \frac{M^2_{ 
\text{{\tiny R}}} C''}{8
\tilde{\omega}_k^5} - \frac{5 M^4_{\text{{\tiny R}}} 
(C')^2}{32 \tilde{\omega}_k^7} \right],
\label{eq-onaregfluc}
\end{equation}
in terms of auxiliary functions
\begin{equation}
C(\eta) = a^2(\eta),
\end{equation}
and 
\begin{equation}
D(\eta) = \frac{C'(\eta)}{C(\eta)}.
\end{equation}
In Eq.\ (\ref{eq-onaregfluc}) the symbol $\tilde{\omega}_k$ is defined 
as follows
\begin{equation}
\tilde{\omega}_k^2 = k^2 + a^2 M^2_{\text{{\tiny R}}}.
\label{eq-tilomk}
\end{equation}
In the adiabatic prescription, the renormalized variance
$\langle \vphi_{\text{{\tiny H}}}^2 \rangle_{\text{{\tiny R}}}$
appearing in Eq.\ (\ref{eq-reneff}) is defined by
\begin{equation}
\langle \vphi_{\text{{\tiny H}}}^2 \rangle_{\text{{\tiny R}}} =
\langle \vphi_{\text{{\tiny H}}}^2 \rangle_{\text{{\tiny B}}} -
\langle \vphi_{\text{{\tiny H}}}^2 \rangle_{\text{{\tiny ad2}}},
\end{equation}
where the first term on the right-hand side is given by 
Eq.\ (\ref{eq-renbfle}), and the second term on the right-hand 
side is given by Eq.\ (\ref{eq-onaregfluc}).
Everything on the right hand side can be expressed 
in terms of renormalized quantities, so this procedure is well defined.
Written out explicitly, the renormalized variance satisfies the equation
\begin{equation}
\langle \vphi_{\text{{\tiny H}}}^2 \rangle_{\text{{\tiny R}}} =
\frac{\hbar}{C} \int \frac{d^{\,3}k}{(2\pi)^3} \left[
|\tilde{u}_k|^2 - \frac{1}{2\tilde{\omega}_k} - 
\frac{(C')^2 - 2 C C''}{16 C^2 \tilde{\omega}_k^3} +
\frac{M^2_{\text{{\tiny R}}} C''}{16 \tilde{\omega}_k^5} -
\frac{5 M^4_{\text{{\tiny R}}} (C')^2}{64 \tilde{\omega}_k^7}
\right]. 
\label{eq-renrfluc}
\end{equation}
One can use the WKB approximation for $\tilde{u}_k(\eta)$ to compute
the asymptotic series for the integrand in Eq.\ (\ref{eq-renrfluc})
in the limit $k \rightarrow \infty$, and show that the integral is
convergent.
Since $M^2_{\text{{\tiny R}}}$ is contained in the
integrand above, Eq.\ (\ref{eq-renrfluc}) leads to an integral 
equation for the renormalized effective mass
$M_{\text{{\tiny R}}}$,
\begin{equation}
M_{\text{{\tiny R}}}^2 = m_{\text{{\tiny R}}}^2 + \frac{
\lambda_{\text{{\tiny R}}}}{2} \phih^2 + \frac{\hbar \lambda_{
\text{{\tiny R}}}}{2 C} \int \frac{d^{\,3}k}{(2\pi)^3} \left[
|\tilde{u}_k|^2 - \frac{1}{2\tilde{\omega}_k} - 
\frac{(C')^2 - 2 C C''}{16 C^2 \tilde{\omega}_k^3} +
\frac{M^2_{\text{{\tiny R}}} C''}{16 \tilde{\omega}_k^5} -
\frac{5 M^4_{\text{{\tiny R}}} (C')^2}{64 \tilde{\omega}_k^7}
\right].
\label{eq-rem}
\end{equation}
Eqs.\ (\ref{eq-renrfluc}) and (\ref{eq-rencst2})  together define
$\lambda_{\text{{\tiny R}}}$ and $m_{\text{{\tiny R}}}$.  All physical
quantities should now be expressed in terms of the renormalized parameters
$m_{\text{{\tiny R}}}$ and $\lambda_{\text{{\tiny R}}}$ of the theory.
The renormalized mean-field equation now reads
\begin{equation}
\phih'' + \frac{2 a'}{a} \phih' + a^2 M^2_{\text{{\tiny R}}} \phih = 0,
\label{eq-eveqp}
\end{equation}
where $M_{\text{{\tiny R}}}^2$ is given by Eq.\ (\ref{eq-rem}), and the mode
functions in Eq.\ (\ref{eq-rem}) obey the homogeneous equation,
\begin{equation}
\left(\frac{d^2}{d\eta^2} + k^2 + a^2 M_{\text{{\tiny R}}}^2\right)
\tilde{u}_k = 0.
\label{eq-eveqmf}
\end{equation}
The initial conditions for the conformal-mode functions at $\eta_0$ are 
discussed in Sec.~\ref{sec-onbc} below.

Having obtained a renormalized mean-field equation, we now turn our
attention to regularizing the quantum energy-momentum tensor.
As a consequence of Eq.\ (\ref{eq-rencst}), we can substitute 
$M \rightarrow M_{\text{{\tiny R}}}$ and
$\xi \rightarrow \xi_{\text{{\tiny R}}}$ in the equations for the
components of the quantum energy-momentum tensor, 
Eqs.\ (\ref{eq-onfrwetc},
\ref{eq-onfrwetcp}).  Since we wish to study the minimal coupling case,
we set $\xi_{\text{{\tiny R}}}= 0$.  To avoid confusion we denote 
the bare energy-momentum tensor components (\ref{eq-onfrwetc}) and
(\ref{eq-onfrwetcp}) by $(T^{\text{{\tiny Q}}}_{00})_{\text{{\tiny B}}}$
and $({\mathcal T}^{\text{{\tiny Q}}})_{\text{{\tiny B}}}$, respectively.
Let us also relabel the bare constants $b$, $c$, $G$, and $\Lambda$
appearing in the bare semiclassical Einstein equation
(\ref{eq-lpfcstsee2}) as $b_{\text{{\tiny B}}}$,
$c_{\text{{\tiny B}}}$, $G_{\text{{\tiny B}}}$, and
$\Lambda_{\text{{\tiny B}}}$.
Applying the method described above to construct the adiabatic regulator, 
for $T^{\text{{\tiny Q}}}_{00}$ we find
\begin{eqnarray}
\left(T_{00}^{\text{{\tiny Q}}}\right)_{\text{{\tiny ad4}}} = 
\frac{\hbar}{4 C} \int \frac{d^{\,3}k}{(2\pi)^3} 
\Biggl\{ && 2 \tilde{\omega}_k +
\frac{(C')^2}{4 C^2 \tilde{\omega}_k} + \biggl[ \frac{M_{\text{{\tiny R}}}^2 
(C')^2}{4C}
- \frac{9 (C')^4}{64 C^4} \nonumber \\ && + \frac{C' (M_{\text{{\tiny R}}}^2)'}{4} +
\frac{(C')^2 C''}{4 C^3} + \frac{(C'')^2}{16 C^2} - 
\frac{C'C'''}{8 C^2}\biggr] \frac{1}{\tilde{\omega}_k^3} \nonumber \\ && +
\biggl[ \frac{M_{\text{{\tiny R}}}^4 (C')^2}{16} + 
\frac{M_{\text{{\tiny R}}}^2}{32 C^3}
\Bigl( -5 (C')^4 + 4 C^4 (C') (M_{\text{{\tiny R}}}^2)' \nonumber \\ && + 10 C (C')^2 C''
 +
2 C^2 (C'')^2 - 4 C^2 C' C''' \Bigr) \biggr] \frac{1}{\tilde{\omega}_k^5} 
\nonumber \\ && +
\biggl( \frac{M_{\text{{\tiny R}}}^4}{128 C^2} \biggr)\biggl[ -5(C')^4 +
40 (C')^2 C'' + 2 C^2 (C'')^2 \nonumber \\ && 
- 4 C^2 C' C''' \biggr] \frac{1}{\tilde{\omega}_k^7}
+ \frac{7 M_{\text{{\tiny R}}}^6 (C')^2}{128 C} \Bigl( -5 (C')^2 + 2 C C'' 
\Bigr)
\frac{1}{\tilde{\omega}_k^9} \nonumber \\ &&
- \frac{105 M_{\text{{\tiny R}}}^8 (C')^4}{1024} \frac{1}{
\tilde{\omega}_k^{11}} \Biggr\},
\label{eq-onaregt00}
\end{eqnarray}
where $\tilde{\omega}_k$ is defined in Eq.\ (\ref{eq-tilomk}). For ${\mathcal T}$, 
we find
\begin{eqnarray}
({\mathcal T}^{\text{{\tiny Q}}})_{\text{{\tiny ad4}}} = 
\frac{\hbar}{2 C^2} \int \frac{d^{\,3}k}{(2\pi)^3} && 
\Biggl\{ \Bigl( M_{\text{{\tiny R}}}^2 C - \frac{(C')^2}{2 C^2} + 
\frac{C''}{2C}\Bigr) \frac{1}{\tilde{\omega}_k} \nonumber \\ && +
\biggl[ \frac{M_{\text{{\tiny R}}}^2}{8C}\Bigl( -3(C')^2 + 4 C C'' \Bigr) +
\frac{1}{16 C^4} \Bigl( 9(C')^4 + 4 C^4 C' (M_{\text{{\tiny R}}}^2)'\nonumber \\ && -
21 C (C')^2 C'' + 6 C^2 (C'')^2 + 4 C^5 (M_{\text{{\tiny R}}}^2)'' + 8 C^2 C' 
C''' \nonumber \\ &&
- 2 C^3 C''''\Bigr) \biggr]\frac{1}{\tilde{\omega}_k^3} 
+ \biggl[ \frac{M_{\text{{\tiny R}}}^4}{8}\Bigl( -3 (C')^2 + C C''\Bigr) \nonumber \\ && +
\frac{M_{\text{{\tiny R}}}^2}{128 C^3}\Bigl( 87 (C')^4 - 64 C^4 C' 
(M_{\text{{\tiny R}}}^2)' 
- 208 C (C')^2 C'' \nonumber \\ && + 60 C^2 (C'')^2 + 16 C^5 (M_{\text{{\tiny R}}}^2)'' + 
80 C^2 C' C''' - 16 C^3 C'''' \Bigr) \biggr] \frac{1}{\tilde{\omega}_k^5} \nonumber \\ &&
+ \biggl[ -\frac{5 C M_{\text{{\tiny R}}}^6}{32} (C')^2 + 
\frac{M_{\text{{\tiny R}}}^4}{32 C^2}
\Bigl( 15 (C')^4 - 10 C^4 C' (M_{\text{{\tiny R}}}^2)' \nonumber \\ && - 40 C (C')^2 C'' 
+ 15 (C')^2 (C'')^2 + 20 C^2 C' C''' - C^3 C'''' \Bigr) \biggr] 
\frac{1}{\tilde{\omega}_k^7}  \nonumber \\ && + \biggl[ 
\frac{7 M_{\text{{\tiny R}}}^6}{256 C}
\Bigl( 15 (C')^4 - 80 C (C')^2 C'' + 6 C^2 (C'')^2 + 8 C^2 C' C''' \Bigr)
\Biggr] \frac{1}{\tilde{\omega}_k^9} \nonumber \\ && +
\frac{21 M_{\text{{\tiny R}}}^8 (C')^2}{256}\Bigl( 15 (C')^2 - 11 C C'' \Bigr)
\frac{1}{\tilde{\omega}_k^{11}} + \frac{1155 C M_{\text{{\tiny R}}}^{10}}{2048}
(C')^4 \frac{1}{\tilde{\omega}_k^{13}}  \Biggr\}.
\label{eq-onaregtr}
\end{eqnarray}
In the free-field limit ($\lambda_{\text{{\tiny R}}} = 0$), the regulators
(\ref{eq-onaregt00}) and (\ref{eq-onaregtr})
agree with the minimal-coupling, spatially flat limit of the adiabatic
regulators obtained by Bunch \cite{bunch:1980a}.

The renormalization procedure for the semiclassical Einstein equation 
(\ref{eq-lpfcstsee2}) can now be precisely stated.  According to the adiabatic
prescription, we define the quantum energy-momentum tensor by
\begin{equation}
(T_{\mu\nu}^{\text{{\tiny Q}}})_{\text{{\tiny R}}} =
(T_{\mu\nu}^{\text{{\tiny Q}}})_{\text{{\tiny B}}} -
(T_{\mu\nu}^{\text{{\tiny Q}}})_{\text{{\tiny ad4}}}.
\label{eq-tqren}
\end{equation}
It can be checked that the momentum-integral expressions for the two 
independent components of Eq.\ (\ref{eq-tqren}) are convergent.
In terms of $(T^{\text{{\tiny Q}}}_{\mu\nu})_{\text{{\tiny R}}}$,
the total energy-momentum tensor (after renormalization) is
\begin{equation}
\langle T_{\mu\nu} \rangle_{\text{{\tiny R}}} = 
(T^{\text{{\tiny C}}}_{\mu\nu})_{
\text{{\tiny R}}} + (T^{\text{{\tiny Q}}}_{\mu\nu})_{\text{{\tiny R}}}
- \frac{\lambda_{\text{{\tiny R}}}}{8} \left(\langle \vphi_{\text{{\tiny H}}}^2
\rangle_{\text{{\tiny R}}}\right)^2 g_{\mu\nu},
\end{equation}
where $(T^{\text{{\tiny C}}}_{\mu\nu})_{\text{{\tiny R}}}$ stands for
$T^{\text{{\tiny C}}}_{\mu\nu}$, and renormalized quantities are substituted
for bare quantities.  The bare quantities $G_{\text{{\tiny B}}}$,
$\Lambda_{\text{{\tiny B}}}$, $b_{\text{{\tiny B}}}$, and
$c_{\text{{\tiny B}}}$ are now replaced by $G_{\text{{\tiny R}}}$,
$\Lambda_{\text{{\tiny R}}}$, $b_{\text{{\tiny R}}}$, and
$c_{\text{{\tiny R}}}$ in the renormalized semiclassical geometrodynamical
field equation,
\begin{equation}
G_{\mu\nu} + \Lambda_{\text{{\tiny R}}} g_{\mu\nu} +
c_{\text{{\tiny R}}} \, ^{(1)}H_{\mu\nu} +
b_{\text{{\tiny R}}} \, ^{(2)}H_{\mu\nu} = -8\pi 
G_{\text{{\tiny R}}} \langle T_{\mu\nu} \rangle_{\text{{\tiny R}}}.
\label{eq-rscee}
\end{equation}
\subsection{Renormalized semiclassical Einstein equation}
\label{sec-rsee}
Using semiclassical methods to study the dynamics of the inflaton field
in FRW spacetime requires that the Hubble parameter be much less than the
Planck mass, $H \ll \Mpl.$
On dimensional grounds, $c_{\text{{\tiny R}}}$ and $b_{\text{{\tiny R}}}$
are likely to be of order $\hbar^2 \Mpl^{-2}$, in which case
$R \gg c_{\text{{\tiny R}}} R^2,$ 
and $R \gg b_{\text{{\tiny R}}} R^{\alpha\beta}
R_{\alpha\beta}$, provided $R_{\alpha\beta} \neq 0$.
Let us, therefore, set $b_{\text{{\tiny R}}} = 0$ and
$c_{\text{{\tiny R}}} = 0$, and additionally, let us choose
$\Lambda_{\text{{\tiny R}}} = 0$, so that Eq.\ (\ref{eq-rscee}) becomes
the renormalized semiclassical Einstein equation (without cosmological 
constant),
\begin{equation}
G_{\mu\nu} = -8\pi G_{\text{{\tiny R}}}\left[
(T_{\mu\nu}^{\text{{\tiny C}}})_{\text{{\tiny R}}} +
(T_{\mu\nu}^{\text{{\tiny Q}}})_{\text{{\tiny R}}} -
\frac{\lambda_{\text{{\tiny R}}}}{8} (\langle \vphi_{\text{{\tiny H}}}^2
\rangle_{\text{{\tiny R}}})^2 \right].
\label{eq-newsee}
\end{equation}
Taking the trace of Eq.\ (\ref{eq-newsee}) in spatially flat FRW spacetime, 
we find
\begin{equation}
\frac{6a''}{a^3} = 8\pi G_{\text{{\tiny R}}}\left[
({\mathcal T}^{\text{{\tiny C}}})_{\text{{\tiny R}}} +
({\mathcal T}^{\text{{\tiny Q}}})_{\text{{\tiny R}}} -
\frac{\lambda_{\text{{\tiny R}}}}{2} \langle (\vphi_{\text{{\tiny H}}}^2
\rangle_{\text{{\tiny R}}})^2 \right].
\label{eq-eveqa}
\end{equation}
Recalling that $\xi_{\text{{\tiny R}}} = 0$, and using Eq.\ (\ref{eq-onemtclb}),
the classical part of the
trace of the renormalized energy-momentum tensor is given by
\begin{equation}
({\mathcal T}^{\text{{\tiny C}}})_{\text{{\tiny R}}} =
\frac{1}{a^2} \left[ -(\phih')^2 + 2 \left( m_{\text{{\tiny R}}}^2
+ \frac{\lambda_{\text{{\tiny R}}}}{4} \phih^2 \right) \phih^2 \right],
\end{equation}
and the quantum trace of the renormalized energy-momentum tensor is given
by
\begin{eqnarray}
({\mathcal T}^{\text{{\tiny Q}}})_{\text{{\tiny R}}} =
-\frac{\hbar}{a^4} \int \frac{d^{\,3}k}{(2\pi)^3} \Biggl[
|\tilde{u}'_k|^2 - (k^2 - 2 a^2 M_{\text{{\tiny R}}}^2)|\tilde{u}_k|^2
- \frac{a'}{a}\left[ (\tilde{u}_k')^{\star} \tilde{u}_k +
\tilde{u}_k' \tilde{u}_k^{\star}\right] && + \frac{(a')^2}{a^2}
|\tilde{u}_k|^2 \Biggr] \nonumber \\ && - ({\mathcal T}^{\text{{\tiny Q}}})_{
\text{{\tiny ad4}}},
\end{eqnarray}
where $({\mathcal T}^{\text{{\tiny Q}}})_{\text{{\tiny ad4}}}$ is
defined in Eq.\ (\ref{eq-onaregtr}).  As discussed in Sec.~\ref{sec-idfrw},
the 00 component of the semiclassical Einstein equation is a constraint,
and is given by
\begin{equation}
\frac{3 (a')^2}{a^2} = 8\pi G_{\text{{\tiny R}}} \left[
(T^{\text{{\tiny C}}}_{00})_{\text{{\tiny R}}} +
(T^{\text{{\tiny Q}}}_{00})_{\text{{\tiny R}}} -
\frac{\lambda_{\text{{\tiny R}}}}{8} a^2 (\langle \vphi_{\text{{\tiny H}}}^2
\rangle_{\text{{\tiny R}}})^2 \right].
\label{eq-cneqa}
\end{equation}
From Eq.\ (\ref{eq-onemtcla}), the expression for the classical part of the 00
component of the renormalized energy-momentum tensor is given by
\begin{equation}
(T^{\text{{\tiny C}}}_{00})_{\text{{\tiny R}}} =
\frac{1}{2} (\phih')^2 + \frac{1}{2} a^2 \left( m_{\text{{\tiny R}}}^2
+ \frac{\lambda_{\text{{\tiny R}}}}{4} \phih^2 \right) \phih^2,
\end{equation}
and the quantum part of the 00 component of the renormalized energy-momentum
tensor is given by
\begin{equation}
(T^{\text{{\tiny Q}}}_{00})_{\text{{\tiny R}}} = \frac{\hbar}{2a^2}
\int \frac{d^{\,3}k}{(2\pi)^3} \left[ |\tilde{u}'_k|^2 +
\left( k^2 + a^2 M_{\text{{\tiny R}}}^2 \right)|\tilde{u}_k|^2
-\frac{a'}{a}\left[ (\tilde{u}_k')^{\star}\tilde{u}_k +
\tilde{u}_k' \tilde{u}_k^{\star}\right]\right] -
(T^{\text{{\tiny Q}}}_{00})_{\text{{\tiny ad4}}},
\end{equation}
where $(T^{\text{{\tiny Q}}}_{00})_{\text{{\tiny ad4}}}$ is defined in
Eq.\ (\ref{eq-onaregt00}).

Eqs.\ (\ref{eq-eveqa}) and (\ref{eq-eveqp}) are coupled differential
equations for $a$ and $\phih$, involving complex homogeneous
conformal-mode functions $\tilde{u}_k$ which satisfy Eq.\ (\ref{eq-eveqmf}).
The conformal mode functions are related to the variance
$\langle \vphi_{\text{{\tiny H}}}^2 \rangle_{\text{{\tiny R}}}$ by
Eq.\ (\ref{eq-renrfluc}).  This is a closed, time-reversal-invariant system of 
equations.  The initial data at $\eta_0$ must satisfy the constraint equation
(\ref{eq-cneqa}).  We now drop all ``R'' subscripts, because we
will henceforth work only with renormalized quantities.

\subsection{Reduction of derivative orders}
The adiabatic regulators (\ref{eq-onaregfluc}),
(\ref{eq-onaregt00}), (\ref{eq-onaregtr}) 
for the variance and energy-momentum
tensor contain derivatives of up to
fourth order in $a$ and up to second order in $\phih^2$ and 
$\langle \vphi_{\text{{\tiny H}}}^2 \rangle$.  The presence of the former
can be understood as resulting in part from the well-known trace anomaly for
a quantum field in curved spacetime \cite{hu:1979a}, which contains 
higher-derivative local geometric terms, e.g., $\square R$.
In addition, there are nonanomalous finite terms
which result from the renormalization of the energy-momentum tensor and
the choice of minimal coupling.

The effect of higher derivatives in the semiclassical Einstein equation
has been much studied in the literature
\cite{anderson:1985a,suen:1987a,suen:1987b,simon:1990a,parker:1993a}.
The higher-derivative evolution equations for $a$ and $\phih$ have
a much larger solution space than the classical Einstein and mean-field
equations, and in general, the higher-derivative semiclassical Einstein
equation is expected to have many solutions which are unphysical.  In addition,
the semiclassical Einstein equation (which is fourth order in $a$) requires
more initial data than the classical Einstein equation in order to uniquely 
specify a solution.  However, Simon and Parker \cite{simon:1990a,parker:1993a},
following the methods of Ja\'{e}n, Llosa, and Molina \cite{jaen:1986a},
have shown that in one-loop semiclassical gravity,
there exists a procedure for consistently removing the unphysical solutions
within the perturbative ($\hbar$) expansion in which the equations are
derived.   The procedure corresponds to the addition of perturbative 
constraints, thereby
yielding second-order equations which require the same amount of initial data 
as does the classical Einstein equation.
Their method involves reducing the order of the $a'''$ and $a''''$ terms
in the semiclassical Einstein equation using strict perturbation theory
in $\hbar$.

In this study we follow the approach of Simon and Parker 
to reduce the order of the equations for $\phih$, $a$, and 
$\langle \vphi_{\text{{\tiny H}}}^2
\rangle$ to second order.  We replace all expressions involving
$a'''$ and $a''''$ with expressions $a_{\text{{\tiny cl}}}'''$
and $a_{\text{{\tiny cl}}}''''$ obtained from the {\em classical\/}
Einstein equation, i.e., Eq.\ (\ref{eq-eveqa})
with $\hbar = 0$.  This procedure is physically justifiable in this model for
the following reason:  At early times, the dominant contribution to the 
energy-momentum tensor is classical, 
$T^{\text{{\tiny C}}}_{\mu\nu}$.  Therefore, the deviations
$a''' - a'''_{\text{{\tiny cl}}}$ and $a'''' - a''''_{\text{{\tiny cl}}}$,
which are entirely quantum in origin and $\propto \hbar$, are at
early times expected to be very small.  In addition, at late times
the Universe is expected to become asymptotically radiation dominated,
in which case $a''' = a'''' = 0$.  The classical approximations to the
late-time behavior of $a'''$ and $a''''$ should also have this property,
regardless of whether the mean-field oscillations are harmonic or elliptic.
This procedure is, therefore, 
physically justifiable in the system studied here.

\section{Analysis}
\label{sec-analysis}
Having derived coupled dynamical equations (\ref{eq-eveqp}),
(\ref{eq-eveqa}), (\ref{eq-eveqmf}) for the mean field $\phih$, 
scale factor $a$, and conformal-mode functions $\tilde{u}_k$, 
respectively, we now proceed to solve them.
\subsection{Initial conditions}                              
\label{sec-onbc}
At the Cauchy hypersurface at $\eta_0$, 
we specify initial conditions on the conformal-mode functions
$\tilde{u}_k$ which
correspond to a choice of quantum state for the fluctuation field
$\vphi_{\text{{\tiny H}}}$.  
Based on the analysis in Sec.~\ref{sec-initcond},
we choose boundary conditions at $\eta_0$ which correpsond to the
adiabatic vacuum state for $\vphi_{\text{{\tiny H}}}$ at $\eta \rightarrow 
-\infty$.
From the semiclassical Einstein equation (\ref{eq-seesimp}),
the slow-roll condition (\ref{eq-sras}), the
potential-dominated condition (\ref{eq-pdas}), and assuming that 
the variance $\langle \vphi_{\text{{\tiny H}}}^2 \rangle$ satisfies
\begin{equation}
\frac{\lambda}{2}\langle \vphi_{\text{{\tiny H}}}^2 \rangle \ll
m^2 + \frac{\lambda}{2} \phih^2
\end{equation}
for $\eta < \eta_0$, it follows that the spacetime is asymptotically 
de~Sitter at conformal-past infinity.
Using the approximate solution (\ref{eq-asol}) for the scale factor 
for $\eta < \eta_0$, we can solve the mode function equation (\ref{eq-eveqmf})
for $\eta < \eta_0$ at the same (0th) adiabatic order.  The general
solution is of the form
\begin{equation}
\tilde{u}_k(\eta) \simeq \left( \frac{\pi(\eta - H^{-1} - \eta_0)}{4}\right)^{
\frac{1}{2}} \Bigl[ c^{1}_{k} H_{\nu}^{(1)}\bigl\{k(\eta - H^{-1}(\eta) 
- \eta_0)
\bigr\} + c^{2}_{k} H_{\nu}^{(2)}\bigl\{k(\eta - H^{-1}(\eta) - \eta_0)\bigr\}
\Bigr],
\end{equation}
where $H^{(1)}$ and $H^{(2)}$ are the Hankel functions of first and second
kind, respectively \cite{gradshteyn:1964a}, and $\nu$ is defined by
\begin{equation}
\nu^2 = \frac{9}{4} - \frac{M^2}{H^2}.
\end{equation}
The function $H(\eta)$ is defined as in Eq.\ (\ref{eq-defhinv}),
\begin{equation}
H(\eta) = \sqrt{\frac{8 \pi G \rho_{\text{{\tiny C}}}}{3}},
\end{equation}
where now $\rho_{\text{{\tiny C}}} = a^2 T^{\text{{\tiny C}}}_{00}$.
The Hubble parameter must be slowly varying for this approximation to hold,
i.e., the expansion rate nonadiabaticity parameter \cite{hu:1993d}
\begin{equation}
\bar{\Omega}_H \equiv \frac{H'}{H^2} \ll 1.
\label{eq-areq}
\end{equation}
The Wronskian condition on the mode functions (which comes from
the canonical field commutation relations) requires that
\begin{equation}
|c^1_k|^2 + |c^2_k|^2 = 1.
\end{equation}
By choosing $c^1_k$ and $c^2_k$, different vacua are obtained.
The 0th-order adiabatic vacuum (matched at $\eta = -\infty$)
is constructed by choosing $c^1_k$ and $c^2_k$ so that $\tilde{u}_k$
smoothly matches the positive-frequency 0th-order WKB mode
function at $\eta = -\infty$.  This corresponds to $c^2_k = 1$
and $c^1_k = 0$, for all $k$.
Using the asymptotic properties of the Hankel function, the adiabatic limit
$k, |\eta| \rightarrow \infty$, can be derived, and verified to have the
correct form,
\begin{equation}
\lim_{k,|\eta| \rightarrow \infty} 
\tilde{u}_k \simeq \frac{1}{\sqrt{2 k}} e^{-i k \eta}.
\end{equation}
In addition, the high-momentum, flat-space limit
($k, H^{-1} \rightarrow \infty$) gives the
same result.  The initial conditions for the $\tilde{u}_k$ at $\eta_0$
are then defined by demanding that the $\tilde{u}_k$ functions 
smoothly match the 
approximate adiabatic mode function solutions (for $\eta < \eta_0$)
at $\eta = \eta_0$.  This leads to the following initial conditions
for the conformal-mode functions:
\begin{mathletters}
\begin{eqnarray}
\tilde{u}_k(\eta_0) &=& \left( \frac{-\pi}{4H_0} \right)^{1/2}
H_{\nu}^{(2)}(-k H^{-1}_0),  \\
\tilde{u}'_k(\eta_0) &=& \frac{d}{d\eta} \left[ \left( \frac{\pi \eta}{4}
\right)^{1/2} H_{\nu}^{(2)}(k\eta)\right]_{|\eta = -H^{-1}_0},
\end{eqnarray}
\end{mathletters}
where $H_0 = H(\eta_0)$.  The above conditions are valid only at 0th order
in the above-defined adiabatic approximation, where terms of order $H'/H$
are discarded.  It is straightforward to show that Eq.\ (\ref{eq-areq}) is valid
given the slow-roll (\ref{eq-sras}) and inflation (\ref{eq-pdas}) assumptions.
In addition to the initial conditions for $\tilde{u}_k$ at $\eta_0$,
we may freely choose initial values for $\phih(\eta_0)$ and
$\phih'(\eta_0)$, subject to the constraint that $\phih'$ must be
small enough that conditions (\ref{eq-pdas}) and
(\ref{eq-sras}) are valid.  We are already assuming that $a(\eta_0) = 1$.
The initial value of $a'(\eta_0)$ is fixed by the constraint
equation (\ref{eq-cneqa}).

\subsection{Numerical solution}
\label{sec-numerics}
In this section we describe the methods we used to 
solve the coupled evolution equations for $\phih$ [Eq.\ (\ref{eq-eveqp})],
$a$ [Eq.\ (\ref{eq-eveqa})], and $\tilde{u}_k$ [Eq.\ (\ref{eq-eveqmf})]
numerically.\footnote{Henceforth, we set $\hbar=1$ and work in units of energy
where $m = 1$.}
We evolved a representative sampling of mode functions
$\tilde{u}_k$ for the region of momentum space
$0 \leq k \leq K a$, where $K$ is a physical upper momentum cutoff.\footnote{
A finite momentum cutoff is necessary due to the triviality of the
theory when the cutoff is taken to infinity \cite{cooper:1994a}.}
Employing a physical \cite{ringwald:1987a}, as opposed to comoving, momentum
cutoff is necessary because a comoving cutoff would require the
use of the renormalization group
equation to track how the renormalized parameters flow as the 
scale factor $a$ increases at each time step.  
For a comoving cutoff the quadratic divergence in the variance would be 
proportional to $1/a^2$, requiring a time-dependent renormalization
(see \cite{boyanovsky:1994a}, for example).
The use of a physical upper
momentum cutoff yields a quadratic divergence which can be removed by
a non-time-dependent mass renormalization \cite{ringwald:1987a}.

We chose a variety of values of $K/m$ between $50$ and $70$.
The sampling of momentum-space is carried out with a uniform binning,
with total number of bins $N_{\text{{\tiny bins}}}$.  Various values of 
$N_{\text{{\tiny bins}}}$ were used, all greater than $10^4$.  Eq.\
(\ref{eq-rem}) was solved by iteration, and the momentum space integrations
were performed numerically using the 
$O(1/N_{\text{{\tiny bins}}}^4)$
extended Simpson rule.  The differential
equations (\ref{eq-eveqa}) and (\ref{eq-eveqp}) were evolved using
4th-order Runge-Kutta with adaptive step-size control; the target precision 
for the time steps varied between $10^{-6}$ and $10^{-8}$.  Cutoff independence
was verified {\em a posteriori\/} by explicitly checking that
the results of the numerical solution were insensitive to a rescaling
of $K/m$.  The solutions were computed to a conformal-time scale of 
$400 \; m^{-1}$.  A typical solution computed according  to the
above methods required on the order of 300 h of CPU time on a modern  
workstation.

\subsection{Results}
\label{sec-results}
\begin{figure}[htb]
\begin{center}
\epsfig{file=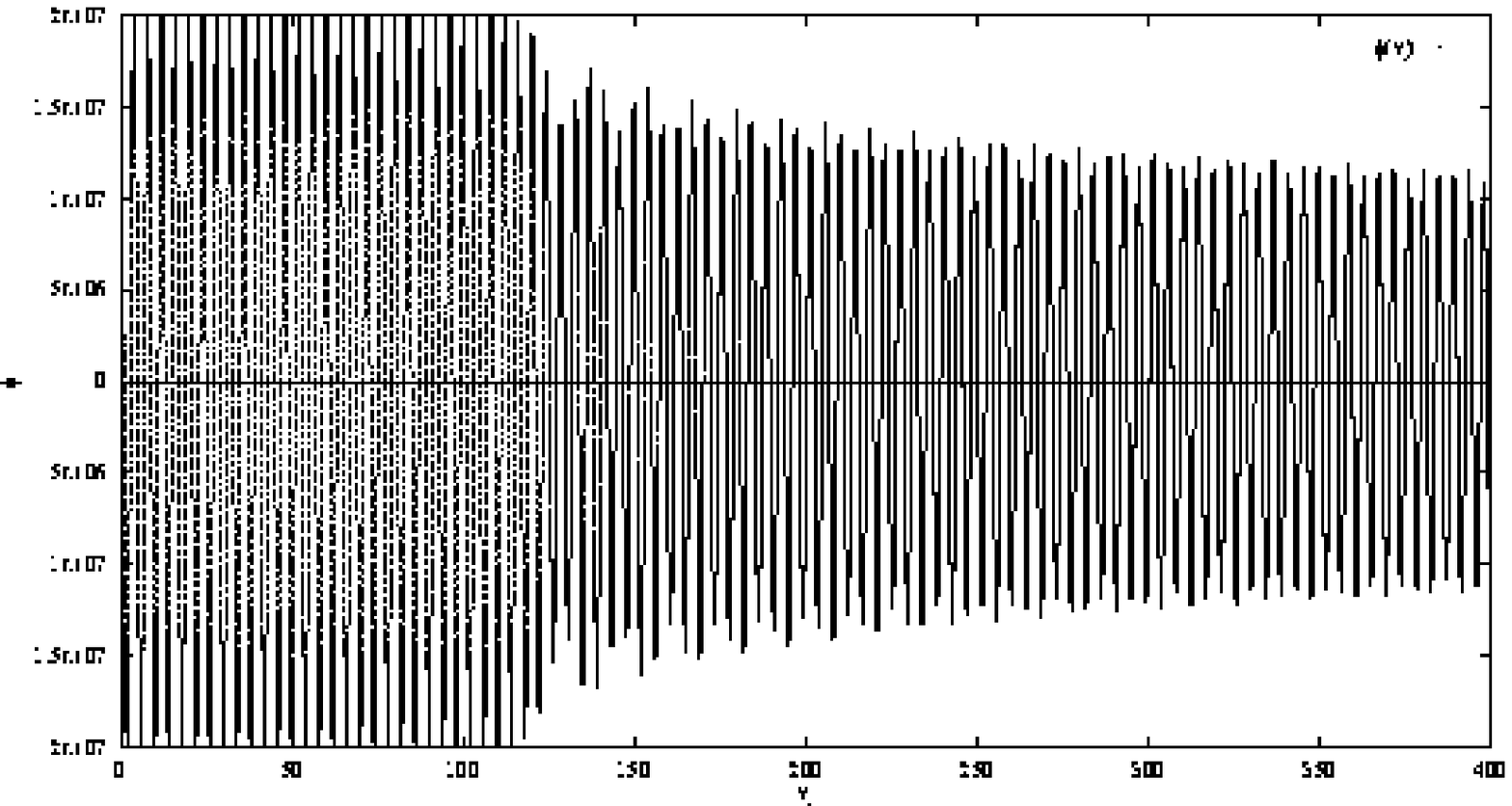,width=6.00in} 
\end{center}
\caption{Plot of $\phi$ vs $\eta$, with $\Mpl/m = 1.0 \times 10^{14}$.}
\label{fig-run19phi}
\end{figure}

\begin{figure}[htb]
\begin{center}
\epsfig{file=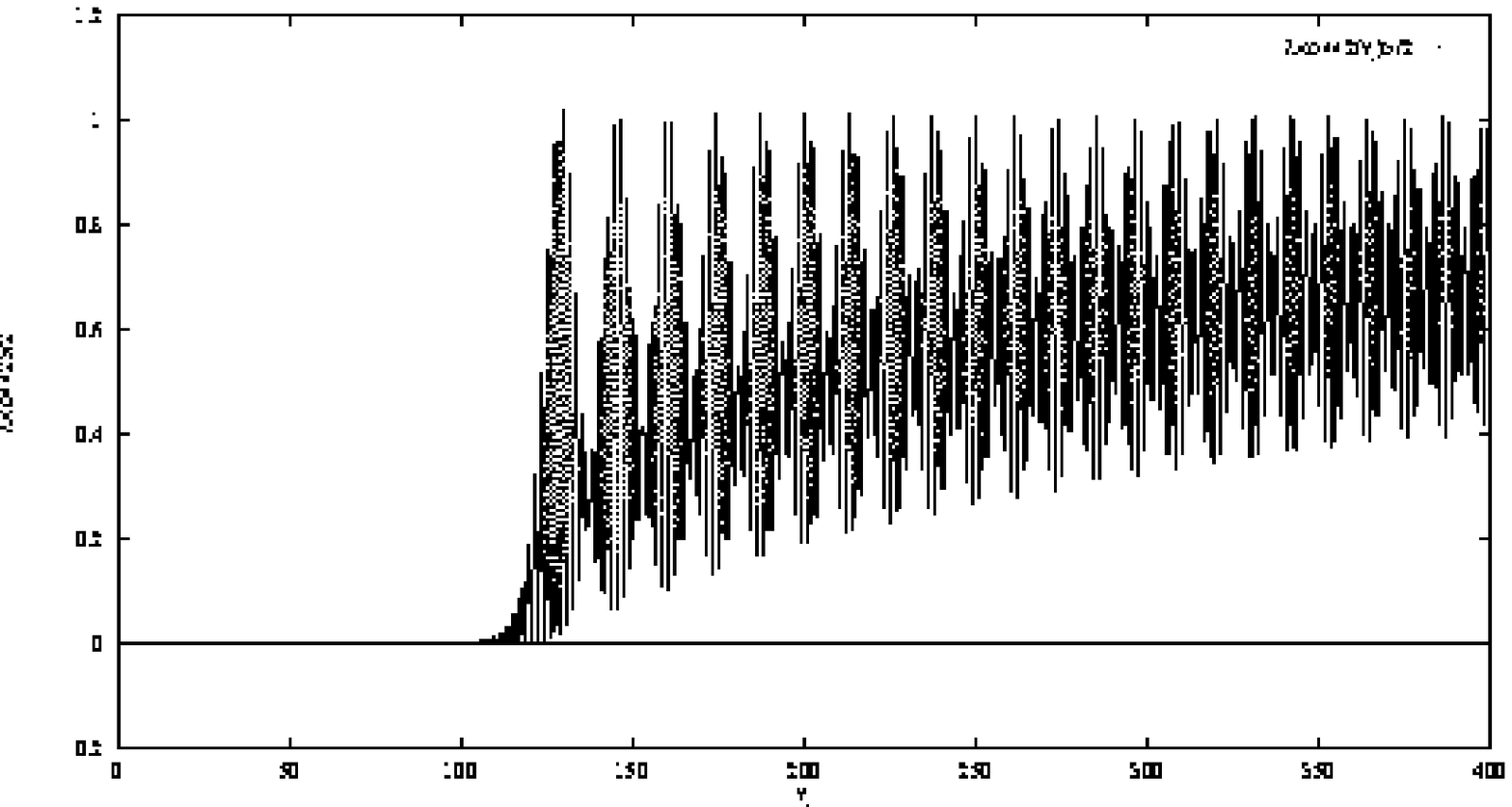,width=6.00in} 
\end{center}
\caption{Plot of $\lambda \langle \vphi^2 \rangle/2$ vs $\eta$, 
with $\Mpl/m = 1.0 \times 10^{14}$.}
\label{fig-run19fluc}
\end{figure}

\begin{figure}[htb]
\begin{center}
\epsfig{file=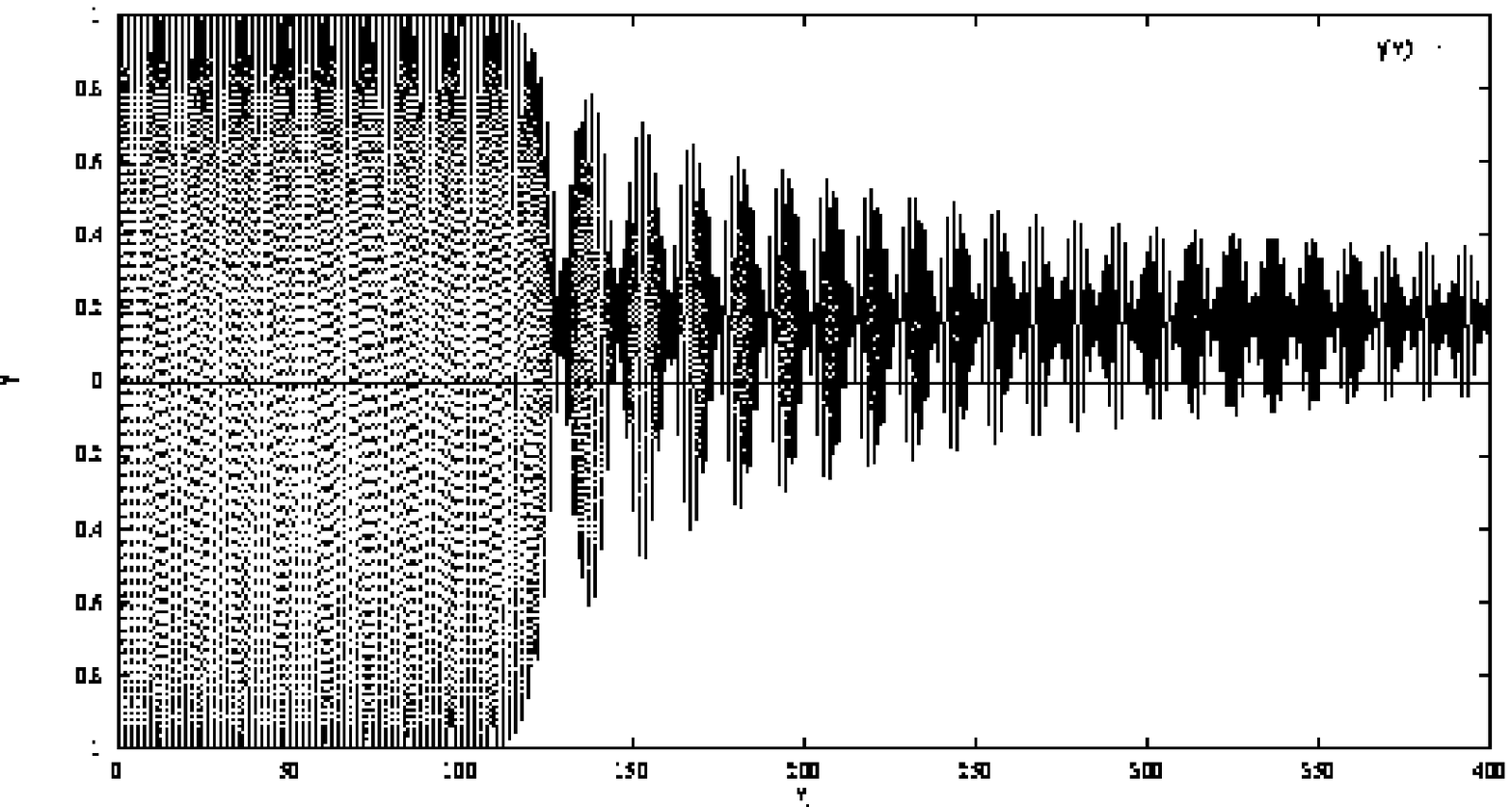,width=6.00in} 
\end{center}
\caption{Plot of $\gamma$ vs $\eta$, 
with $\Mpl/m = 1.0 \times 10^{14}$.}
\label{fig-fun19eos}
\end{figure}

\begin{figure}[htb]
\begin{center}
\epsfig{file=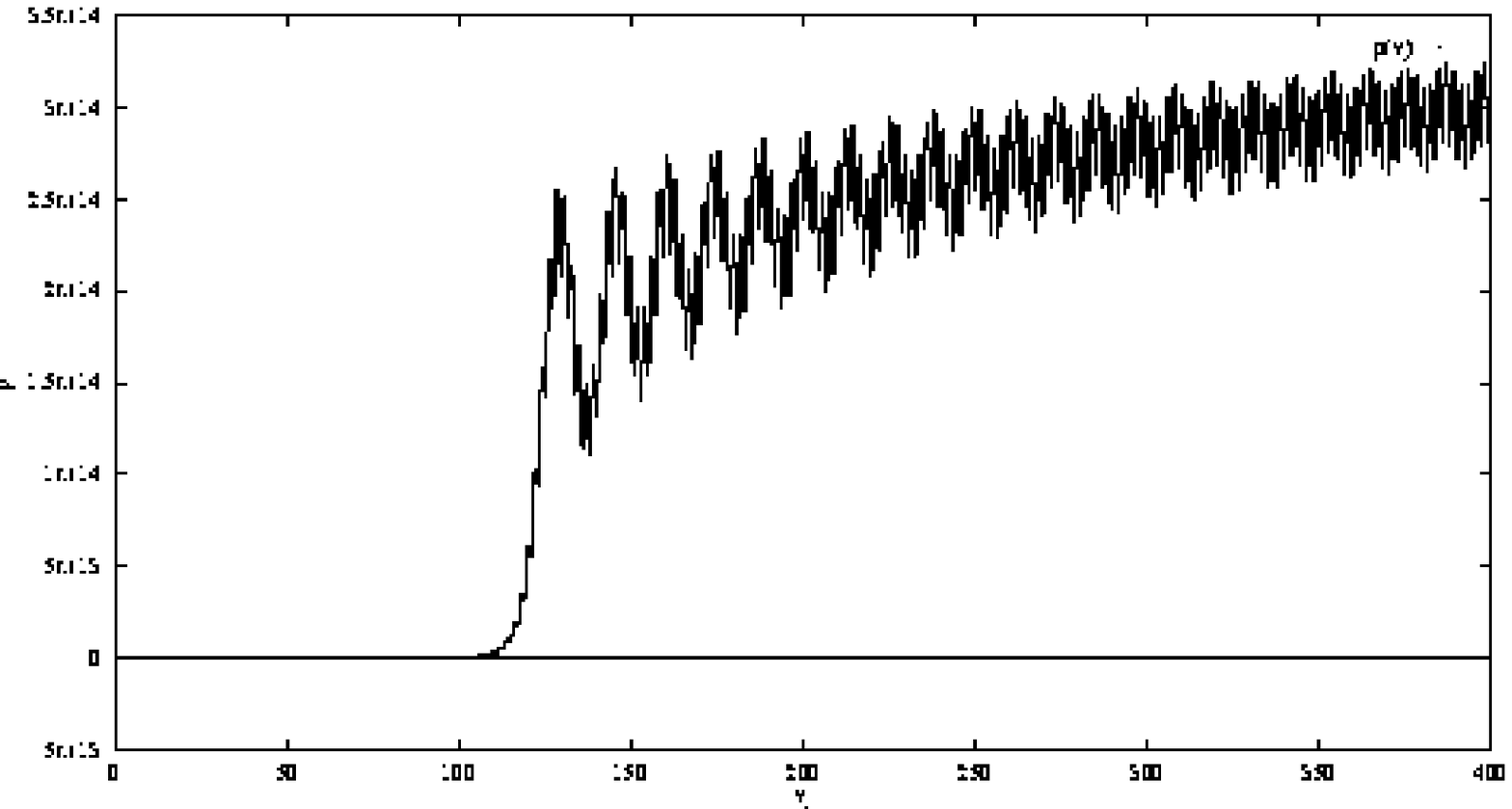,width=6.00in} 
\end{center}
\caption{Plot of $\rho_{\text{{\tiny Q}}}$ vs $\eta$, 
with $\Mpl/m = 1.0 \times 10^{14}$.}
\label{fig-run19rhoq}
\end{figure}

\begin{figure}[htb]
\begin{center}
\epsfig{file=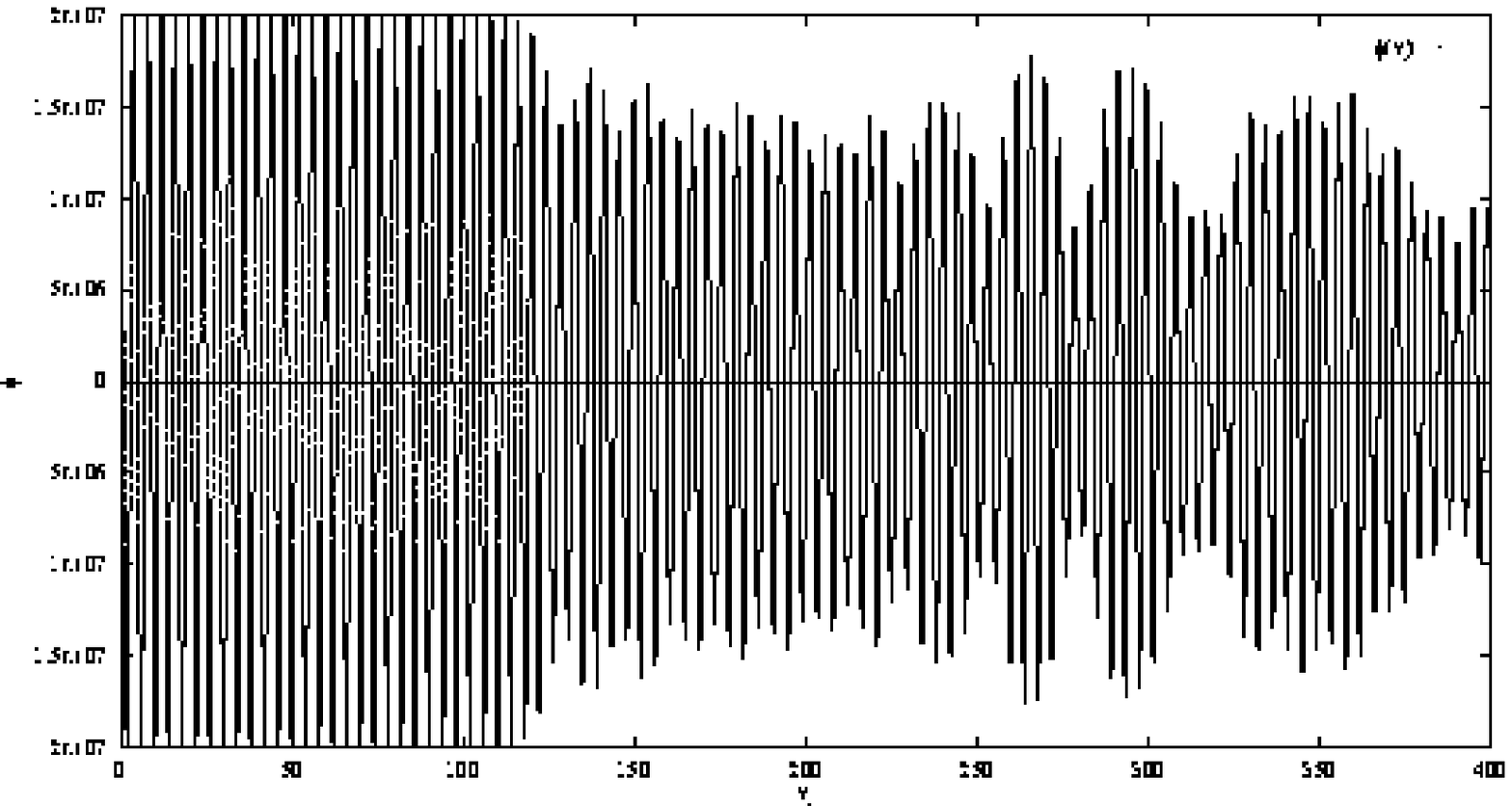,width=6.00in} 
\end{center}
\caption{Plot of $\phi$ vs $\eta$, 
with $\Mpl/m = 1.0 \times 10^{12}$.}
\label{fig-run33phi}
\end{figure}

\begin{figure}[htb]
\begin{center}
\epsfig{file=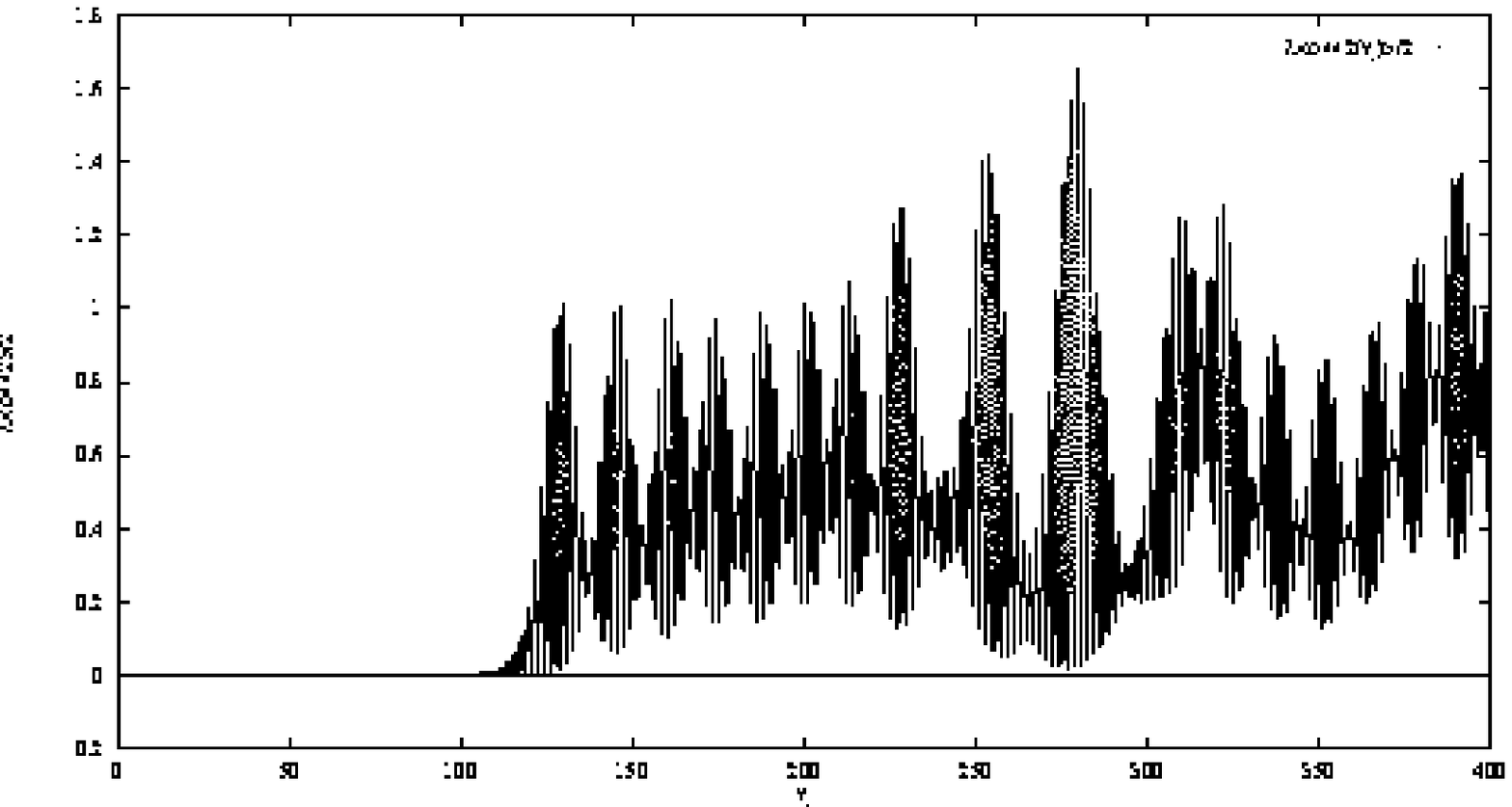,width=6.00in} 
\end{center}
\caption{Plot of $\lambda \langle \vphi^2 \rangle/2$ vs $\eta$,
with $\Mpl/m = 1.0 \times 10^{12}$.}
\label{fig-run33fluc}
\end{figure}

\begin{figure}[htb]
\begin{center}
\epsfig{file=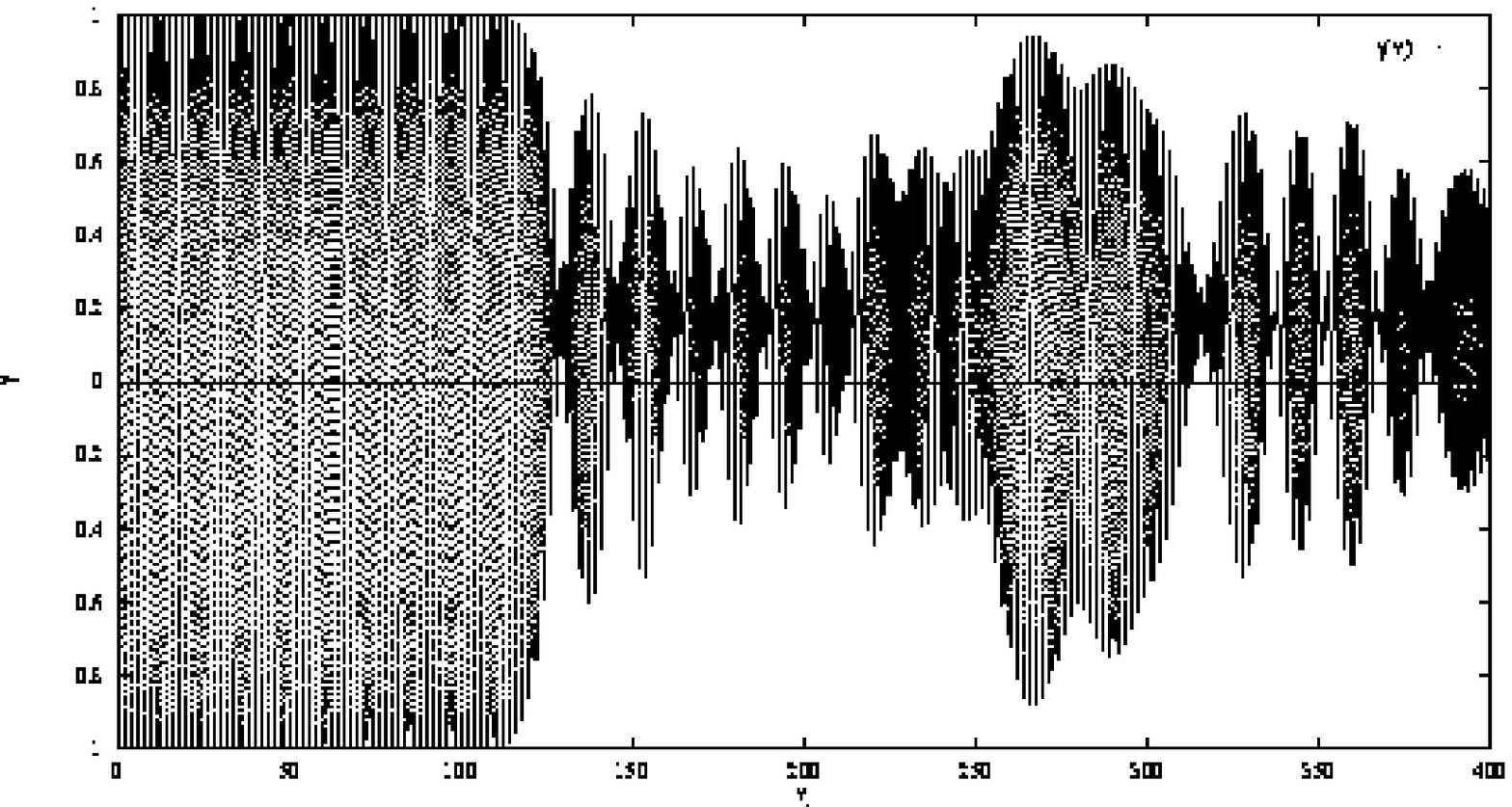,width=6.00in} 
\end{center}
\caption{Plot of $\gamma$ vs $\eta$, 
with $\Mpl/m = 1.0 \times 10^{12}$.}
\label{fig-run33eos}
\end{figure}

\begin{figure}[htb]
\begin{center}
\epsfig{file=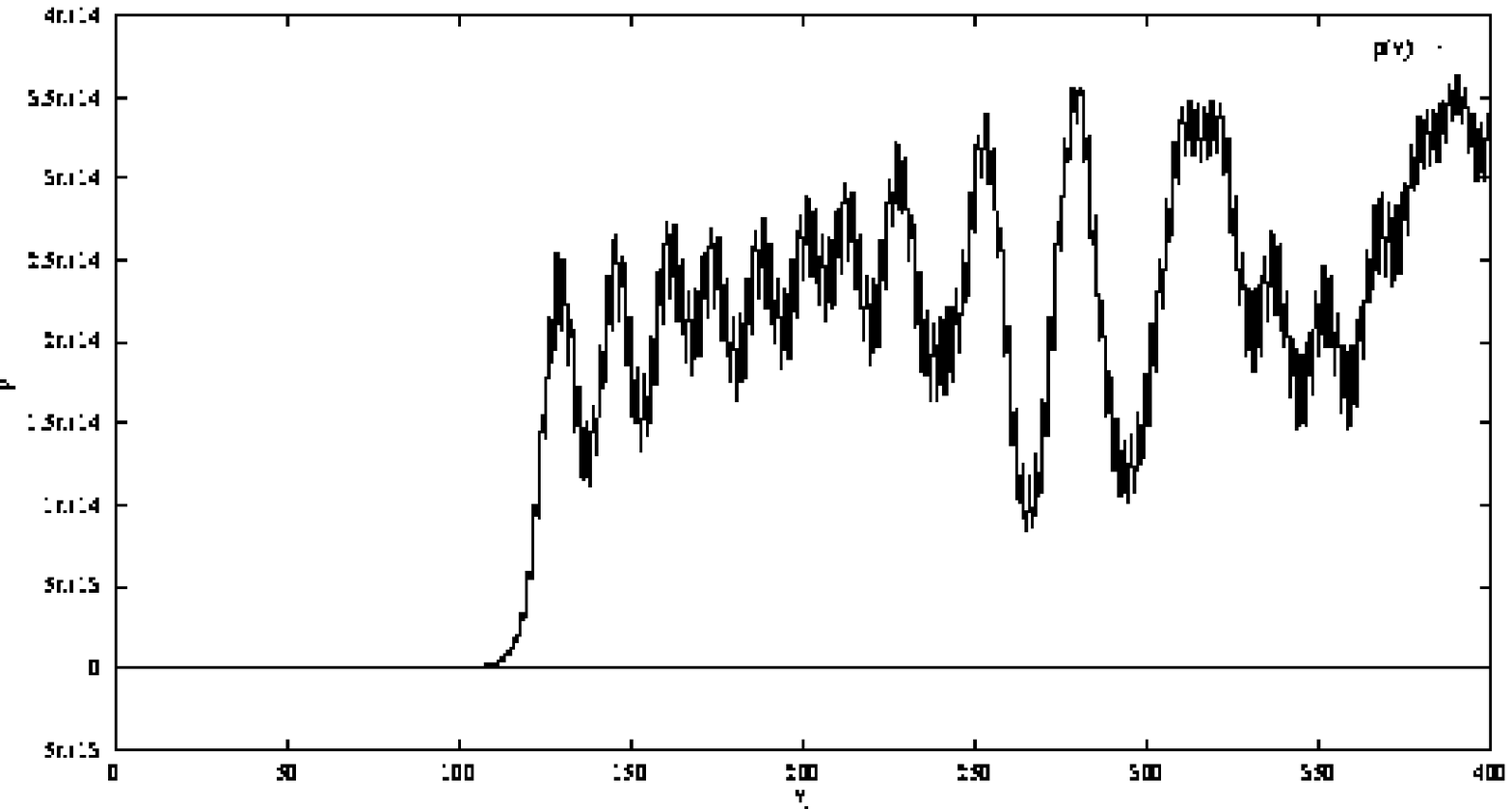,width=6.00in} 
\end{center}
\caption{Plot of $\rho_{\text{{\tiny Q}}}$ vs $\eta$, 
with $\Mpl/m = 1.0 \times 10^{12}$.}
\label{fig-run33rhoq}
\end{figure}

\begin{figure}[htb]
\begin{center}
\epsfig{file=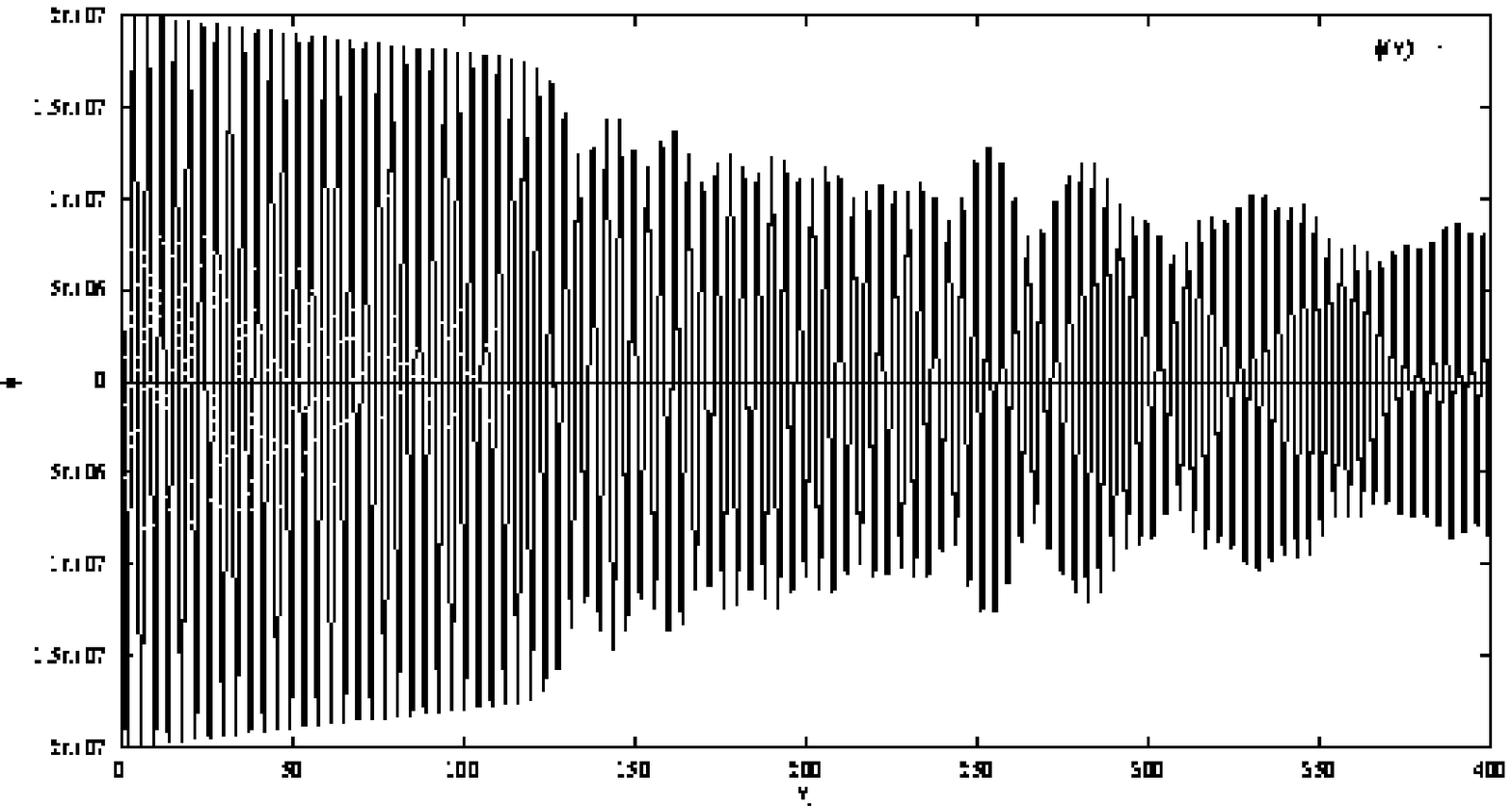,width=6.00in} 
\end{center}
\caption{Plot of $\phi$ vs $\eta$, 
with $\Mpl/m = 6.0 \times 10^{10}$.}
\label{fig-run18phi}
\end{figure}

\begin{figure}[htb]
\begin{center}
\epsfig{file=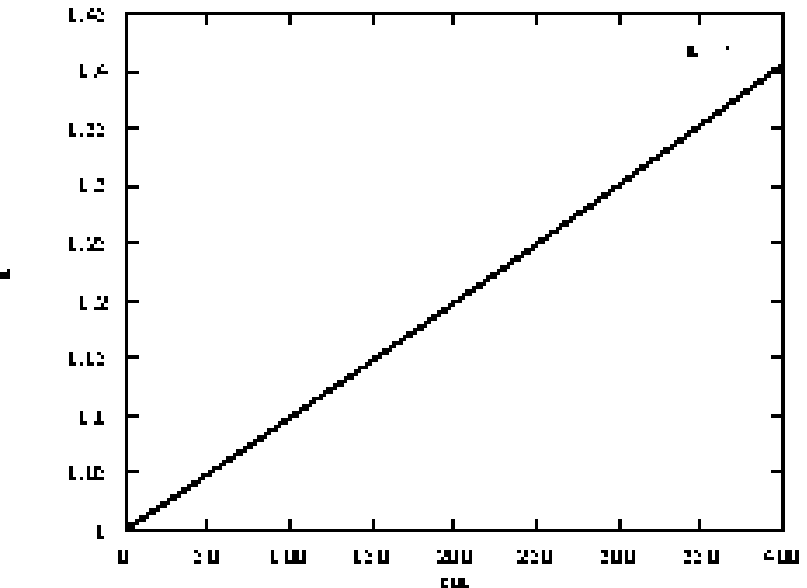,width=3.375in} 
\end{center}
\caption{Plot of $a$ vs $\eta$, 
with $\Mpl/m = 6.0 \times 10^{10}$.}
\label{fig-run18a}
\end{figure}

\begin{figure}[htb]
\begin{center}
\epsfig{file=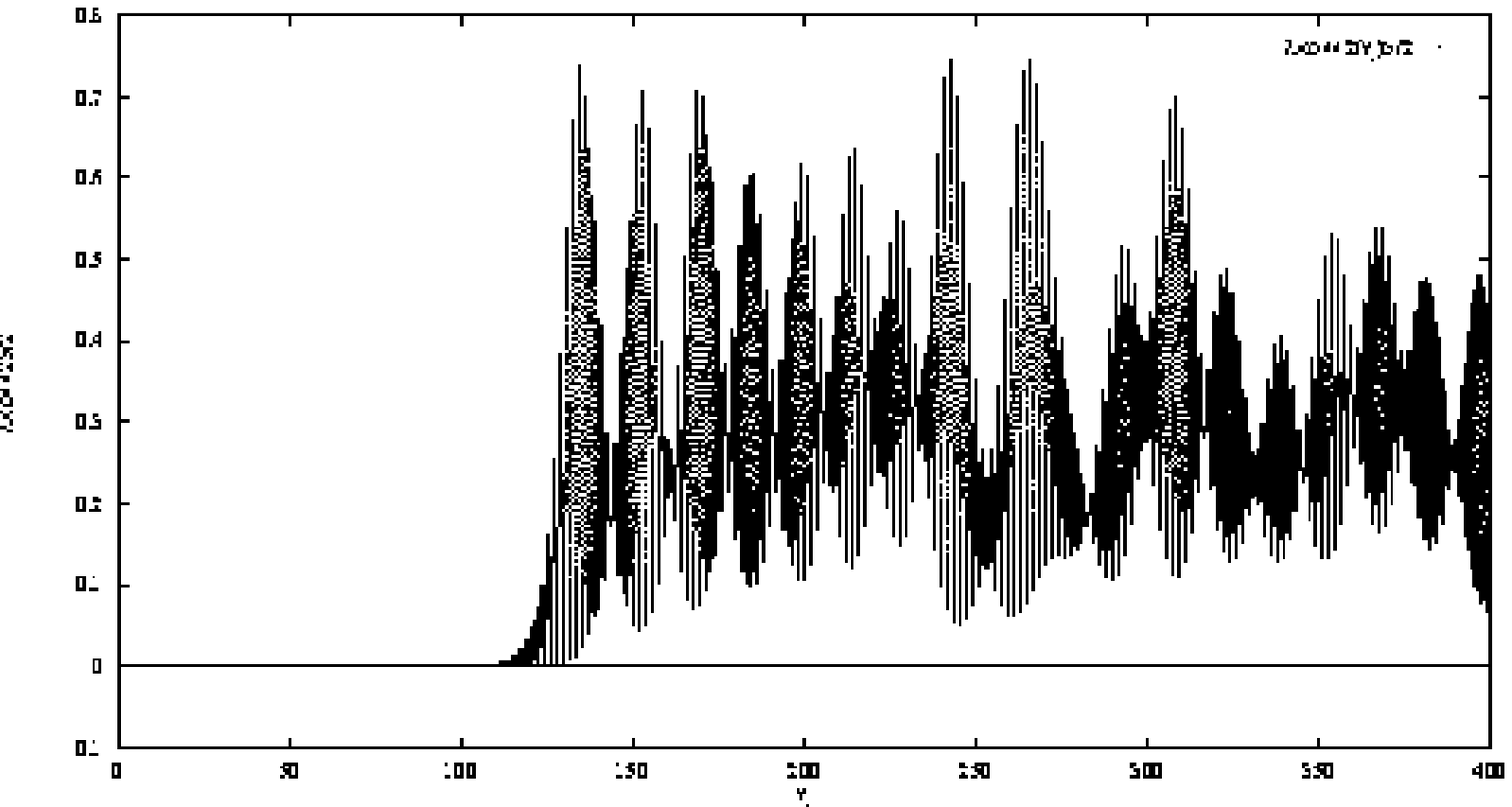,width=6.00in} 
\end{center}
\caption{Plot of $\lambda \langle \vphi^2 \rangle/2$ vs $\eta$,
with $\Mpl/m = 6.0 \times 10^{10}$.}
\label{fig-run18fluc}
\end{figure}

\begin{figure}[htb]
\begin{center}
\epsfig{file=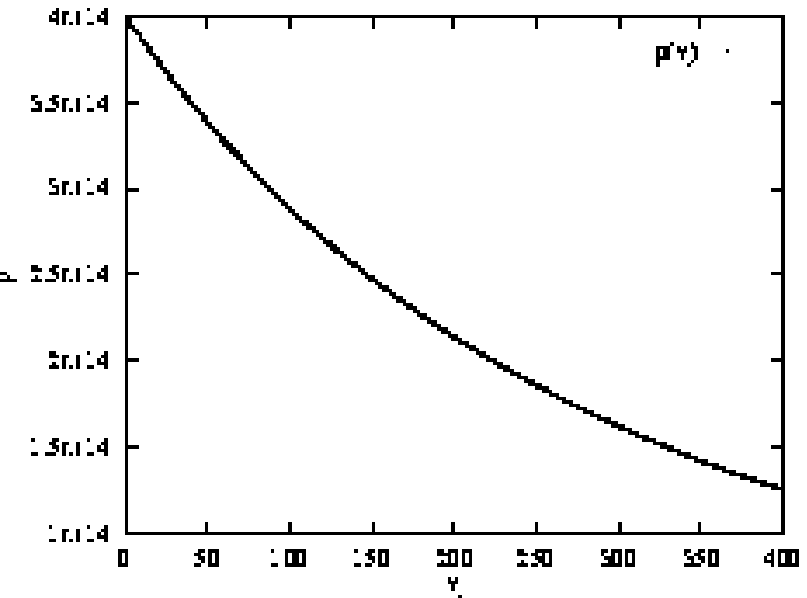,width=3.375in} 
\end{center}
\caption{Plot of $\rho$ vs $\eta$,
with $\Mpl/m = 6.0 \times 10^{10}$.}
\label{fig-run18rho}
\end{figure}

\begin{figure}[htb]
\begin{center}
\epsfig{file=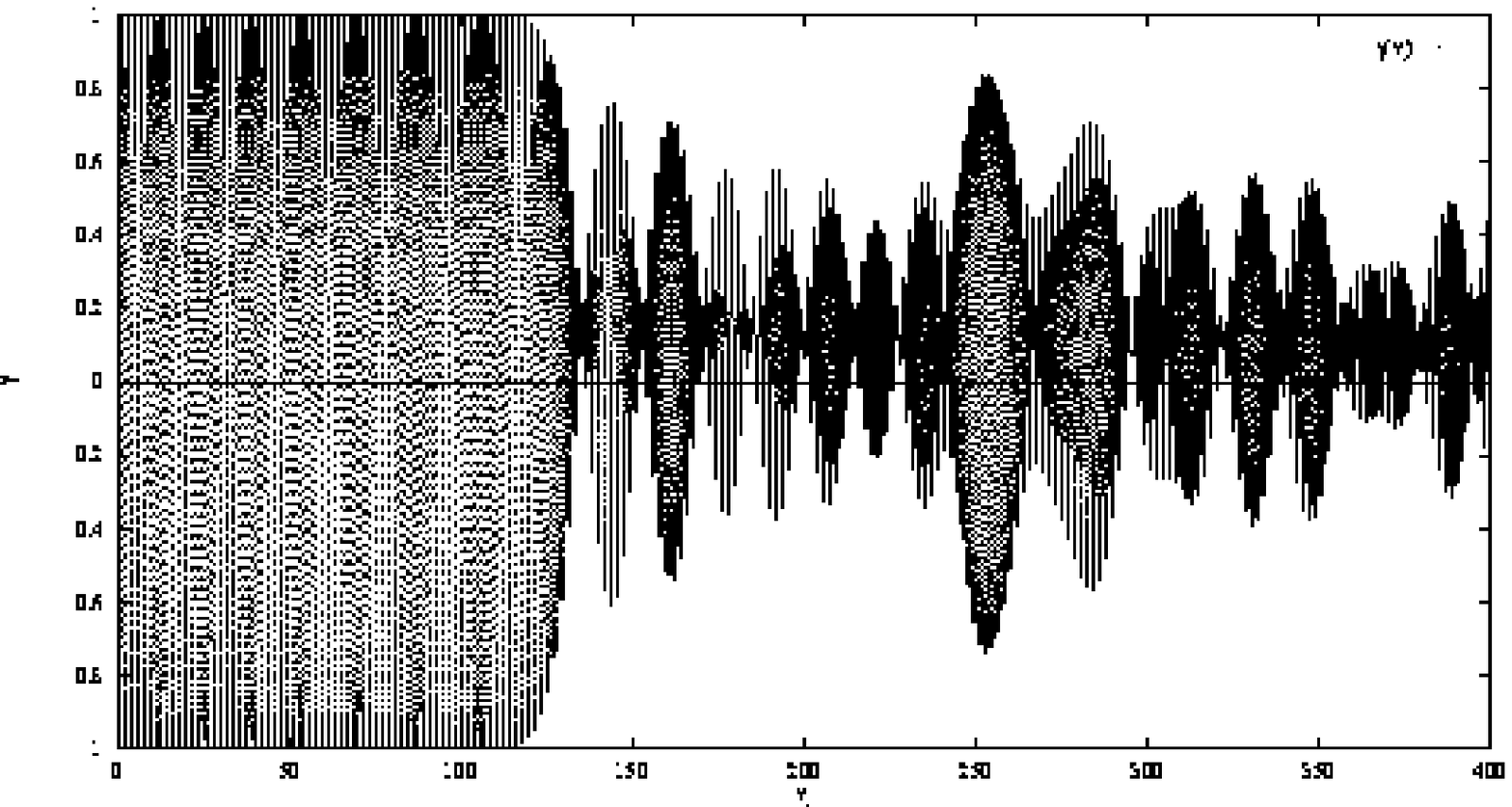,width=6.00in} 
\end{center}
\caption{Plot of $\gamma$ vs $\eta$,
with $\Mpl/m = 6.0 \times 10^{10}$.}
\label{fig-run18eos}
\end{figure}

\begin{figure}[htb]
\begin{center}
\epsfig{file=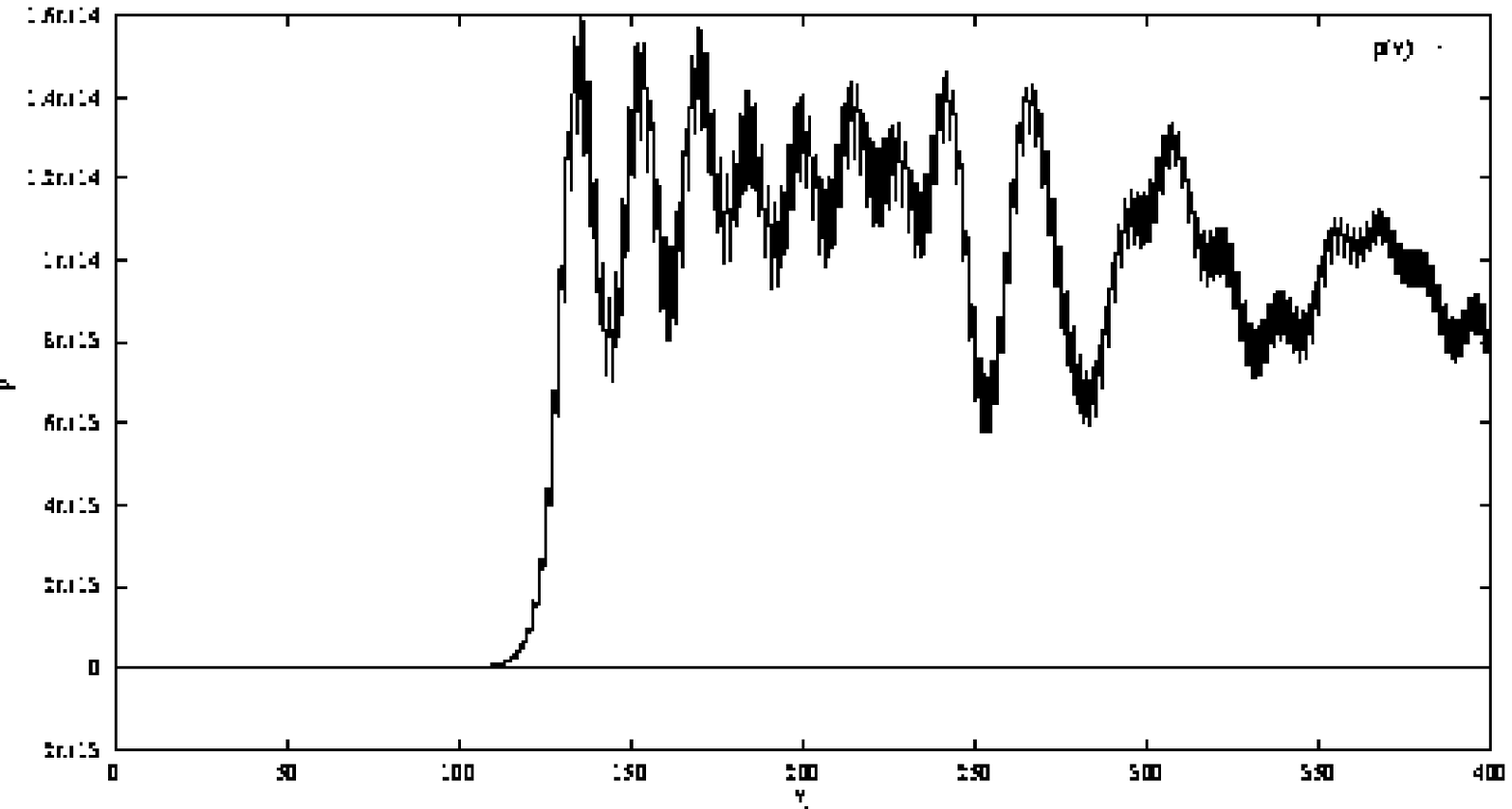,width=6.00in} 
\end{center}
\caption{Plot of $\rho_{\text{{\tiny Q}}}$ vs $\eta$,
with $\Mpl/m = 6.0 \times 10^{10}$.}
\label{fig-run18rhoq}
\end{figure}

\begin{figure}[htb]
\begin{center}
\epsfig{file=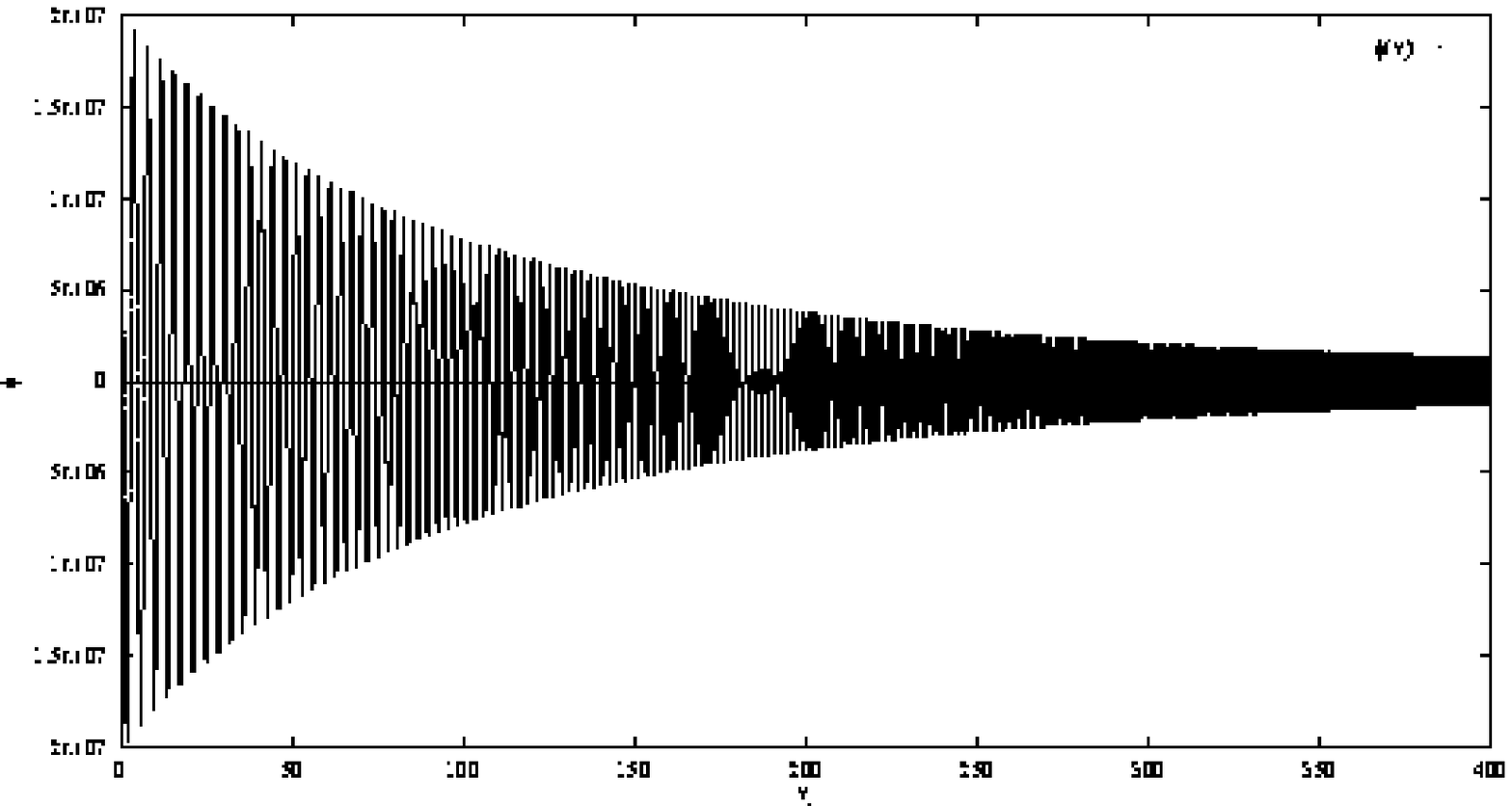,width=6.00in} 
\end{center}
\caption{Plot of $\phi$ vs $\eta$,
with $\Mpl/m = 6.0 \times 10^{9}$.}
\label{fig-run14phi}
\end{figure}

\begin{figure}[htb]
\begin{center}
\epsfig{file=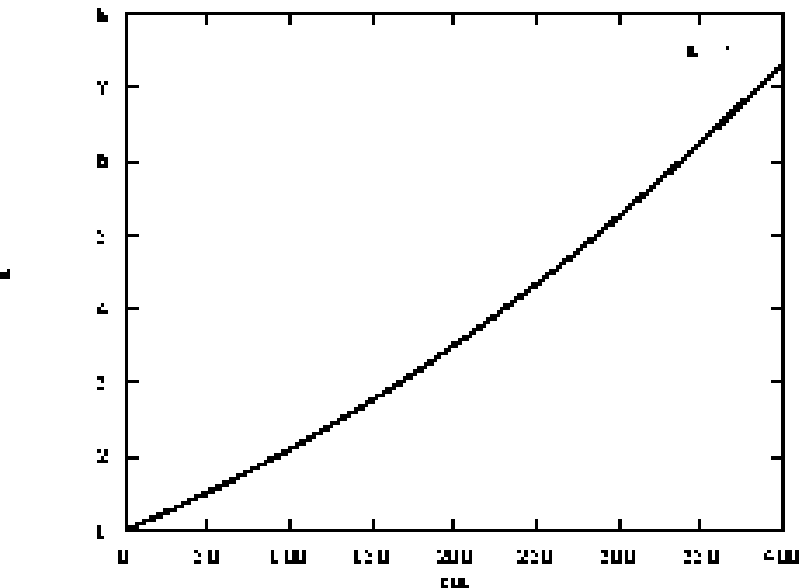,width=3.375in} 
\end{center}
\caption{Plot of $a$ vs $\eta$,
with $\Mpl/m = 6.0 \times 10^{9}$.}
\label{fig-run14a}
\end{figure}

\begin{figure}[htb]
\begin{center}
\epsfig{file=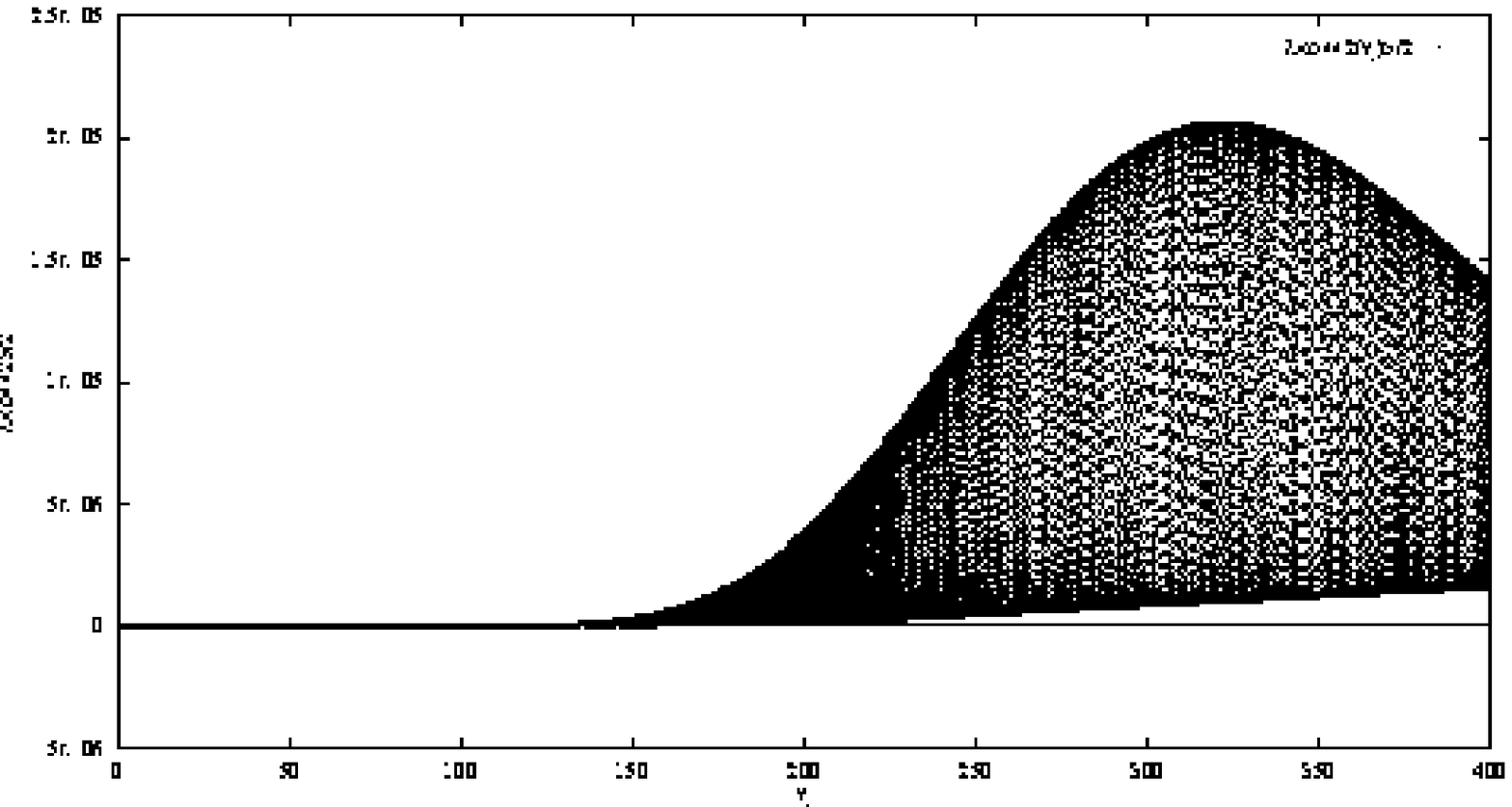,width=6.00in} 
\end{center}
\caption{Plot of $\lambda \langle \vphi^2 \rangle/2$ vs $\eta$,
with $\Mpl/m = 6.0 \times 10^{9}$.}
\label{fig-run14fluc}
\end{figure}

\begin{figure}[htb]
\begin{center}
\epsfig{file=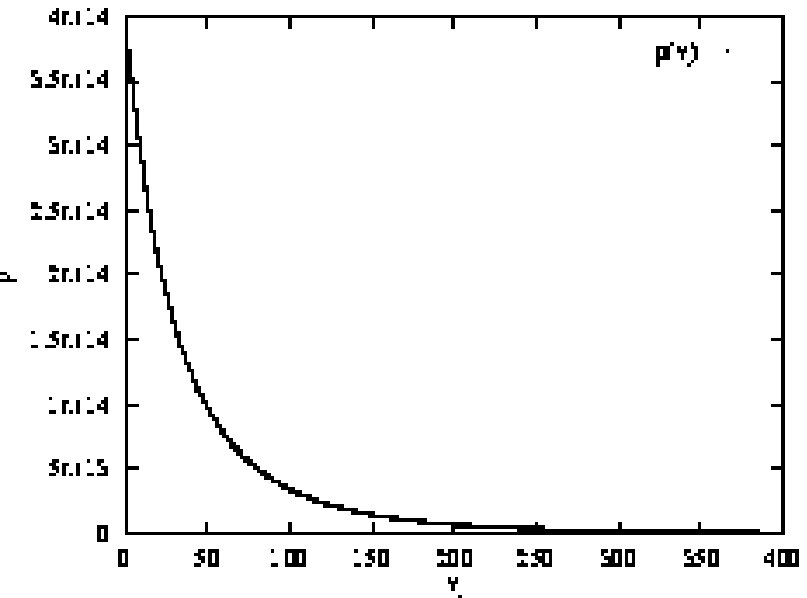,width=3.375in} 
\end{center}
\caption{Plot of $\rho$ vs $\eta$,
with $\Mpl/m = 6.0 \times 10^{9}$.}
\label{fig-run14rho}
\end{figure}

\begin{figure}[htb]
\begin{center}
\epsfig{file=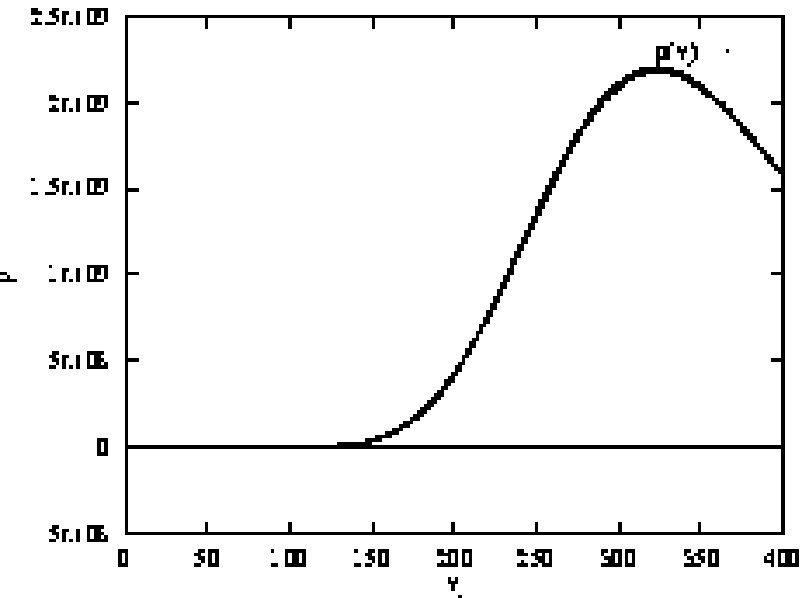,width=3.375in} 
\end{center}
\caption{Plot of $\rho_{\text{{\tiny Q}}}$ vs $\eta$,
with $\Mpl/m = 6.0 \times 10^{9}$.}
\label{fig-run14rhoq}
\end{figure}

A primary goal of this work is the quantitative study of the effect of 
spacetime dynamics on the parametric resonance energy-transfer mechanism
in nonequilibrium zero-mode oscillations of a quantum field.  As discussed in 
Sec.~\ref{sec-reheating}, this energy transfer, and the corresponding 
damping of the mean field due to back reaction, occur on a time scale of order 
$\tau_1$ defined in Eq.\ (\ref{eq-deftau1}).  We numerically evolved the 
evolution equations for $a$, $\phih$, and 
$\langle \vphi_{\text{{\tiny H}}}^2 \rangle$ for various values of $\Mpl/m$, 
ranging from very large values (corresponding to Minkowski space),
to small values (corresponding to a strong-curvature, rapid-expansion regime). 
Figs.~\ref{fig-run19phi}--\ref{fig-run14rhoq} show the resulting time 
dependences for the mean field $\phih$, the scale factor $a$, 
the variance, $\lambda \langle \vphi_{\text{{\tiny H}}}^2 \rangle/2$, the 
energy density $\rho$, the energy density in quantum modes 
$\rho_{\text{{\tiny Q}}}$ [defined in Eq.\ (\ref{eq-rhoqdef})], and the 
pressure-to-energy-density ratio $\gamma$.
The different solutions plotted correspond to different values of $\Mpl/m$,
with $\lambda = 10^{-14}$, $K/m = 50$, and $\phi(\eta_0)/m = 2.0 \times
10^7$.  As discussed in Sec.~\ref{sec-numerics}, a physical momentum
cutoff $K$ was used. The values chosen for $\Mpl/m$ were $10^{14}$,
$10^{12}$, $6 \times 10^{10}$, and $6 \times 10^9$.  
The choice of $\phih(\eta_0)$ and $\lambda$ fixes $\eta_0$ by Eq.\ (\ref{eq-asol})
and $H_0$ by Eq.\ (\ref{eq-defhinv}).  
Table~\ref{table-runparms} shows the values of $\Mpl/m$, the
inverse Hubble constant $H^{-1}(\eta_0)$,
and the figure numbers in which the corresponding solutions are
plotted.   

The time scales defined in Sec.~\ref{sec-reheating} can now be explicitly
computed. Using Eqs.\ (\ref{eq-deff}) and (\ref{eq-defhinv}), we have 
$f(\eta_0) = \sqrt{2}$, $\rho_0 = \phih^2_0$, and 
\begin{equation}
H(\eta_0) = \sqrt{\frac{8 \pi \phih_0^2}{3\Mpl^2}} \equiv H_0.
\end{equation}
Using Eq.\ (\ref{eq-deftau0}), we find $\tau_0 \simeq 4.118 \, 32 \; m^{-1}$.  
The value of $\tau_1$ predicted by Eq.\ (\ref{eq-deftau1}) is 132.624 $m^{-1}$,
which is very close to the value predicted by Eq.\ (\ref{eq-adtau1}), 
132.759 $m^{-1}$.  For the cases $\Mpl/m = 10^{14}$ and $10^{12}$, it is clear 
from Table~\ref{table-runparms} that $H^{-1}_0 \gg \tau_1$, so that
the effect of cosmic expansion is expected to be insignificant on the
preheating time scale $\tau_1$.  For the case $\Mpl = 6 \times 10^{10}$,
$1 / ( H_0 \tau_1) \sim 7.8$, so that the effect of cosmic expansion should
be apparent and non-negligible.  For the case $\Mpl = 6 \times 10^9$,
$1 / (H_0 \tau_1) \sim 0.78$, and cosmic expansion should have a 
significant effect on parametric amplification of quantum fluctuations.

Figs.~\ref{fig-run19phi}--\ref{fig-run33rhoq} show the dynamics of the
mean field and variance in the regime of very weak cosmic expansion,
$H^{-1} \ll \tau_1$.  As expected, under the influence of the elliptically
oscillating mean field, the variance $\langle \vphi_{\text{{\tiny H}}}^2
\rangle$ grows exponentially in time until $\lambda \langle
\vphi_{\text{{\tiny H}}}^2 \rangle/2$ is of the same order as
$m^2 + \lambda \phih^2/2$, at which point back reaction shuts off the
resonant transfer of energy to the inhomogeneous modes.  The time scale
for the variance to become of order unity can be clearly seen to be
$\sim \tau_1$.  As seen previously in studies of preheating dynamics in 
Minkowski
space \cite{boyanovsky:1996b}, on the time scale $\sim \tau_1$, the
mean field decouples from its own fluctuations and oscillates with
an asymptotically finite amplitude, given by \cite{boyanovsky:1996b}
$\lambda \bar{\phih^2}/ (2 m^2) = 0.914$.  In the Minkowski space limit,
corresponding to $\Mpl/m \rightarrow \infty$, covariant conservation
of the energy-momentum tensor implies that $d{\rho}/dt = 0$.  This was 
verified for the case of $\Mpl/m = 10^{14}$, where no change in $\rho$ was 
detected to within the numerical precision of our algorithm,
as expected from dimensional analysis of Eq.\ (\ref{eq-eveqa}).  The increase
in the scale factor for these cases was within a few parts in $10^6$ of
the initial value $a(\eta_0) = 1$.  The asymptotic equation of state
plotted in Fig~\ref{fig-run18eos} is observed to be $\bar{\gamma} \sim
0.18$.  This is exactly what would be predicted for a two-fluid model
consisting of a mean field with equation of state given by
Eq.\ (\ref{eq-aseos}),
$\bar{\gamma}_{\text{{\tiny C}}} \simeq 0.0288$, and a relativistic gas
corresponding to the energy density of the $\vphi$ field, with
$\bar{\gamma}_{\text{{\tiny Q}}} \simeq 0.3333$.  The average 
$\bar{\gamma}_{\text{{\tiny Q}}} + \bar{\gamma}_{\text{{\tiny C}}} = 0.182$.

For the case $\Mpl/m = 6 \times 10^{10}$, the effect of cosmic expansion
is clearly visible in Figs.~\ref{fig-run18phi}--\ref{fig-run18rhoq}.
In Fig.~\ref{fig-run18phi}, the coherent oscillations of the mean field
for the time period $0 < \eta - \eta_0 < \; \sim 27 \tau_0$ are clearly seen
to be redshifted by the usual $1/a$ factor expected from the
Hubble damping term in Eq.\ (\ref{eq-eveqp}).  The expected asymptotic equation
of state (taking into account cosmic expansion) computed from a
simple two-fluid model is $\sim 0.133$, in agreement
with Fig.~\ref{fig-run18eos}.

Figs.~\ref{fig-run14phi}--\ref{fig-run14rhoq} show the solution
for $\Mpl/m = 6 \times 10^9$.  In this case, $1 /(H_0 \tau_1)
\sim 0.781$.  From Fig.~\ref{fig-run14fluc}, we clearly see that
cosmic expansion renders parametric amplification of quantum fluctuations 
an inefficient mechanism
of energy transfer to the inhomogeneous modes.  The very rapid oscillations
of the mean field at late times are due to the conformal time scale used
here, in which the oscillation period of the mean field decreases inversely
with $a$.  Damping of the mean field due to cosmic expansion is the dominant 
effect in Fig.~\ref{fig-run14phi}.  The power-law decrease in energy density
consistent with matter having an effective equation of state
$\bar{\gamma} \simeq 0.0288$ can be seen in
Fig.~\ref{fig-run14rho}.  At $\eta = 300 \; m^{-1}$, the ratio
$\rho_{\text{{\tiny Q}}} / \rho \sim 0.0002$, so the fraction of
energy density in the inhomogeneous modes is negligible in comparison
to the classical, mean-field contribution.   
Since the variance $\langle \vphi_{\text{{\tiny H}}}^2\rangle$ is never
large enough that it dominates the effective mass $M$, the mode functions
approximately obey the one-loop equation, in which the effective
frequency is $k^2 + a^2 (m^2 + \lambda \phih^2)$, neglecting the $a''/a$ term.
The width of the resonance can then be shown to be approximately
given by $k^2 \leq \lambda \phih^2_0 /2$.  The variance is damped by
$1/a^2$ due to cosmic redshift, so when $H^{-1} \sim \tau_1$, the variance 
never grows to be of order unity.

In addition to varying $\Mpl$, the coupling $\lambda$ was varied, with
results in agreement with Eq.\ (\ref{eq-deftau1}), showing a logarithmic
dependence of $\tau_1$ on $\lambda^{-1}$.  

\section{Discussion}
\label{sec-discussion}
The main objective of this series of papers is to investigate the 
nonperturbative and nonequilibrium quantum processes in the reheating
epoch of inflationary cosmology.  In this paper we use the minimally coupled,
quartically self-interacting scalar O$(N)$ field theory as a model for the
inflaton field, and study its nonequilibrium dynamics nonperturbatively
in a spatially flat FRW spacetime whose evolution is driven by the quantum
field.  We solve the coupled, self-consistent semiclassical Einstein
equation, mean-field equation, and conformal-mode-function equations 
numerically.  Our goal in this paper is to study the effects of spacetime
dynamics on the mean field, and parametric amplification of quantum 
fluctuations. This process of energy transfer from the mean field to the 
inhomogeneous modes is inherently nonperturbative and nonequilibrium.
It requires the use of the closed-time-path formalism
and the two-particle-irreducible effective action.
As our focus in this paper is on the parametric amplification of quantum 
fluctuations, we assume unbroken symmetry.  Our analysis is, therefore, most 
relevant to reheating in chaotic inflation scenarios. We use  the two-loop,
covariant equations for the mean field and the two-point function for the 
fluctuation field derived in the preceding paper \cite{ramsey:1997a},
and study the case of leading order in the
$1/N$ expansion, an approximation which is valid on time scales much shorter 
than the mean-free time for multiparticle scattering ($\tau_2$).  
For FRW spacetimes, we use the well-established adiabatic regularization
procedure to obtain finite expressions for the renormalized variance and 
energy-momentum tensor which enter into the mean-field equation, conformal-mode
function equations, and the semiclassical Einstein equation. In our
approach, covariant conservation of the energy-momentum tensor is preserved
at all times, as it should be.  (It should not and
need not be put in by hand, as was done in a recent study of reheating in 
a fixed background FRW spacetime \cite{boyanovsky:1997c}.)
We use the adiabatic vacuum construction (matched at conformal-past infinity)
to define the quantum state for the fluctuation field in FRW spacetime with 
asymptotic de~Sitter initial conditions; this is the most physical vacuum 
construction given the decidedly nonadiabatic conditions which prevail at 
the end of inflation. The instantaneous Hamiltonian diagonalization 
constructions used in earlier studies of reheating in curved spacetime 
\cite{boyanovsky:1996c,khlebnikov:1997a} are known to be problematic
\cite{fulling:1989a}.

We evolved the coupled dynamical equations for the mean field, variance, 
and scale factor using standard numerical methods, for time scales 
of 400 $m^{-1}$, where the initial period of mean-field oscillations is
4.11832 $m^{-1}$.
Several regimes for the parameters of the system were investigated.
From the solutions of the dynamical equations we studied the
behavior of the scale factor $a$, the mean field $\phih$, the energy
density $\rho$, pressure-to-energy-density ratio $\gamma$, and the
inhomogeneous-mode (fluctuation-field) energy density 
$\rho_{\text{{\tiny Q}}}$.  The solutions of the dynamical equations
were analyzed for a variety of values for $\Mpl \tau_0$, the parameter
which controls
the rate of cosmic expansion relative to the time scale for mean-field
oscillations in the model.  In the case of negligible cosmic expansion, 
corresponding to very small initial inflaton amplitude, the dynamics is 
identical to that seen in the group 2B (see Sec.~\ref{sec-backissu})
studies of O$(N)$ preheating in Minkowski space \cite{boyanovsky:1996b}.  
In particular, the conservation of energy and logarithmic dependence of the 
preheating time scale $\tau_1$ on the inverse coupling $\lambda^{-1}$ [as
shown in Eq.\ (\ref{eq-deftau1})] are confirmed.  For the case of 
moderate cosmic expansion, $H(\eta_0) \tau_1 \sim 10$ [where $H(\eta_0)$ is
the Hubble parameter at the initial time $\eta_0$], energy transfer
via parametric amplification of quantum fluctuations
is still efficient, and the dynamics can be understood
using the analytic results of \cite{boyanovsky:1996b} (for Minkowski
space), in terms of the conformally transformed mean-field amplitude 
$\tilde{\phih} = \phih/a$ and oscillation period $\tilde{\tau}_0 = \tau_0/a$. 
The asymptotic effective equation of state is found to be consistent with 
the prediction of a simple two-fluid description of the late-time behavior
of the system.

The most significant physical result concerns the case of rapid cosmic
expansion, where $H^{-1}(\eta_0) \simeq \tau_1$.  In this case
we find that parametric amplification of quantum fluctuations (via parametric 
resonance) is an inefficient mechanism of energy transfer to the inhomogeneous 
modes of the inflaton, because the parametric resonance effect
is inhibited both by redshifting of the mean-field amplitude and by
the redshifting of the physical momenta of the modes out of the resonance 
band.  The energy density of particles produced through parametric resonance is
in this case redshifted so rapidly that, in our model,
the term $\lambda \langle \vphi_{\text{{\tiny H}}}^2\rangle/2$ 
never grows to be of the order of the tree-level effective mass, $m^2 +
\lambda \phih^2/2$.  As the mean-field amplitude is damped ($\propto 1/a$)
due to cosmic expansion,
eventually the resonant particle production ceases, and the mean field
oscillates with a damped envelope at late times.  This leads us to the
following conclusions:  (i) On the physical level, in chaotic inflation 
scenarios with a $\lambda \Phi^4$ inflaton minimally coupled to gravity and 
with a large initial inflaton amplitude at the end of slow roll, parametric 
amplification of the inflaton's {\em own\/} quantum fluctuations is not a 
viable mechanism 
for reheating the Universe, unless the self-coupling is significantly 
increased.\footnote{In recent work on galaxy 
formation from quantum fluctuations, Calzetta, Hu, and Matacz 
\cite{calzetta:1995a,matacz:1996a} report that
$\lambda$ can be as high as $\sim 10^{-5}$.}  
This does not imply that the phenomenon of parametric
amplification of quantum fluctuations does not play a vital role in the 
``preheating'' period of inflationary cosmology, for different models 
and/or couplings.
The interesting case of a $\phi^2 \chi^2$ model will be addressed 
in a future publication \cite{ramsey:1997d}.
(ii) On a more methodological level, we conclude that a correct study of the 
reheating period in a chaotic inflation model with large inflaton amplitude 
at the onset of reheating {\em must\/} take into account the effects of 
spacetime dynamics.  This should be carried out {\it self-consistently}
using the coupled semiclassical Einstein equation and matter-field 
equations, so that no {\em ad hoc\/} assumptions need be made about 
the effective equation of state and/or the relevant time scales involved.

A full two-loop treatment of the unbroken symmetry
mean-field dynamics of the O$(N)$ field theory [which involves solving the
nonlocal, integro-differential equations (5.24) 
and (5.25) of Ref.\ \cite{ramsey:1997a}] 
includes multiparticle scattering processes, which
provide a mechanism for reheating; but they are of a qualitatively different
nature than the parametric resonance energy-transfer mechanism studied here.
In addition, the nonlocal nature of the gap equation in the full two-loop
analysis makes numerical solution of the coupled Einstein and matter equations
difficult.  In our model, multiparticle scattering occurs on a time scale 
$\tau_2$ which is significantly longer than the time scale $\tau_1$ 
for parametric amplification of quantum fluctuations.  Therefore, in this model
the leading-order, large-$N$ (collisionless) approximation is sufficient
for a study of parametric amplification of quantum fluctuations.
In addition, realistic models of inflation invariably involve couplings
of the inflaton to other fields, which provides additional mechanisms
of energy transfer away from the inflaton mean field, and into 
its (or other fields') quantum modes. 

The issues involved in a systematic study of the thermalization stage
of post-inflationary physics are more complex.  A quantum kinetic field
theory treatment taking into account multiparticle scattering is required.
The two-loop, 2PI effective action is the simplest and most generally
applicable rigorous formalism which contains the leading-order multiparticle 
scattering processes.  The leading-order, $1/N$ approximation is a 
collisionless subcase of the two-loop, 2PI effective action; it is employed 
in this study in order to obtain local dynamical equations which can be
solved numerically, and is adequate for a study of parametric amplification
of quantum fluctuations.  In addition, the growth of entropy must be understood
within the context of a physically meaningful coarse graining of the
full time-reversal-invariant quantum dynamics of the field theory.
This, together with a first-principles analysis of the thermalization stage,
is currently under investigation 
\cite{ramsey:1997c,ramsey:1997d,ramsey:1997b}.

\section{Acknowledgments}
This work was supported in part by NSF Grant No.\ PHY94-21849\@.  
Part of this work was carried out at the Institute for Advanced Study, 
Princeton, NJ where B.L.H. was a Dyson Visiting Professor.
We enjoyed the hospitality of Los Alamos National Laboratory during the
Santa Fe workshop on {\em Nonequilibrium Phase Transitions\/} sponsored by
the Center for Nonlinear Studies, and thank Drs.\ D.~Boyanovsky, E.~Calzetta,
and E.~Mottola for general discussions.  
S.A.R. wishes to thank G.~Stephens, J.~Simon, A.~Raval,
and Professor T.~Jacobson for illuminating discussions.  
Finally, we wish to thank Professor E.~Williams at the University of Maryland
and D.~Benton at the University of Pennsylvania for the use of their
respective groups' computer facilities.


\begin{table}[htb]
\begin{center}
\begin{tabular}{|l||l|l|l|l||l|}
Figures & $\phih(\eta_0)$ & $\lambda$ & $\Mpl$ & $K$ & $H^{-1}(\eta_0)$ \\ 
\hline
\ref{fig-run19phi}--\ref{fig-run19rhoq} & $2 \times 10^7$ & $1 \times
10^{-14}$ & $1 \times 10^{14}$ & 50.0 & $1.7275 \times 10^6$ \\
\ref{fig-run33phi}--\ref{fig-run33rhoq} & $2 \times 10^7$ & $1 \times
10^{-14}$ & $1 \times 10^{12}$ & 50.0 & $1.7275 \times 10^4$ \\
\ref{fig-run18phi}--\ref{fig-run18rhoq} & $2 \times 10^7$ & $1 \times
10^{-14}$ & $6 \times 10^{10}$ & 50.0 & $1.0364 \times 10^3$ \\
\ref{fig-run14phi}--\ref{fig-run14rhoq} & $2 \times 10^7$ & $1 \times
10^{-14}$ & $6 \times 10^9$ & 50.0 & 103.65 
\end{tabular}
\end{center}
\caption{Values of parameters for numerical solutions of Eqs.\
(\ref{eq-eveqp}), (\ref{eq-eveqa}), (\ref{eq-eveqmf}) in units where $m=1$,
and the corresponding figures in which the solutions are plotted.  The
$H^{-1}(\eta_0)$ column is the initial inverse Hubble constant, which
gives the initial time scale for cosmic expansion.
Figs.~\ref{fig-run19phi}--\ref{fig-run14rhoq} plot 
the resulting solutions.}
\label{table-runparms}
\end{table}


\end{document}